\documentclass{egpubl} 

 \JournalSubmission    
 \electronicVersion 

\ifpdf \usepackage[pdftex]{graphicx} \pdfcompresslevel=9
\else \usepackage[dvips]{graphicx} \fi

\PrintedOrElectronic

\usepackage{t1enc,dfadobe}

\usepackage{egweblnk}
\usepackage{cite}

\usepackage{verbatim} 
\usepackage{pifont} 

\usepackage{amssymb,float,amsmath,tikz,rotating,bm}
\usepackage[normalem]{ulem}
\usepackage[utf8]{inputenc} 

\graphicspath{{./arxiv_figures/},{./}}

\usepackage{xcolor}
\definecolor{tc}{cmyk}{0, 0.4, 1, 0}
\definecolor{tr}{cmyk}{0.8, 0, 0.8, 0}
\definecolor{nr}{rgb}{0, 0.4, 0.1}

\newcommand\redsout{\bgroup\markoverwith{\textcolor{red}{\rule[0.5ex]{2pt}{0.4pt}}}\ULon}
\newcommand\greensout{\bgroup\markoverwith{\textcolor{green}{\rule[0.5ex]{2pt}{0.4pt}}}\ULon}

\DeclareMathOperator*{\argmin}{argmin}

\newcommand{\eg}{{e.g. }}
\newcommand{\etal}{{et al.}}
\newcommand{\ie}{{i.e. }}

\title[On Demand Solid Texture Networks]%
      {On Demand Solid Texture Synthesis Using Deep 3D Networks}

\author[J. Gutierrez \& J. Rabin \& B. Galerne \& T. Hurtut]
{\parbox{\textwidth}{\centering  
J. Gutierrez$^{1}$ \ J. Rabin$^{2}$ \ B. Galerne$^{3}$ \ T. Hurtut$^{1}$ } \\
{\parbox{\textwidth}{\centering $^1$Polytechnique Montr\'{e}al, Canada\\
         $^2$Normandie Univ. UniCaen, ENSICAEN, CNRS, GREYC, France\\
        $^3$Institut Denis Poisson, Universit\'{e} d'Orl\'{e}ans, Universit\'{e} de Tours, CNRS, France }}
}

\begin{document}


\maketitle
\begin{abstract}
This paper describes a novel approach for on demand volumetric texture synthesis based on a deep learning framework that allows for the generation of high quality 3D data at interactive rates.
Based on a few example images of textures, a generative network is trained to synthesize coherent portions of solid textures of arbitrary sizes that reproduce the visual characteristics of the examples along some directions.
To cope with memory limitations and computation complexity that are inherent to both high resolution and 3D processing on the GPU, only 2D textures referred to as ``slices'' are generated during the training stage. 
These synthetic textures are compared to exemplar images \emph{via} a perceptual loss function based on a pre-trained deep network.
The proposed network is very light (less than 100k parameters), therefore it only requires sustainable training (\ie few hours) and is capable of very fast generation (around a second for $256^3$ voxels) on a single GPU.
Integrated with a spatially seeded PRNG the proposed generator network directly returns an RGB value given a set of 3D coordinates.
The synthesized volumes have good visual results that are at least equivalent to the state-of-the-art patch based approaches. They are naturally seamlessly tileable and can be fully generated in parallel.



\end{abstract}

\section{Introduction}

{\let\thefootnote\relax\footnote{{This document is a lightweight preprint version of the journal article published in Computer Graphics Forum. DOI:10.1111/cgf.13889 (\URL{https://doi.org/10.1111/cgf.13889}) 
Another preprint version with uncompressed images is available here: \URL{https://hal.archives-ouvertes.fr/hal-01678122v3}.
}}}

2D textures are ubiquitous in 3D graphics applications. Their visual complexity combined with a widespread availability allows for the enrichment of 3D digital objects' appearance at a low cost. In that regard, solid textures, which are the 3D equivalent of stationary raster images, offer several visual quality advantages over their 2D counterparts. Solid textures eliminate the need for a surface parametrization and its accompanying visual artifacts. They produce the feeling that the object was carved from the texture material. Additionally, the availability of consistent volumetric color information allows for the interactive manipulation including object fracturing or cut-away views to reveal internal texture details.
However, unlike scanning a 2D image, digitization of volumetric color information is impractical. 
As a result, most of the existent solid textures are synthetic.

One early  way to generate solid textures is by \emph{procedural generation}~\cite{peachey1985solid,perlin1985image}. 
In procedural methods the color of a texture at a given point only depends on its coordinates. This allows for a localized evaluation to generate only the required portions of texture at a given moment. We refer to this characteristic as \emph{on demand} evaluation. 
Procedural methods are indeed fast and memory efficient. Unfortunately finding the right parameters of a procedural model to synthesize a given texture requires a high amount of expertise and trial and error. Photo-realistic textures with visible elemental patterns are particularly hard to generate by these methods.

In order to give up the process of empirically tuning the model for a given texture,  several \emph{by-example} solid texture synthesis methods have been proposed~\cite{heeger1995pyramid,kopf2007solid,dong2008lazy,chen2010highquality}. These methods are able to generate solid textures that share the visual characteristics of a given \emph{target} 2D texture example through all the cross-sections in a given set of slicing directions (inferring the 3D structure given the constrained directions). 
Although they do not always deliver perfect results, by-example methods can be used to approximate the characteristics of a broad set of 2D textures. 
One convenient approach to synthesize textures by-example is called \emph{lazy synthesis}~\cite{dong2008lazy,zhang2011sketch}. It consists of an \emph{on demand} synthesis \ie synthesizing only voxels at a specific location in contrast to generating a whole volume of values. 
%
%
Current lazy synthesis methods tend to deliver lower visual quality.

Several solid texture synthesis methods arise as the extrapolation of a 2D model. While some approaches can be trivially expanded (\eg procedural), others require a more complex strategy, \eg pre-computation of 3D neighborhoods~\cite{dong2008lazy} or averaging of synthesized slices~\cite{heeger1995pyramid}. Currently, 2D texture synthesis methods based on Convolutional Neural Networks (CNN)~\cite{gatys2015cnn,Ulyanov17improved} roughly define the state-of-the-art in the 2D literature.
\emph{Texture networks} methods~\cite{ulyanov2016texturenets,Ulyanov17improved,li2017diversified} stand out thanks to their fast synthesis times. 

In this work we introduce a  CNN generator capable of synthesizing 3D textures on demand.
On demand generation, also referred to as \emph{local evaluation} in the literature (\eg\cite{wei2003order,lefebvre2005parallel,dong2008lazy}), 
is critical for solid texture synthesis because storing the values of a whole solid texture at a useful resolution is prohibitive for most applications.
It provides the ability to generate on demand only the needed portions of the solid texture. This speeds up texturing surfaces and saves memory.
It was elegantly addressed for patch based methods for 2D and solid textures \cite{wei2003order,lefebvre2005parallel, dong2008lazy, zhang2011sketch}. 
However, none of the aforementioned CNN based methods study this property.

There are substantial challenges in designing a CNN to accomplish such task.
On demand evaluation requires the association of the synthesis to a coordinate system and determinism to enforce spatial coherence.
For the training, first we need to devise a way to evaluate the quality of the synthesized samples. Since the seminal work of~\cite{gatys2015cnn,johnson2016Perceptual}, most 2D texture synthesis methods such as ~\cite{ulyanov2016texturenets,Ulyanov17improved,zhou2018nonstationary,yu2019texture} use the activations in the hidden layers of VGG-19~\cite{Simonyan14c} as a \emph{descriptor} network to characterize the generated samples and evaluate their similarity with the example. There is, however, not a volumetric equivalent of such image classification network that we could use off-the-shelf. 
Second, we need a strategy to surmount the enormous amount of memory demanded by the task.

We propose a compact solid texture generator model based on CNN capable of on demand synthesis at interactive rates. On training, we assess the samples' appearance via a volumetric loss function that compares  slices of the generated textures to the target image.
We exploit the stationarity of the model to propose a fast and memory efficient single-slice training strategy. This allows us to use target examples at higher resolutions than those in the current 3D literature. 
The resulting trained network is simple, lightweight, and powerful at reproducing the visual characteristics of the example on the cross sections of the generated volume along one or up to three directions.

\section{Related works}\label{Sec:Related_work}

To the best of our knowledge, the method proposed here is the first to employ a CNN to generate solid textures. 
Here we briefly outline the state-of-the-art on solid texture synthesis, then we describe some successful applications of using CNNs to perform 2D texture synthesis. 
Finally we mention relevant CNN methods that use 2D views to generate 3D objects. 
\subsection{Solid texture synthesis}
Procedural methods~\cite{perlin1985image,peachey1985solid} are quite convenient for computer graphics applications thanks to their real time computation and on demand evaluation capability.
Essentially one can add texture to a 3D surface by directly evaluating a function  given the coordinates of only the required (visible) points in the 3D space. 
The principle is as follows for texture generation on a surface: a colormap function, such as a simple mathematical expression, is evaluated at each of the  visible 3D points.
In~\cite{perlin1985image}, authors use pseudo-random numbers that depend on the coordinates of the local neighborhood of the evaluated point which ensures both the random aspect  of the texture and its spatial coherence.
Creating realistic textures with a procedural noise is a trial and error process that necessitates technical and artistic skills.
Some procedural methods alleviate this process by automatically estimating their parameters from an example image~\cite{ghazanfarpour1995Spectral,GLLD_Gabor_noise_by_example_2012,Gilet_et_al_2014_local_random_phase_noise_2014,galerne2017texton}. 
However, they only deal with surface textures and most photo realistic textures are still out of the reach of these methods.

Most example-based solid texture synthesis methods aim at generating a full block of voxels whose cross-sections are visually similar to a respective texture example. These methods are in general fully automated and they are able to synthesize a broader set of textures.  
The texture example being only 2D, a  prior model is required to infer 3D data. Some methods~\cite{heeger1995pyramid,dischler1998anisotropic,qin2007aura} employ an iterative \emph{slicing} strategy where they alternate between independent 2D synthesis of the slices and 3D aggregation.
Another strategy starts with a solid noise and then iteratively modifies its voxels by assigning them a color value depending on a set of \emph{coherent} pixels from the example. Wei~\cite{wei2003multiplesources} uses the 2D neighborhood of a pixel also called \emph{patch} to assess coherence, then, the set of contributing pixels is formed by the closest patch along each axis of the solid. Finally, the assigned color is the average of the contributing pixels.
Dong~\etal~\cite{dong2008lazy} determine coherence using three overlapping 2D neighborhoods (\ie forming a 3D neighborhood) around each voxel and find only the best match among a set of precomputed candidates.

The example-based solid texture synthesis methods that achieve the best visual results in a wider category of textures \cite{kopf2007solid,chen2010highquality} involve a patch-based global optimization framework \cite{kwatra2005texture} governed by a statistical matching strategy. 
This improves the  robustness to failure.
These methods, however, require high computation times, which limits them to low resolution textures (typically $256^2$ input examples), and are incapable of on demand evaluation which limits their usability. 
Regarding speed, the  patch-based method proposed by Dong~\etal~\cite{dong2008lazy} allows for fast on demand evaluation, thus allowing for visual versatility and practicality. Here the patch-matching strategy is accelerated via the pre-processing of compatible 3D neighborhoods accordingly to the examples given along some axis. This preprocessing is a trade-off between visual quality and speed as it reduces the richness of the synthesized textures. 
Thus, their overall visual quality is less satisfactory than the one of the  optimization methods previously mentioned.

\subsection{Neural networks on texture synthesis}~%
Our method builds upon previous work on example-based 2D texture synthesis using convolutional neural networks. We distinguish two types of approaches: \emph{image optimization} and \emph{feed-forward texture networks}. 

\paragraph*{Image optimization}~
Image optimization methods were inspired from previous statistical matching approaches~\cite{portilla2000parametric} and use a variational framework which aims at generating a new image that matches the features of an example image. 
The role of CNN in this class of methods is to deliver a powerful characterization of the images. 
It typically comes in the form of feature maps at the internal layers of a pretrained deep CNN~\cite{Simonyan14c} namely VGG-19.
Gatys \etal~\cite{gatys2015cnn} pioneered this approach for texture synthesis, by considering the discrepancy between the feature maps of the synthesized image and the example ones. 
More precisely for texture synthesis where one has to take into account spatial stationarity, the corresponding \emph{perceptual loss} as coined later by \cite{johnson2016Perceptual} is the Frobenius distance of the Gram matrices of CNN features at different layers. 
Starting from a random initialization, the input image is then optimized via a stochastic gradient descent algorithm, where the gradient is computed using back-propagation through the CNN.
Since then, several variants have built on this framework to improve the quality of the synthesis:
for structured textures by adding a Fourier spectrum discrepancy~\cite{liu2016texture}
or using co-occurence information computed between the feature maps and their translation~\cite{berger2017incorporating};
for non-local patterns by considering spatial correlation and smoothing~\cite{sendik2017deep};
for stability by considering a histogram matching loss and smoothing\cite{wilmot2017stable}.
These methods deliver good quality and high resolution results as they can process images of resolutions up to $1024^2$ pixels. 
Their main drawback comes from the optimization process itself, as it requires several minutes to generate one image. 
Implementing local evaluation on these methods is infeasible since they use a global optimization scheme as for patch-based texture optimization methods~\cite{kwatra2005texture,kopf2007solid}. 
Extension to dynamic texture synthesis has also been proposed in~\cite{tesfaldet2018}.
The textured video is optimized using a perceptual loss combined with a loss based on estimated optical flow to take into account the time dimension.

\paragraph*{Feed-forward texture networks}~ 
Feed-forward  networks approaches were introduced by Johnson~\etal~\cite{johnson2016Perceptual} for style transfer and Ulyanov~\etal~\cite{ulyanov2016texturenets} for texture synthesis. In the latter, they train an actual generative CNN to synthesize texture samples that produce the same visual characteristics as the example.
These methods use the loss function in \cite{gatys2015cnn} to compare the visual characteristics between the generated and example images.
However, instead of optimizing the generated output image, the training aims at tuning the parameters of the generative network. 
Such optimization can be more demanding since there is no spatial regularity shared across iterations as for the optimization of a single image. 
However this training phase only needs to be done once for a given input example.
This is achieved in practice using back propagation and a gradient-based optimization algorithm using batches of noise inputs. Once trained, the generator is able to quickly generate samples similar to the input example by forward evaluation. 
Originally these methods train one network per texture sample. Li \etal~\cite{li2017diversified} proposed a large network architecture and a training scheme to allow one network to have the capacity to generate and mix several different textures. 
By improving the capacity of the generator network Li \etal~\cite{li2017diversified} and Ulyanov~\etal~\cite{Ulyanov17improved} methods reached a modestly higher visual quality but found that the synthesized textures did not change sufficiently for different inputs.
In order to prevent the generator from producing identical images they were forced to incorporate a term that encourages diversity in the objective function.
Feed-forward texture networks methods generally produce results with slightly lower visual quality than the image optimization methods, however they exhibit faster generation times.
Visual quality aside, the earlier architecture model proposed by Ulyanov~\etal~\cite{ulyanov2016texturenets} holds several advantages over more recent feed-forward methods: it does not require a diversity term, it is able to generate images of arbitrary sizes and it is significantly smaller in terms of number of parameters.
Moreover, as we show in Section~\ref{Sec:Generator} this framework can be customized to allow on demand evaluation.

Other approaches achieve texture synthesis as a feed-forward evaluation of a generator network using a different training strategy.
Bergmann~\etal~\cite{bergmann17apsGAN} train a generator network without using the perceptual loss. Instead they employ a generative adversarial network (GAN) framework \cite{goodfellow2014generative} with a purely convolutional architecture \cite{radford2016unsupervised} to allow for flexibility on the size of the samples to synthesize. This method shares the advantages of feed-forward networks regarding evaluation but is based in a more complex training scheme where two cascading networks have to be optimized using an \emph{adversarial loss} which can affect the quality on different texture examples. 
Zhou~\etal~\cite{zhou2018nonstationary} use a combination of perceptual and adversarial loss and achieve impressive results for the synthesis of non-stationary textures. This is a more general problem seldom addressed in the literature and it requires extra assumptions about the behavior of the texture at hand.
Similarly, Yu~\etal~\cite{yu2019texture} train a CNN to perform texture mixing using a hybrid approach, combining the perceptual loss and adversarial loss to help the model produce plausible new textures.
Finally, Li~\etal~\cite{li2017universalstyle} proposed another strategy that leverages auto-encoders~\cite{yann1987modeles,bourlard1988auto}. They use truncated versions of VGG-19 network as encoders that map images to a feature space. Then they design decoder networks that generate images from such features. During the training stage, this generator is optimized to invert the encoder by trying to generate images that match encoded images from a large dataset of natural images.
During synthesis, a random image and an example image are first encoded; random features are matched to the target ones using first and second order moment matching, and then fed to the decoder to generate a random texture, without requiring a specific training for this example.
While very appealing, this approach is difficult to adapt to 3D texture synthesis where such large dataset is not available.
Moreover, the quality is not as good as for previously mentioned methods~\cite{gatys2015cnn,ulyanov2016texturenets}.

\subsection{Neural networks for volumetric object generation}~%
A related problem in computer vision is the generation of binary volumetric objects from 2D images~\cite{gwak2017weakly2,jimenez2016unsupervised,yan2016perspective}. 
Similarly to our setting, these approaches rely on unsupervised training with a loss function comparing 3D data to 2D examples. 
However these methods do not handle color information and only produce low resolution  volumes ($32^3$).

\section{Overview}\label{Sec:Overview}

\begin{figure*}[ht]
\centering
\includegraphics[width = 16cm]{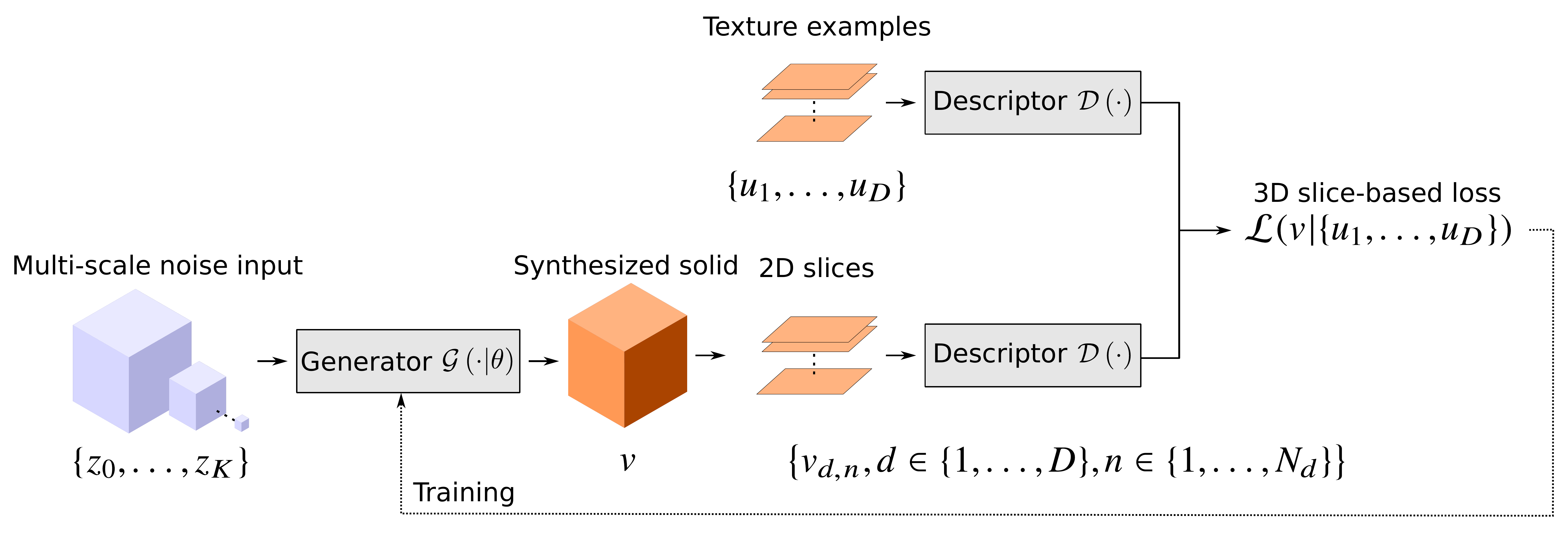}
\caption{Training framework for the proposed CNN Generator network $\mathcal{G(\cdot|\mathit{\theta)}}$ with parameters $\theta$. The Generator
processes a multi-scale noise input $Z$ to produce a solid texture $v$. The loss $\mathcal{L}$ compares, for each direction $d$, the feature statistics induced by the example $u_d$ in the layers of the pre-trained Descriptor network $\mathcal{D}(\cdot)$ to those induced by each slice of the set $\left\{v_{d,1},\ldots,v_{d,N_d}\right\}$. The training iteratively updates the parameters $\theta$  to reduce the loss. We show in Section~\ref{Sec:Training} that we can perform training by only generating single-slice solids instead of full cubes.}
\label{Fig:Overview}
\end{figure*}

Figure~\ref{Fig:Overview} outlines the proposed method. 
We perform solid texture synthesis using the convolutional neural \emph{generator} network $\mathcal{G}$ detailed in Section~\ref{Sec:Generator}. The generator with learnable parameters $\theta$, takes a multi-scale volumetric noise input $Z$ and processes it to produce a color solid texture $v = \mathcal{G}(Z|\theta)$. 
The proposed model is able to perform on demand evaluation which is a critical property for solid texture synthesis algorithms. On demand evaluation spares computations and memory usage as it allows the generator to only synthesize the voxels that are visible. 

The desired appearance of the samples $v$ is specified in the form of a view dissimilarity term for each direction.
The 3D generated samples $v$ are compared to $D\in\{1,2,3\}$ example images $\{u_1,\ldots,u_D\}$ that correspond to the desired view along $D$ directions among the 3 canonical directions of the Cartesian grid.
The generator \emph{learns} to sample solid textures 
from the visual features extracted in the examples \emph{via} the optimization of its parameters $\theta$.
To do so, we formulate a volumetric slice-based loss function $\mathcal{L}$. It measures how appropriate the appearance of a solid sample is by comparing its 2D slices $v_{d,n}$ ($n$\textsuperscript{th} slice in along the $d$\textsuperscript{th} direction) to each corresponding example $u_d$. 
Similar to previous methods, the comparison is carried out in the space of features from the descriptor network $\mathcal{D}$ based on VGG-19.   

The training scheme, detailed in Section~\ref{Sec:Training}, involves the generation of batches of solid samples which would \emph{a priori} require a prohibitive amount of memory if relying on classical optimization approach for CNN. We overcome this limitation thanks to the stationarity properties of the model. We show that training the proposed model only requires the generation of single slice volumes along the specified directions.
Section~\ref{Sec:Results} presents experiments and comparative results. 
Finally, in Section~\ref{Sec:Limitations} we discuss the current limitations of the model.

\section{On demand evaluation enabled CNN generator}\label{Sec:Generator}

The architecture of the proposed CNN generator is summarized in Figure~\ref{Fig:Generator2} and detailed in Subsection~\ref{SubSec:Architecture}. 
The generator applies a series of convolutions to a multi-scale noise input to produce a single solid texture. It is inspired by Ulyanov~\etal~\cite{ulyanov2016texturenets} model, which stands out for on demand evaluation thanks to its small number of parameters and its local dependency between the output and input. 
It is based on a multi-scale approach inspired itself from the human visual system that has been successfully used in many computer vision applications, and in particular for texture synthesis\cite{heeger1995pyramid,debonet1997multiresolution,wei_levoy_2000,portilla2000parametric,kwatra2005texture,rabin2011wasserstein,galerne2018texture}.

This fully convolutional generator allows the generation of on demand box-shaped/rectangular volume textures of an arbitrary size (down to a single voxel) controlled by the size of the input. Formally, given an infinite noise input it represents an infinite texture model. 
A first step to achieve on demand evaluation is to control the size of the generated sample. 
To do so, we unfold the contribution of the values in the noise input to each value in the output of the generator. 
This dependency is described in Subsection~\ref{SubSec:Spatial_dependency}.
Then, on demand voxel-wise generation is  achieved
thanks to the multi-scale shift compensation detailed in Subsection~\ref{SubSec:Shift_compensation}.  
The resulting generator is able to synthesize coherent and expandable portions of a theoretical infinite texture.

\subsection{Architecture}\label{SubSec:Architecture}
\begin{figure*}[ht]
\centering
\includegraphics[width=17.5cm]{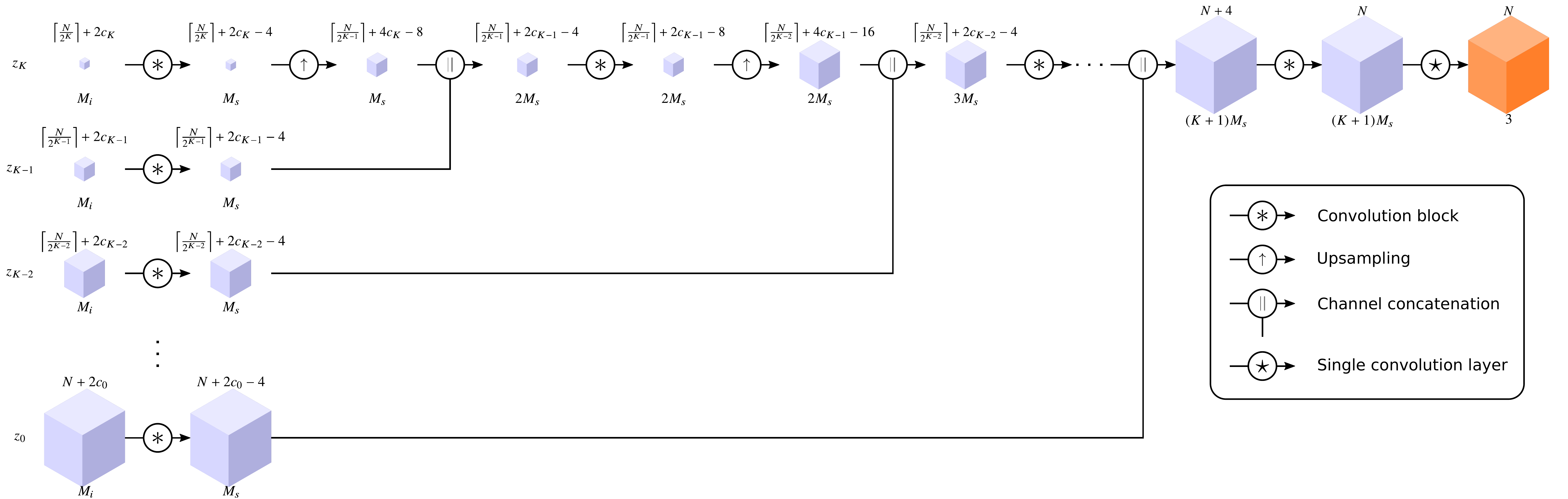}
\caption{Schematic of the network architecture. 
Noise input $Z = \{z_0,\ldots,z_K\}$ at $K+1$ different scales is processed using convolution operations and non-linear activations. 
The information at different scales is combined using upsampling and channel concatenation. $M_i$ indicates the number of input channels and $M_s$ controls the number of channels at intermediate layers.  For simplicity we consider a cube shaped output with spatial size of $N_1=N_2=N_3=N$.
For each intermediate cube the spatial size is indicated above 
and the number of channels below.}
\label{Fig:Generator2}
\end{figure*}
The generator produces a solid texture $v = \mathcal{G}(Z|\theta)$ 
from a set of multi-channel volumetric white noise  inputs $Z = \{z_0,\ldots,z_K\}$. 
The spatial dimensions of $Z$ directly control the size of the generated sample.
The process of transforming the noise $Z$ into a solid texture $v$ is depicted in Figure~\ref{Fig:Generator2}. It follows a multi-scale  architecture built upon three main operations: convolution, concatenation, and upsampling.
Starting at the coarsest scale, the 3D noise sample is processed with
a set of \emph{convolutions} followed by an \emph{upsampling} to reach the next scale. 
It is then concatenated with the independent noise sample from the next scale, itself also processed with a set of convolutions.
This process is repeated $K$ times before a final single convolution layer that maps the number of channels to three to get a color texture.  
We now detail the three different blocks of operations used in the generator network. 

\paragraph*{Convolution block}~A convolution block groups a sequence of three ordinary 3D convolution layers, each of them followed by a batch-normalization and a leaky rectified linear unit function. Considering $M_{in}$ and $M_{out}$  channels in the input and output respectively, the first convolution layer carries out the shift from $M_{in}$ to $M_{out}$. The following two layers of the block have
$M_{out}$ channels in both the input and the output. The size of the kernels is $3\times3\times3$ for the first two layers and  $1\times1\times1$ for the last. Contrary to \cite{ulyanov2016texturenets} and in order to enable on demand evaluation (see Subsection~\ref{SubSec:Spatial_dependency}), here the convolutions are computed densely and without padding, thus discarding the edge's values. Applying one convolution block with these settings to a volume reduces its size by 4 values per spatial dimension. 

\paragraph*{Upsampling}~An upsampling performs a 3D nearest neighbor upsampling by a factor of 2 on each spatial dimension (\ie each voxel is replicated 8 times). 

\paragraph*{Channel concatenation}~ This operation first applies a batch normalization operation and then concatenates the channels of two multi-channel volumes having the same spatial size. If different, the biggest volume is cropped to the size of the smallest one. 

The learnable parameters $\theta$ of the generator are: the convolution's kernels and bias, and the batch normalization layers' weight, bias, mean and variance.  The training of these parameters is discussed in Section~\ref{Sec:Training}.

\subsection{Spatial dependency}\label{SubSec:Spatial_dependency}
Forward evaluation of the generator is deterministic and local, \ie each value in the output only depends on a small number of neighboring values in the multi-scale input. By handling the noise inputs correctly, we can feed the network separately with two \emph{contiguous} portions of noise to  synthesize textures that can be tiled seamlessly. 
Current 2D CNN methods~\cite{ulyanov2016texturenets,li2017diversified} perform padded convolutions, not addressing on demand evaluation capabilities.
Let  us notice that a perfect tiling between two different samples can only be achieved by using convolutions without padding. Therefore we discard the values on the borders where the operation with the kernel cannot be carried out completely.

When synthesizing a sample, the generator is fed with an input that takes into account the neighboring dependency values. Those extra values are progressively processed and discarded in the convolutional layers: for an output of size $N_1\times N_2\times N_3$ the size of the input at the $k$-th scale has to be $(\frac{N_1}{2^k} + 2c_k) \times (\frac{N_2}{2^k} + 2c_k) \times (\frac{N_3}{2^k} + 2c_k)$ 
where $c_k$ denotes the additional values along each spatial dimension required due to the dependency.
The size in any spatial dimension $N$ can be any positive integer (provided the memory is large enough for synthesizing the volume).

These additional values $c_k$ depend on the network architecture. In our case, thanks to the symmetry of the generator, the coefficients $c_k$ are the same along each spatial dimension.
Each convolutional block requires additional support of two values on each side along each dimension and each upsampling cuts down the dependency by two (taking the smallest following  integer when the result is fractional).
At scale $k=0$ there are two convolution blocks, therefore $c_0 = 4$. For subsequent scales $c_k = \lceil(c_{k-1}-2)/2\rceil + 4$, 
except for the coarsest scale $K$ where there is only one convolution block and therefore $c_K = \lceil(c_{K-1}-2)/2\rceil + 2$. 
For example, in order to generate a single voxel, the spatial dimensions of the noise inputs must be $9^3$ for $z_0$, $11^3$ for $z_1$, $13^3$ for $z_2$ to $z_4$, and $9^3$ for $z_5$, which totals $9380$ random values.

\subsection{Multi-scale shift compensation for on demand evaluation}\label{SubSec:Shift_compensation}
On demand generation is a standard issue in procedural synthesis~\cite{perlin1985image}.
The purpose is to generate consistently any part of an infinite texture model.
It enables the generation of small texture blocks separately, whose localization depends on the geometry of the object to be texturized, instead of generating directly a full volume containing the object.
For procedural noise, this is achieved using a reproducible Pseudo-Random Number Generator (PRNG) seeded with spatial coordinates.
 
In our setting, we enforce spatial consistency by generating the multi-scale noise inputs using a xorshift PRNG algorithm~\cite{marsaglia2003xorshift} seeded with values depending on the volumetric coordinates, channel and scale, similarly to~\cite{galerne2017texton}. Thus, our model only requires the set of reference 3D coordinates and the desired size to generate spatially coherent samples. Given a reference coordinate $n_0$ at the finest scale in any dimension, the corresponding coordinates at the $k$-th scale are computed as $n_k = \lfloor\frac{n_0}{2^k}\rfloor$. 
These corresponding noise coordinates need to be aligned in order to ensure the coherence between samples generated separately.

Feeding the generator with the precise set of coordinates of noise at each scale is only a first step to successfully synthesize compatible textures.
Recall that the model is based on combinations of transformed noises at different scales (see Figure~\ref{Fig:Generator2}), therefore requiring  special care regarding upsampling to preserve the coordinate alignment across scales, \ie which coordinate $n_k$ at scale $k$ must be associated to a given coordinate $n_0$ at the finest scale $k=0$.
Indeed, after every upsampling operation, observe that each value is repeated twice along each spatial dimension, \emph{pushing} the rest of the values spatially. 
Depending on the coordinates of the reference voxel being synthesized, this \emph{shift} of one position can disrupt the coordinate alignment with the subsequent scale. Therefore, the generator network has to compensate accordingly before each concatenation.  
 
For $K=5$ upsamplings, one of the $2^K = 32$ combinations of compensation shifts has to be properly done for each dimension to synthesize a given voxel.
In order to consider these compensation shifts, we make the generator network aware of the coordinate of the sample at hand. In our implementation the reference is on the vertex of the sample closest to the origin.
Given the final reference coordinate $n_0$ of the voxel (in any spatial dimension), the generator deduces the set of shifts recursively  from the following relation,
$n_{k-1} = 2n_k + s_{k}$,
where $n_k$ is the spatial reference coordinate used to generate the noise at scale $k \in \{1,\dots, K\}$, and $s_k \in \{0,1\}$ is the shift value used after the $k$\textsuperscript{th} upsampling operation.
At evaluation time, the generator sequentially applies the set of shifts before every corresponding concatenation operation.

\section{Training}\label{Sec:Training}

Here we detail our approach to obtain the parameters $\theta_u$ that drive the generator network to synthesize solid textures specified by the example $u$. 
Like current texture networks methods, we leverage the power of existing  training frameworks to optimize the parameters of the generator. Typically an iterative gradient-based algorithm is used to minimize a loss function that measures how different the synthetic and target textures are. 

However a first challenge facing the training of the solid texture generator is to devise a discrepancy measure between the solid texture and the 2D examples.
In Subsection~\ref{SubSec:3D_loss} we propose a 3D slice-based loss function that collects the measurements produced by a set of 2D comparisons between 2D slices of the synthetic solid and the examples.
We conduct the 2D comparisons similarly to the state-of-the-art methods,
using the perceptual loss function \cite{gatys2015cnn, johnson2016Perceptual, Ulyanov17improved}. 

The second challenge comes from the memory requirements during training. Typically the optimization algorithm estimates a descent direction by applying backpropagation on the loss function evaluated on a batch of samples. In the case of solid textures, each volumetric sample occupies a large amount of memory, which makes the batch processing impractical. Instead, we show in Subsection~\ref{SubSec:Slice_training} that, thanks to the stationary properties of our generative model, we can carry out the training using batches of single slice solid samples.

\subsection{3D slice-based loss}\label{SubSec:3D_loss}
For a color solid $v\in\mathbb{R}^{N_1\times N_2\times N_3\times 3}$,  we  denote by $v_{d,n}$ the $n$\textsuperscript{th} 2D slice of the solid $v$ orthogonal to the $d$\textsuperscript{th} direction. 
Given a number $D \le 3$ of slicing directions and the corresponding example images $\{u_1,\ldots,u_D\}$, 
we propose the following slice-based loss
\begin{equation} \label{eq:total_slices_loss}
	\mathcal{L}(v | \{u_1, \dots, u_D\}) = 
	\sum_{d = 1}^{D} \frac{1}{N_d} 
	\sum_{n = 0}^{N_d-1}
	\mathcal{L}_{2} (v_{d,n},u_d),
\end{equation}
where  $\mathcal{L}_{2}(\cdot,u)$ is a 2D loss that computes the similarity between an image and the example $u$.  

We use the 2D perceptual loss $\mathcal{L}_2$ from \cite{gatys2015cnn}, which proved successful for training the CNNs~\cite{johnson2016Perceptual,ulyanov2016texturenets}. It compares the Gram matrices of the VGG-19 feature maps of the synthesized and example images.
The feature maps result from the evaluation of the descriptor network $\mathcal{D}$  on an image, \ie $\mathcal{D}:x \in \mathbb{R}^{N_1 \times N_2 \times 3}\mapsto \{F^l(x) \in \mathbb{R}^{N^l\times M^l}\}_{l \in L}$, where $L$ is the set of considered VGG-19 layers, each layer $l$ having $N^l$ spatial values and $M^l$ channels. 
For each layer $l$, the Gram matrix $G^l\in\mathbb{R}^{M^l\times M^l}$ is computed from the feature maps as
\begin{equation}
	G^l(x) = \frac{1}{N^l} F^l(x)^T F^l(x), 
\end{equation}
where $T$ indicates the transpose of a matrix.
The 2D loss between the input example $u_d$ and a slice $v_{d,n}$ is then defined as
\begin{equation}\label{eq:slice_loss}
	\mathcal{L}_{2}(v_{d,n},u_d) = 
	\sum_{l\in L} \ \frac{1}{{(M^{l})}^2} \ {\left\lVert G^l(v_{d,n}) - G^l(u_d)\right\rVert}_F^2, 
\end{equation}
where $\|\cdot\|_{F}$ is the Frobenius norm.
Observe that the Gram matrices are computed along spatial dimensions to take into account the stationarity of the texture.
Those Gram matrices encode both first and second order information of the feature distribution (covariance and mean).

\subsection{Single slice training scheme}\label{SubSec:Slice_training}
Formally, training the generator $\mathcal{G}(Z|\theta)$ with parameters $\theta$ corresponds to minimizing the expectation of the loss in Equation~\ref{eq:total_slices_loss}
given the set of examples $\{u_1,\ldots,u_D\}$,
\begin{equation}\label{eq:training}
	\theta_{u} \in
	\argmin_{\theta} 
	\quad
	\mathbb{E}_{Z}
	\left[\mathcal{L}(\mathcal{G}(Z|\theta), \,\{u_1,\ldots,u_D\})\right],
\end{equation}
where $Z$ is a multi-scale noise, independent and identically distributed from a uniform distribution in the interval $[0,1]$. 
Note that each scale $z_0,\dots,z_K$ of $Z$ is stationary.
The generator on the other hand induces a non stationary behavior on the output due to upsampling operations. 
When upsampling a stationary signal by a factor 2 with a nearest neighbor interpolation the resulting process is only invariant to translations of multiples of two.
Because our model contains $K$ volumetric upsampling operations, the process is translation invariant by multiples of $2^K$ values on each axis. 
Considering the $d$-th axis, for any coordinate at the scale of the generated sample $n = 2^Kp+q$ with
$q\in\{0,\ldots,2^K-1\}$, the statistics of the slice $\mathcal{G}(Z|\theta)_{d,n}$ only depend on the value of $q$, therefore
\begin{equation}
\mathbb{E}_{Z} \left[\mathcal{L}_{2} (\mathcal{G}(Z|\theta)_{d,n},u_d)\right] =
\mathbb{E}_{Z} \left[\mathcal{L}_{2} (\mathcal{G}(Z|\theta)_{d,q},u_d)\right].
\end{equation}
Assuming $N_d$ is a multiple of $2^K$,  we have
\begin{equation}\label{eq:expectation_slice_32}
\begin{split}
&\hspace*{-5mm} \mathbb{E}_{Z} \left[\mathcal{L}(\mathcal{G}(Z|\theta), \,\{u_1,\ldots,u_D\})\right]
\\
&\hspace*{5mm}  = 
\sum_{d=1}^{D} \frac{1}{N_d}
	\sum_{n = 0}^{N_d-1}
	\mathbb{E}_{Z} \left[\mathcal{L}_{2} (\mathcal{G}(Z|\theta)_{d,n},u_d)\right]\\
&\hspace*{5mm} =
\sum_{d=1}^{D} \frac{1}{2^K}
	\sum_{q=0}^{2^K-1} \mathbb{E}_{Z} \left[\mathcal{L}_{2} (\mathcal{G}(Z|\theta)_{d,q},u_d)\right]. 
\end{split}
\end{equation}
As a consequence, instead of using $N_d$ slices per direction  the generator network could be trained using only a set of $2^K$  contiguous slices on each constrained direction.

The GPU memory is a limiting factor during the training process, even cutting down the size of the samples to $2^K$ slices restricts the  training slice resolution. 
For example, training a network for a texture output of size $512\times512\times32$ with $K=5$ and VGG-$19$ would require more than 12GB of memory. 
For that reason we propose to stochastically approximate the inner sum in Equation~\eqref{eq:expectation_slice_32}.  

Considering the slice $n = 2^K p+Q_d$ in the $d$-th axis with a fixed $p \in \{0,\ldots,\frac{N_d}{2^K}-1\}$ and with 
$Q_d\in\{0,\ldots,2^K-1\}$  randomly drawn from a discrete uniform distribution,
\begin{equation}\label{eq:expectation_single_slice2}
\mathbb{E}_{Z,Q_d} \left[\mathcal{L}_{2} (\mathcal{G}(Z|\theta)_{d,Q_d},u_d)\right]
	= \frac{1}{2^K}\sum_{q=0}^{2^K-1} \mathbb{E}_{Z} \left[\mathcal{L}_{2} (\mathcal{G}(Z|\theta)_{d,q},u_d)\right].
\end{equation}
Then using doubly stochastic sampling (noise input values and output coordinates) we have
\begin{equation}\label{eq:double_expectation_slice}
	\mathbb{E}_{Z} \left[\mathcal{L}(\mathcal{G}(Z|\theta), \,\{u_1,\ldots,u_D\})\right] 
	 =
	 \sum_{d=1}^{D} \mathbb{E}_{Z,Q_d} \left[\mathcal{L}_{2} (\mathcal{G}(Z|\theta)_{d,Q_d},u_d)\right],
\end{equation}
which means that we can train the generator using only $D$ single-slice volumes oriented 
according to the constrained directions. 
Note that the whole volume model is impacted since the convolution weights are shared by all slices.

The proposed scheme saves computation time during the training, and more importantly, it also reduces the required amount of memory. In this setting we can use resolutions of up to $1024$ values per size during training (examples and solid samples), a resolution significantly larger than the ones reached in the literature regarding solid texture synthesis by example which are usually limited to $256\times256$.

\section{Results}\label{Sec:Results}

\subsection{Experimental settings} 
\begin{figure}[!htb]
	\centering
	\includegraphics[width=\linewidth]{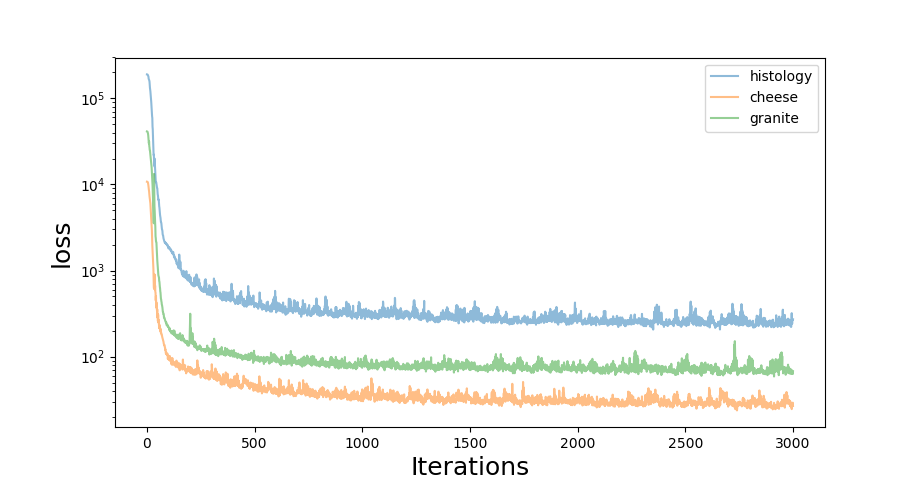}
	\caption{Value of the 3D empirical loss during the training of the generator for the textures \emph{histology}, \emph{cheese} and \emph{granite} of Figure~\ref{Fig:Results_isotropic1}.}
	\label{Fig:Training_curves}
\end{figure}

Unless otherwise specified all the results in this section were generated using the following settings. 
\paragraph*{Generator network}~
We set the number of scales to 6, \ie $K = 5$, which means that each voxel of the input noise at the coarsest scale impacts nearly 300 voxels at the finest scale. We use $M_i=3$ input channels and $M_s = 4$  (number of channels after the first convolution block and \emph{channel step} across scales) which results in the last layer being quite narrow ($6M_s = 24$) and the whole network compact, with $\sim8.5\times10^4$ parameters. 
We include a batch normalization operation after every convolution layer and before the concatenations. As mentioned in previous methods~\cite{ulyanov2016texturenets} we noticed that such a strategy helps stabilizing the training process. 
\paragraph*{Descriptor network}~
Following Gatys~\etal~\cite{gatys2015cnn}, we use a truncated VGG-19~\cite{Simonyan14c} as our descriptor network, with padded convolutions and average pooling. The considered layers for the loss are:  \verb$relu1_1$, \verb$relu2_1$, \verb$relu3_1$, \verb$relu4_1$ and \verb$relu5_1$.
\paragraph*{Training}~
We implemented our approach using \emph{pytorch} (code available in \URL{http://github.com/JorgeGtz/SolidTextureNets}) and we use the pre-trained parameters for VGG-19 available from the BETHGE~LAB~\cite{gatys2015cnn,Gatys_2016_CVPR}. We optimize the parameters of the generator network using Adam algorithm~\cite{kingma2014adam} with a learning rate of 0.1 during 3000 iterations. 
Figure~\ref{Fig:Training_curves} shows the value of the empirical estimation of $\mathcal{L}$ during the training of three of the examples shown below in Figure~\ref{Fig:Results_isotropic1} (\emph{histology, cheese} and \emph{granite}).
We use batches of 10 samples per slicing direction. We compute the gradients individually for each sample in the batch which slows down the training process but allows us to concentrate the available memory on the resolution of the samples. 
With these settings and using 3 slicing directions, the training takes around 1 hour for a $128^2$ training resolution (\ie size of the example(s) and generated slices),  $3.5$ hours for $256^2$ and  $12.5$ hours for $512^2$ using one GPU Nvidia GeForce GTX 1080 Ti.

\paragraph*{Synthesis}~
In order to synthesize large volumes of texture, it is more time efficient to choose the size of the building blocks in a way that best exploits parallel computation.
Considering computation complexity alone, synthesizing large building blocks of texture at once is also more efficient given the spatial dependency (indicated by coefficients $c_k$) shared by neighboring voxels.
In order to highlight the seamless tiling  of on demand generated blocks of texture, most of the samples shown in this work are built by assembling blocks of $32^3$ voxels.
However the generator is able to synthesize box-shaped/rectangular samples of any size, given enough memory is available.

\begin{figure}[!htb]
	\centering
	\raisebox{20mm}{\includegraphics[width = 1.2cm]{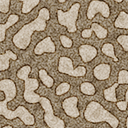}}
	\qquad
	\includegraphics[width = 5cm]{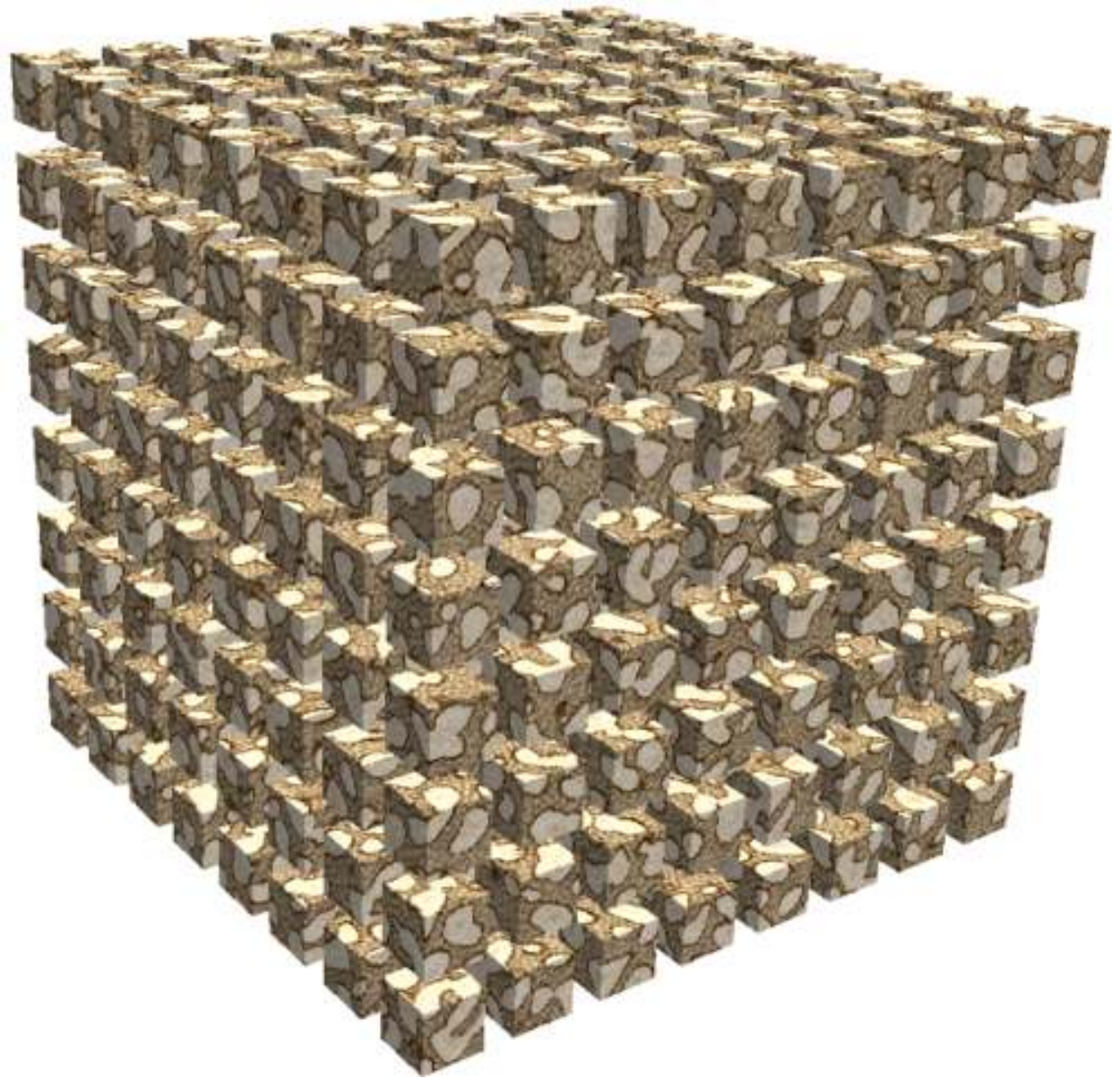}
	\caption{Left, example texture of size $128^2$ pixels, used along the three orthogonal axes. Right $512$ contiguous cubes of $32^3$ voxels generated on demand to form a $256^3$ texture. The gaps are added to better depict the aggregation.}
	\label{Fig:pieces_aggregation}
\end{figure}

\begin{figure}[!htb]
	\centering
	\raisebox{10mm}{\includegraphics[width = 2cm]{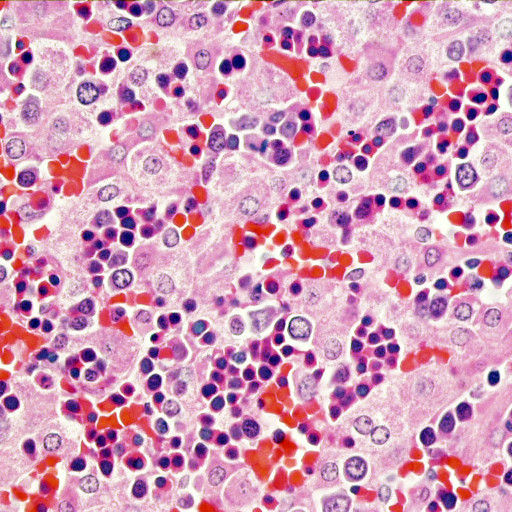}}
	\qquad
	\includegraphics[width = 5cm]{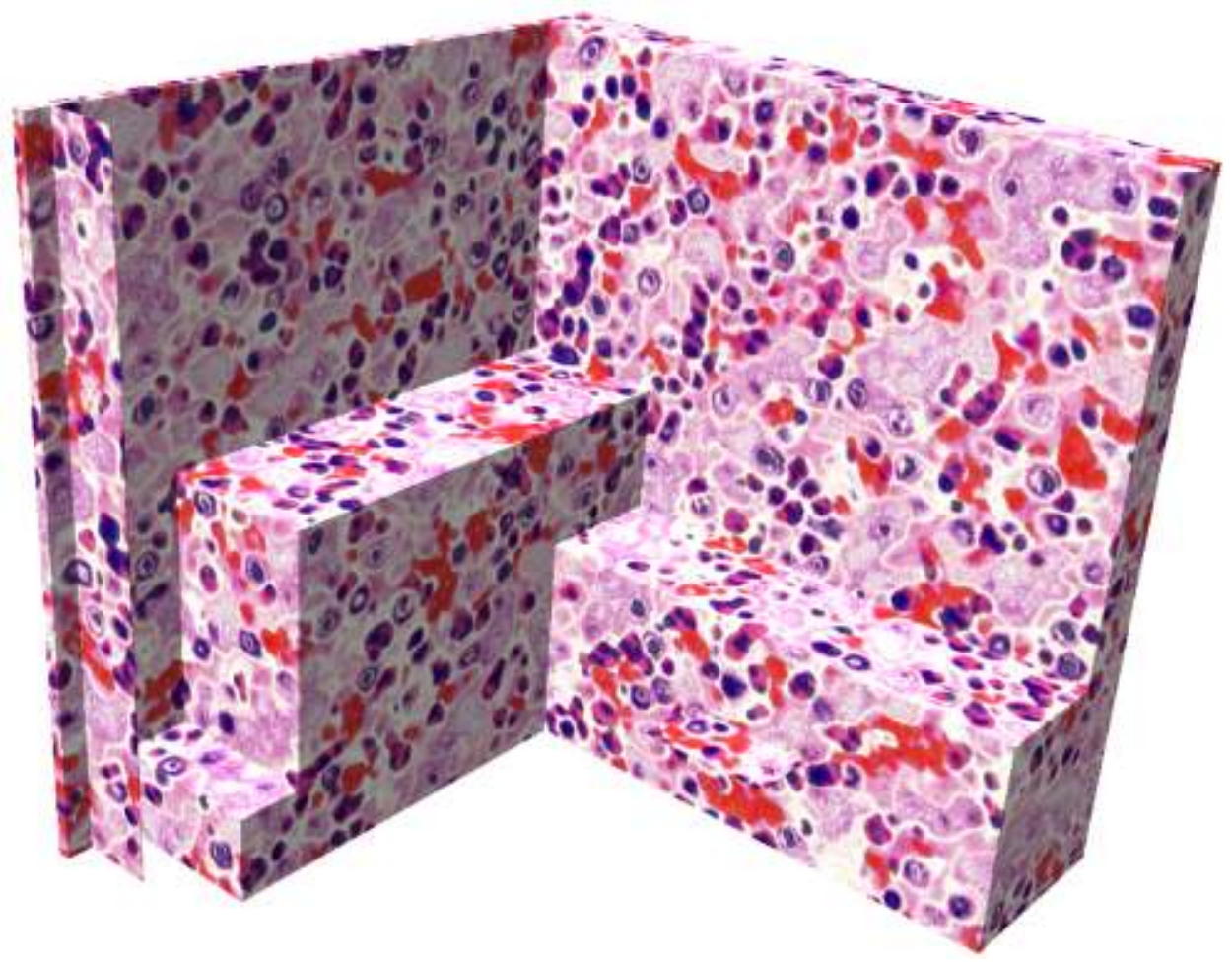}
	\caption{Left, example texture of size $512^2$ pixels, used for training the generator along the three orthogonal axes. Right, assembled blocks of different sizes generated on demand. Note that it is possible to generate blocks of size $1$ in any direction.}
	\label{Fig:pieces_on_demand}
\end{figure}

Figures~\ref{Fig:pieces_aggregation} and~\ref{Fig:pieces_on_demand} depict how the small pieces of texture tile perfectly to form a bigger texture.
It takes nearly 12 milliseconds to generate a block of texture of $32^3$ voxels on a Nvidia GeForce RTX 2080 GPU. However we can use bigger elemental blocks, \eg a cube of $64^3$ voxels takes $\sim24$ milliseconds and one of $128^3$ voxels takes $\sim128$ milliseconds.
For reference, using the method of Dong~\etal~\cite{dong2008lazy} it takes 220 milliseconds to synthesize a $64^3$ volume and 1.7 seconds to synthesize a $128^3$ volume. 

\subsection{Experiments}
\begin{figure*}[!htb]
	\centering
	\includegraphics[width = 17.5cm]{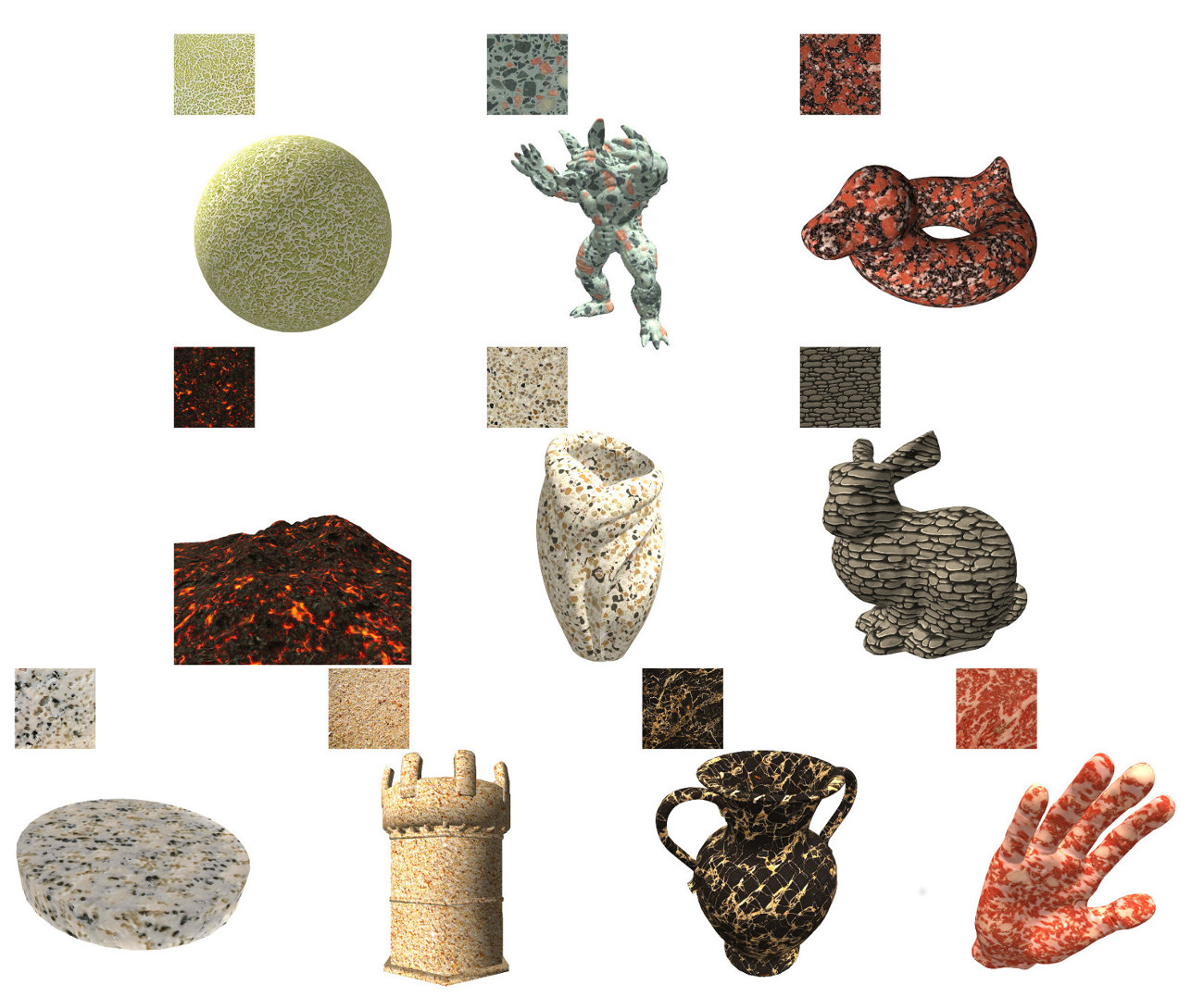}
	
	\caption{
	Texture mapping on 3D mesh models.
	The example texture used to train the generator is show in the upper left corner of each object. When using solid textures the mapping does not require parametrization as they are defined in the whole 3D space. This prevents any mapping induced artifacts. 
	\emph{Sources: the `duck' model comes from Keenan's 3D Model Repository, the `mountain' and `hand' models from free3d.com and the tower and the vase from turbosquid.com.}
	}
	\label{Fig:3Dmodels}
\end{figure*}

In this section we highlight the various properties of the proposed method
and we compare them with state-of-the-art approaches.
\paragraph*{Texturing mesh surfaces}
Figure \ref{Fig:3Dmodels} exhibits the application of textures generated with our model to add texture to 3D mesh models. In this case we generate a solid texture with a fixed size and load it on OpenGL as a regular 3D texture with bilinear filtering. 
Solid textures avoid the need for surface parametrization and can be used on complex surfaces without creating artifacts.

\begin{figure*}[!htb]
	\centering
	\begin{tabular}{c p{3.6cm}@{\hspace{2pt}}|@{\hspace{2pt}}p{2.2cm}@{\hspace{2pt}}|@{\hspace{2pt}}p{2.2cm}@{\hspace{4pt}}p{2.2cm}@{\hspace{4pt}}p{2.2cm}@{\hspace{4pt}}p{2.2cm}}
		 & \centering {Generated} volume
		 & \centering {Examples}
		 & \multicolumn{4}{c}{Generated slices}
		 \\
		 
		 & \centering $v$
		 & \centering $u_1=u_2=u_3$ 
		 & \centering $v_{1,\frac{N_1}{2}}$ 
		 & \centering $v_{2,\frac{N_2}{2}}$ 
		 & \centering $v_{3,\frac{N_3}{2}}$ 
		 & {\centering oblique ($45^{\circ}$) }
		 \\
		\rotatebox{90}{\hspace{1cm} granite}&
		\includegraphics[width=3.6cm]{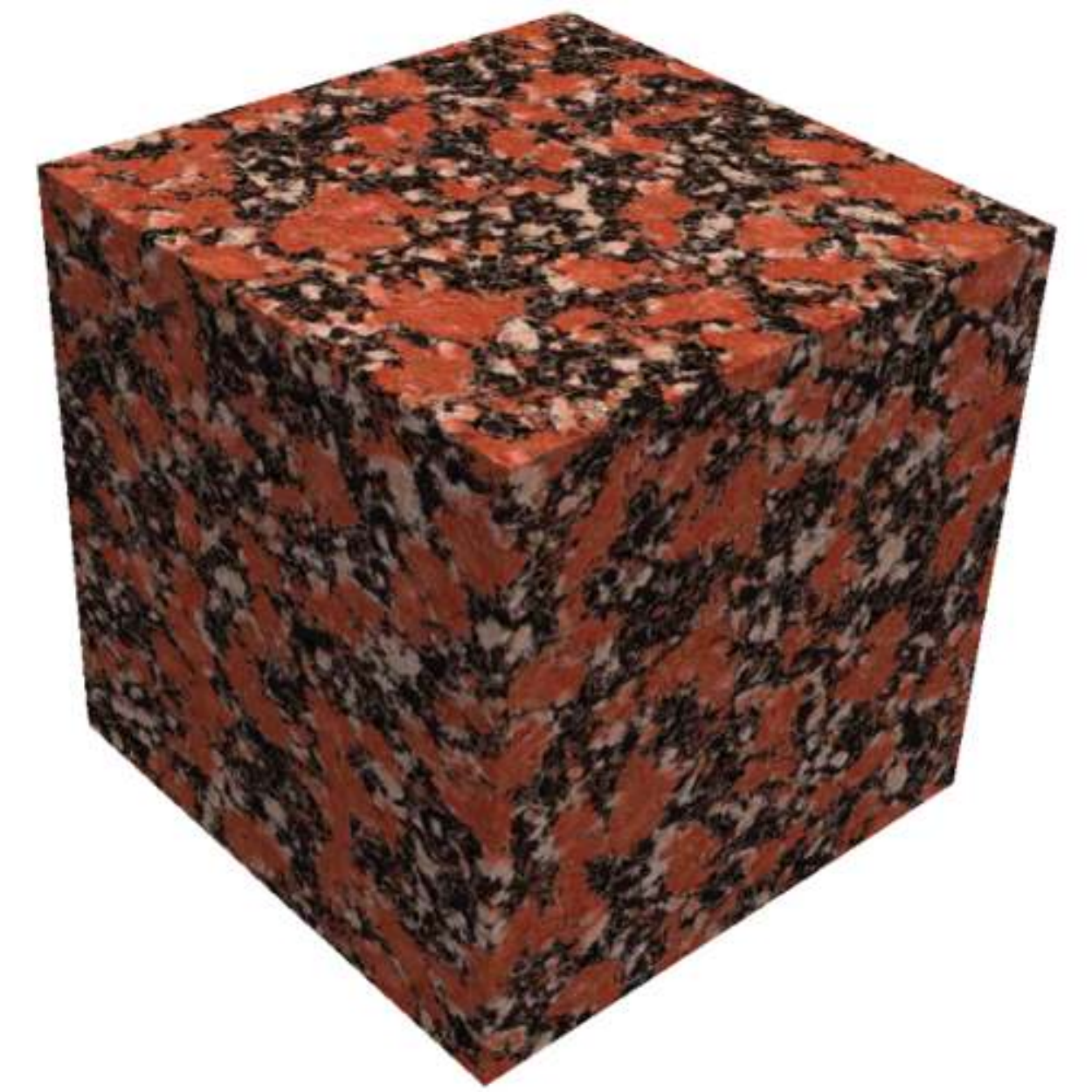}&
		\raisebox{8mm}{\includegraphics[width=2.2cm]{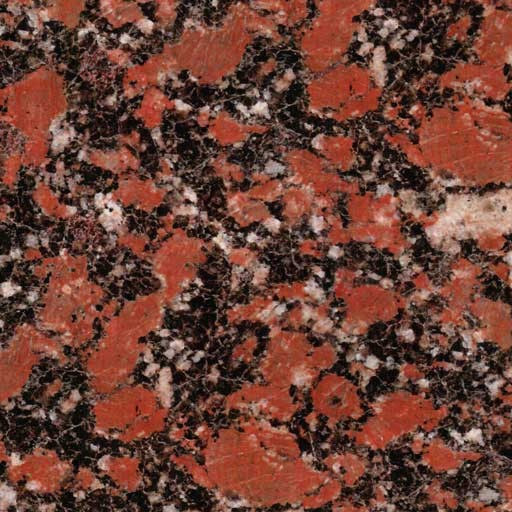}} &
		\raisebox{8mm}{\includegraphics[width=2.2cm]{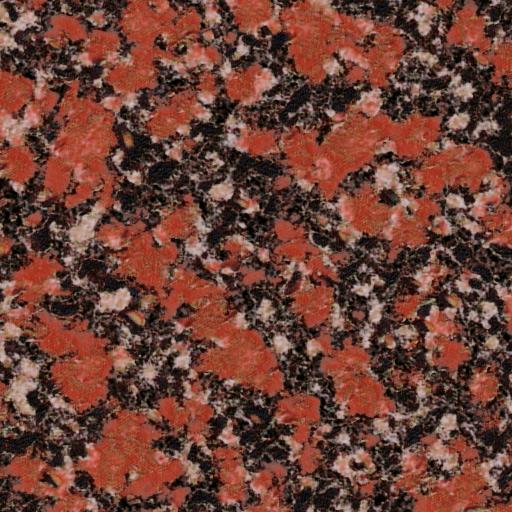}}&
		\raisebox{8mm}{\includegraphics[width=2.2cm]{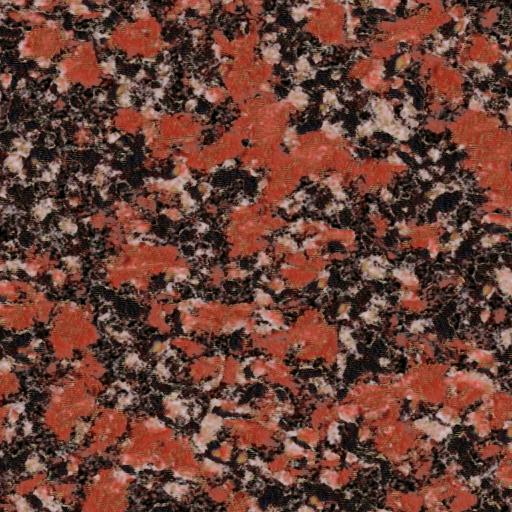}}&
		\raisebox{8mm}{\includegraphics[width=2.2cm]{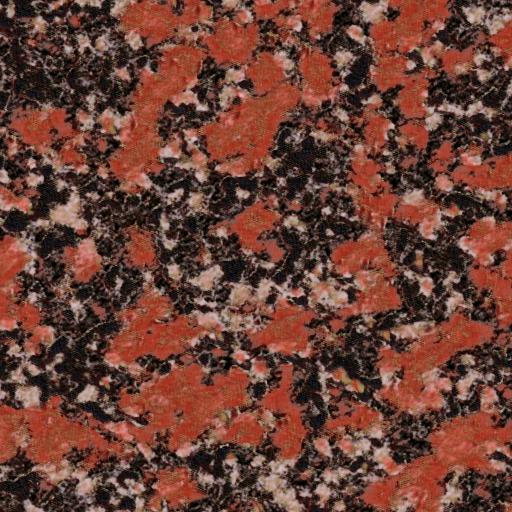}}&
		\raisebox{3mm}{\includegraphics[height=3.111cm, width=2.2cm]{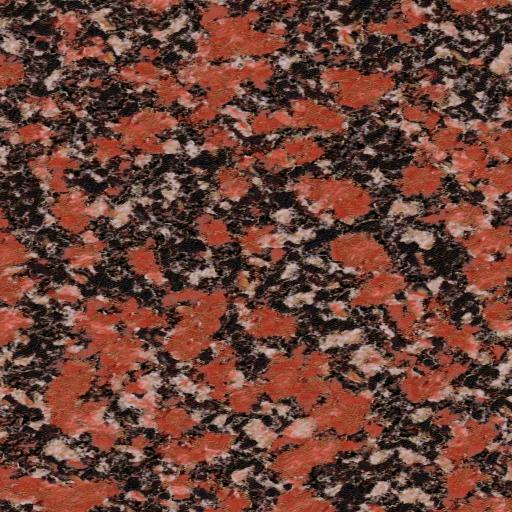}}\\
		\rotatebox{90}{\hspace{1cm} marble}&
		\includegraphics[width=3.6cm]{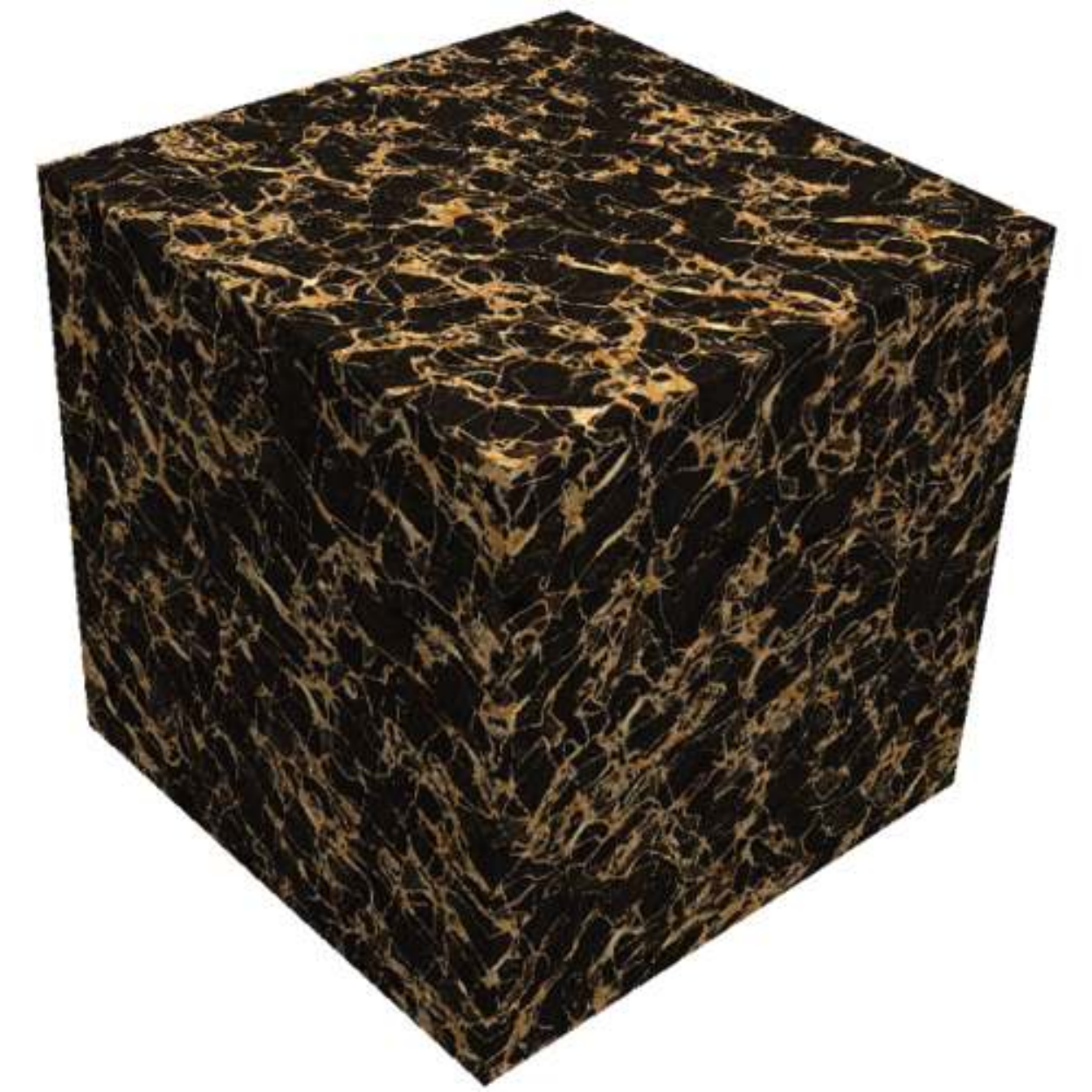}&
		\raisebox{8mm}{\includegraphics[width=2.2cm]{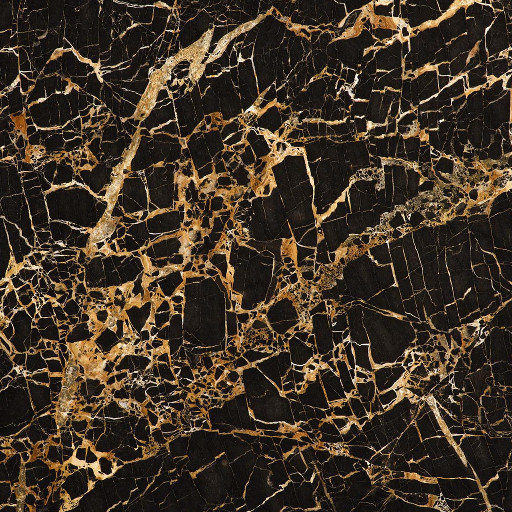}}&
		\raisebox{8mm}{\includegraphics[width=2.2cm]{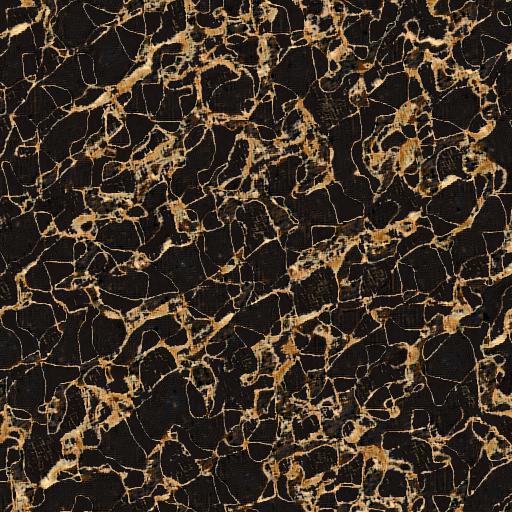}}&
		\raisebox{8mm}{\includegraphics[width=2.2cm]{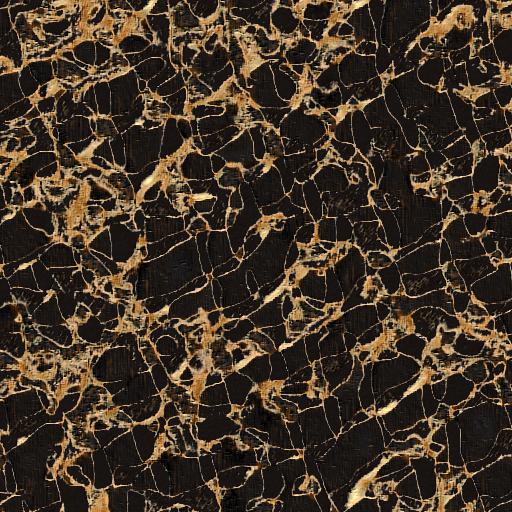}}&
		\raisebox{8mm}{\includegraphics[width=2.2cm]{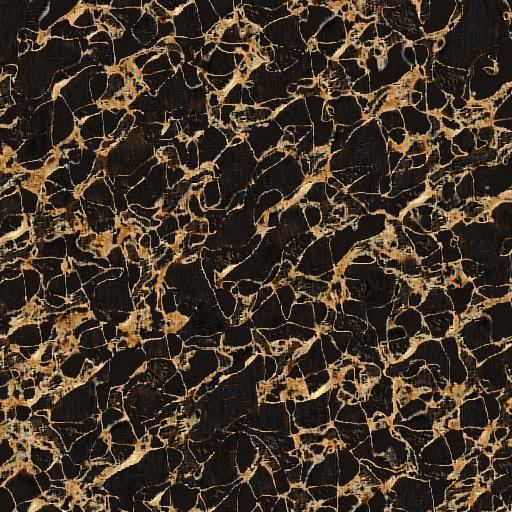}}&
		\raisebox{3mm}{\includegraphics[height=3.111cm, width=2.2cm]{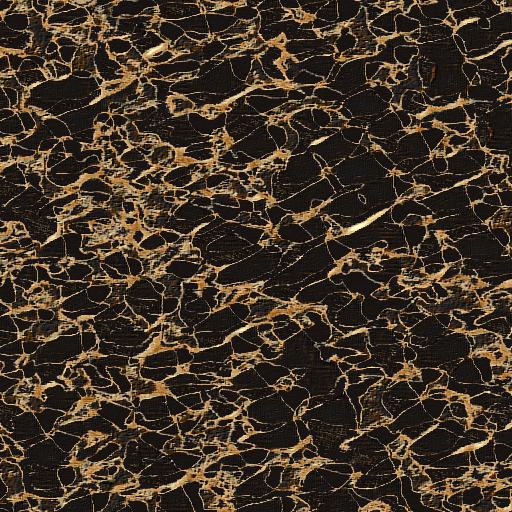}}\\
		\rotatebox{90}{\hspace{1.2cm} beef}&
		\includegraphics[width=3.6cm]{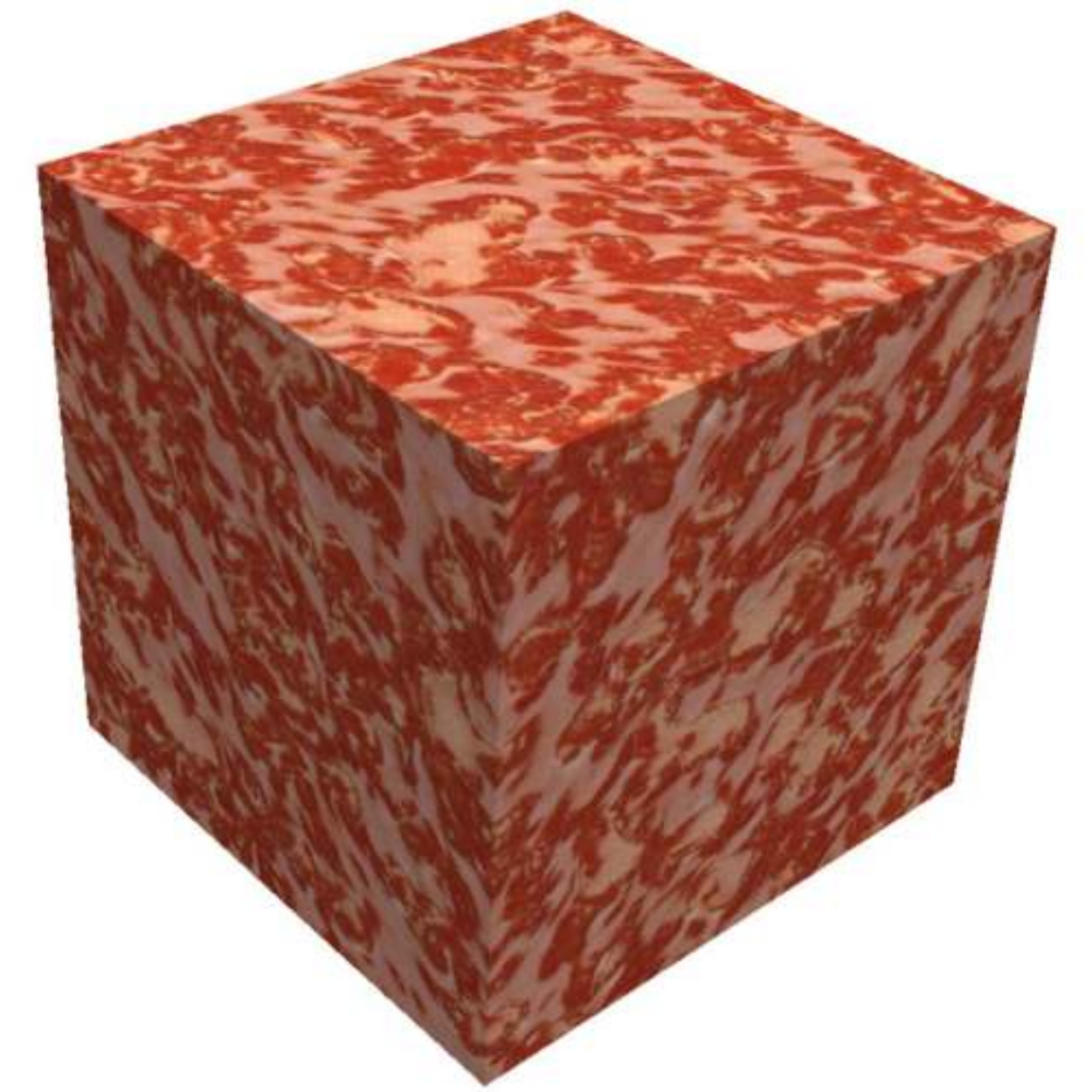}&
		\raisebox{8mm}{\includegraphics[width=2.2cm]{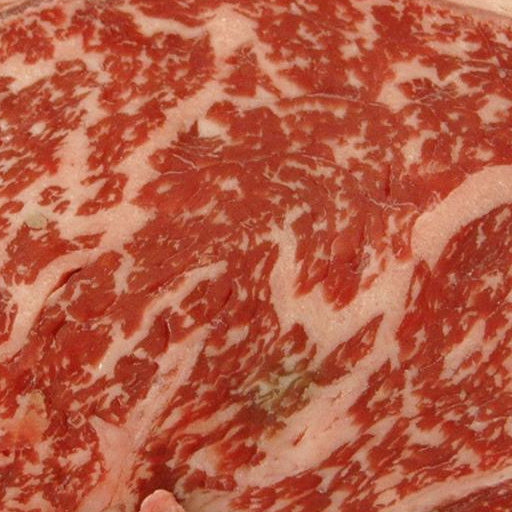}}&
		\raisebox{8mm}{\includegraphics[width=2.2cm]{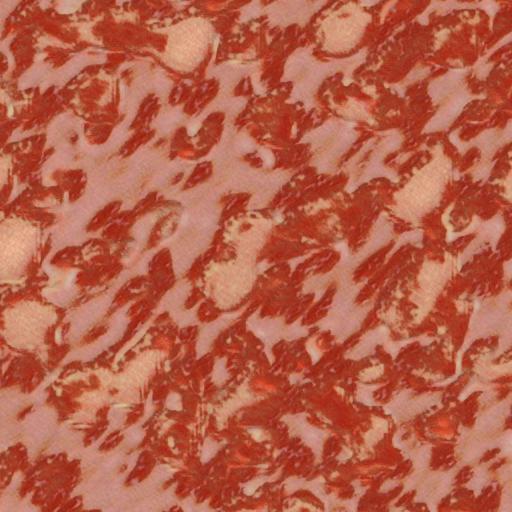}}&
		\raisebox{8mm}{\includegraphics[width=2.2cm]{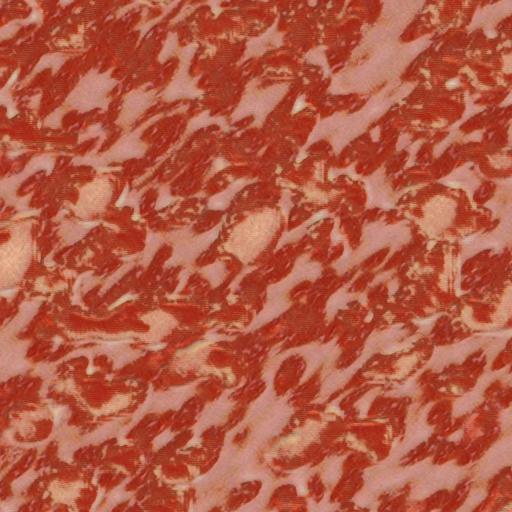}}&
		\raisebox{8mm}{\includegraphics[width=2.2cm]{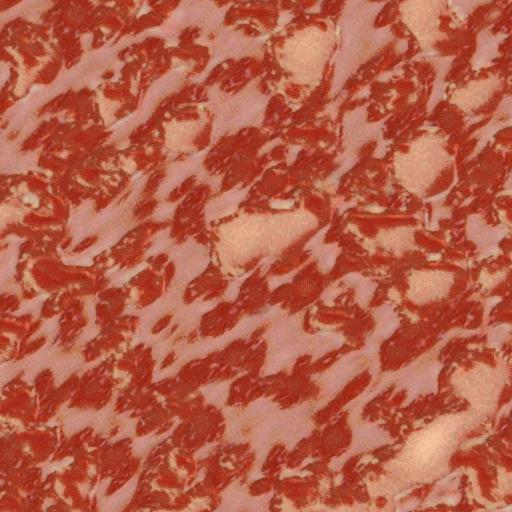}}&
		\raisebox{3mm}{\includegraphics[height=3.111cm, width=2.2cm]{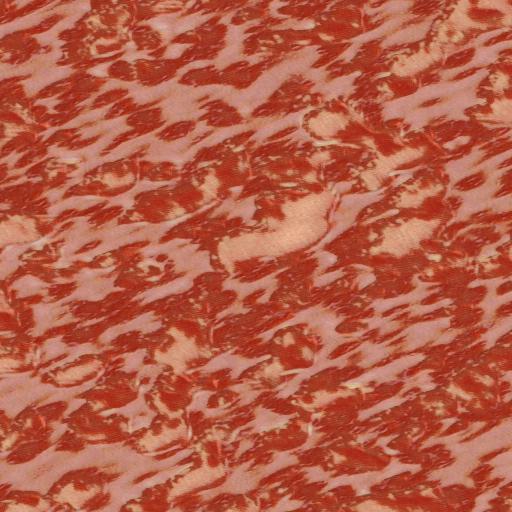}}\\
		\rotatebox{90}{\hspace{1.2cm} cheese}&
		\includegraphics[width=3.6cm]{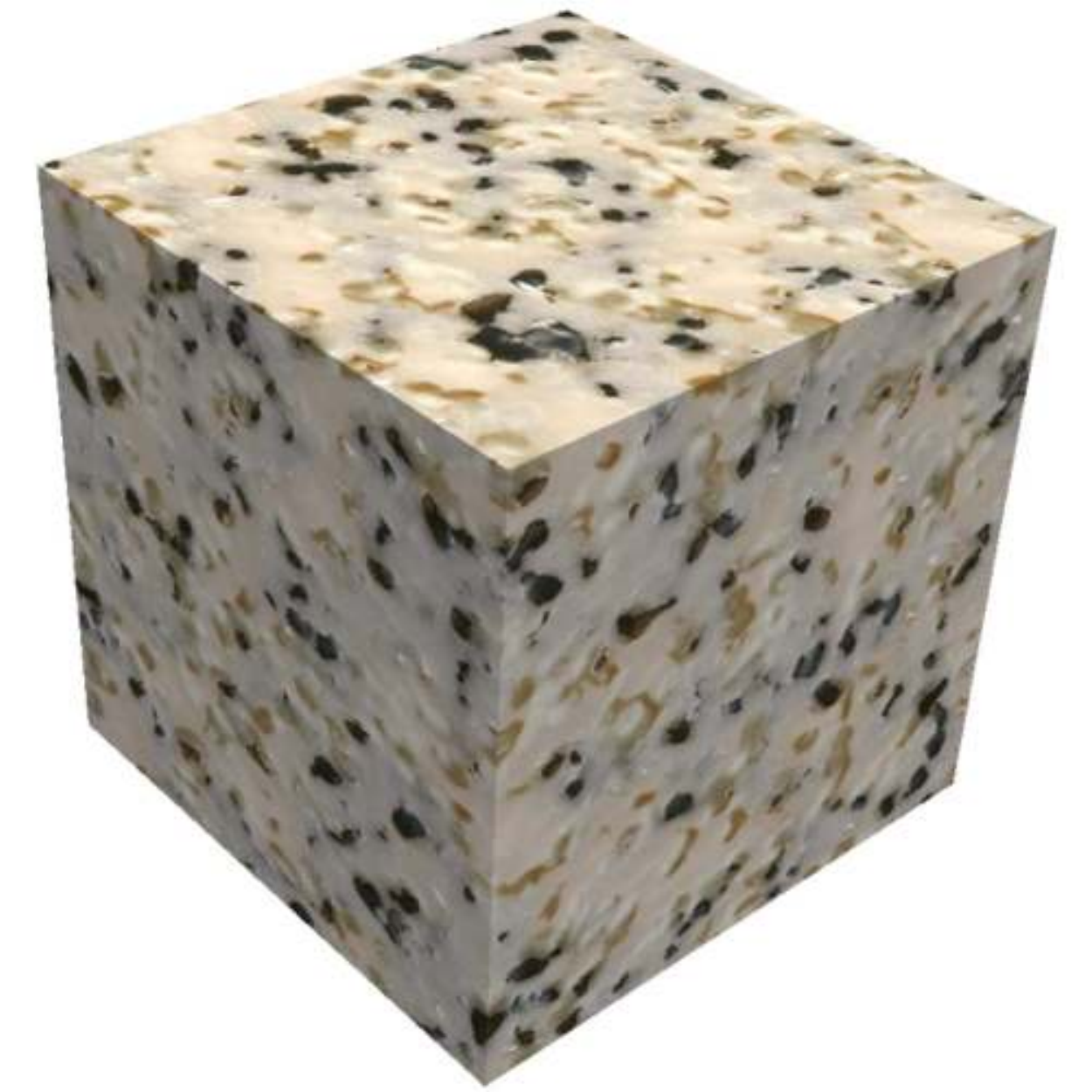}&
		\raisebox{8mm}{\includegraphics[width=2.2cm]{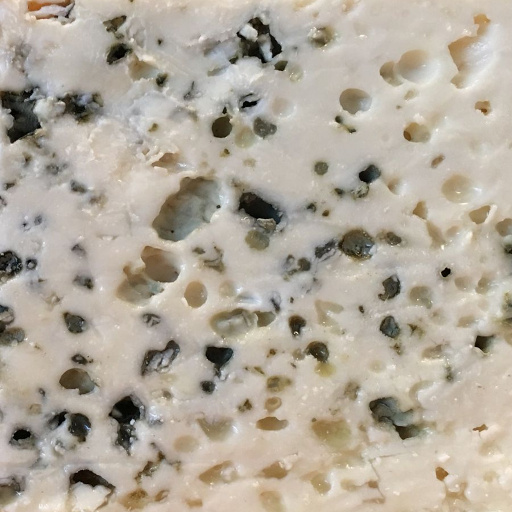}}&
		\raisebox{8mm}{\includegraphics[width=2.2cm]{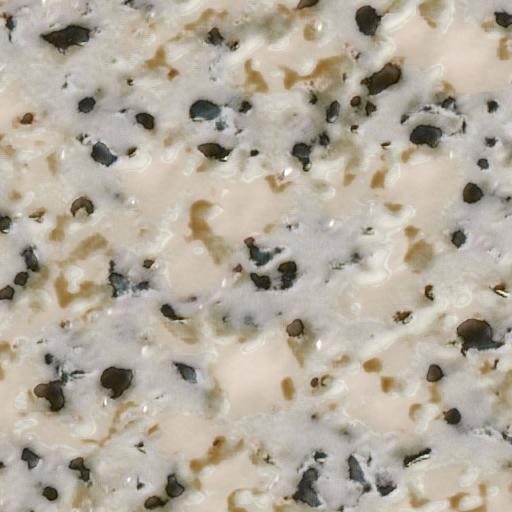}}&
		\raisebox{8mm}{\includegraphics[width=2.2cm]{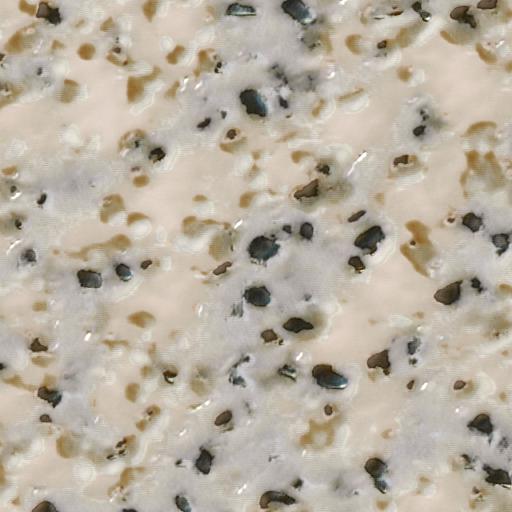}}&
		\raisebox{8mm}{\includegraphics[width=2.2cm]{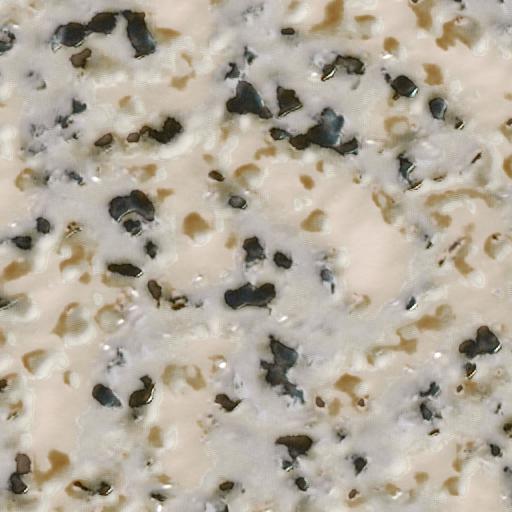}}&
		\raisebox{3mm}{\includegraphics[height=3.111cm, width=2.2cm]{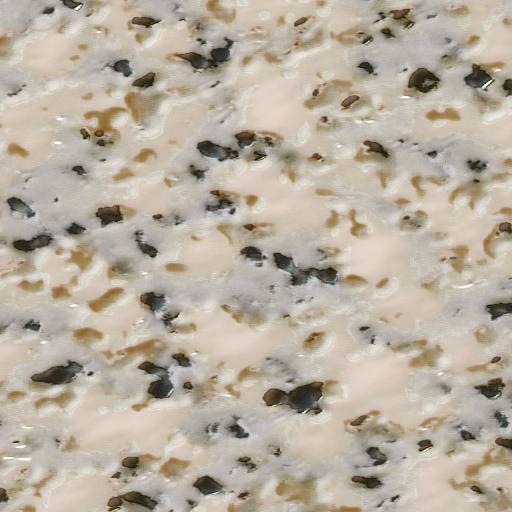}}\\
		\rotatebox{90}{\hspace{1cm} histology}&
		\includegraphics[width=3.6cm]{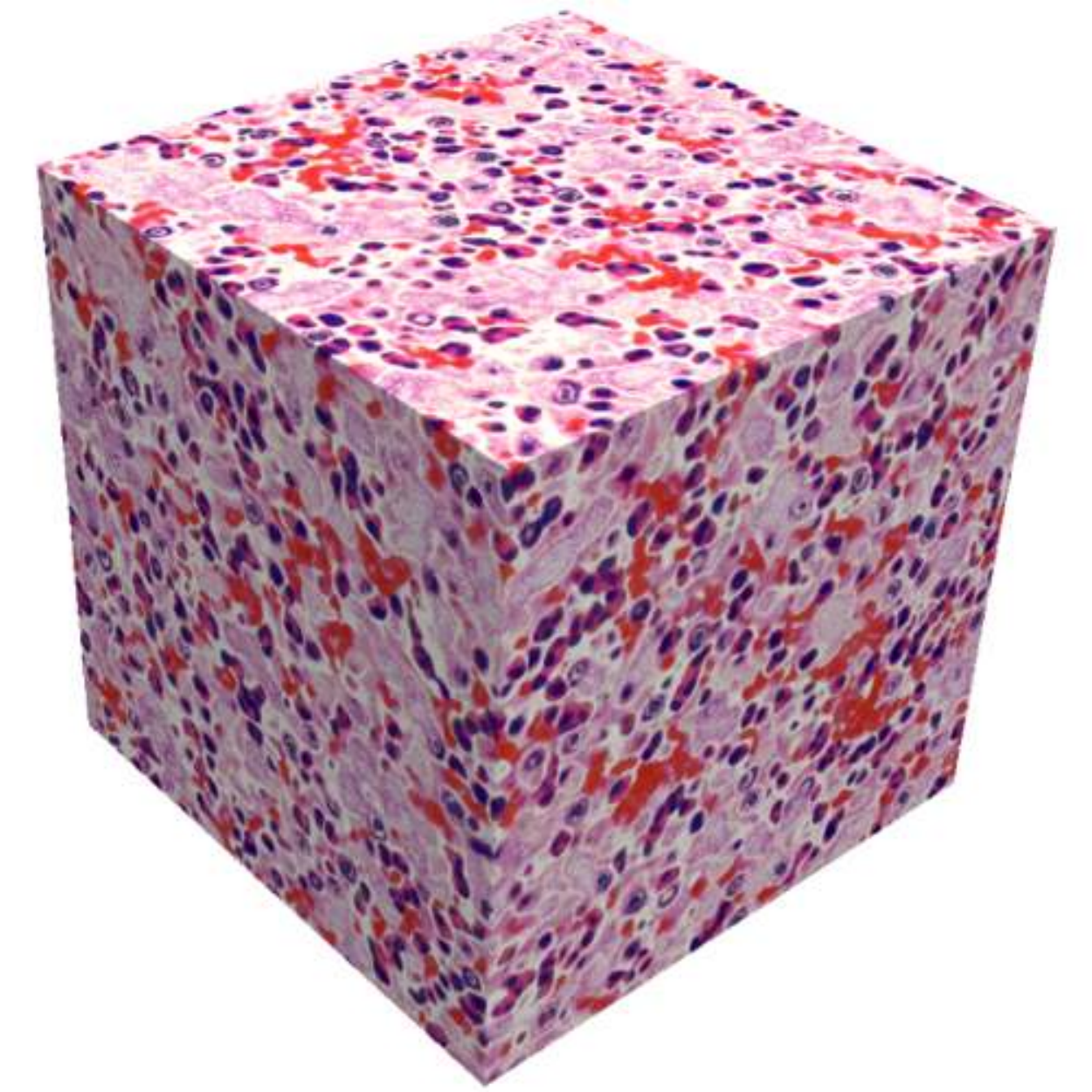}&
		\raisebox{8mm}{\includegraphics[width=2.2cm]{bio1_512}}&
		\raisebox{8mm}{\includegraphics[width=2.2cm]{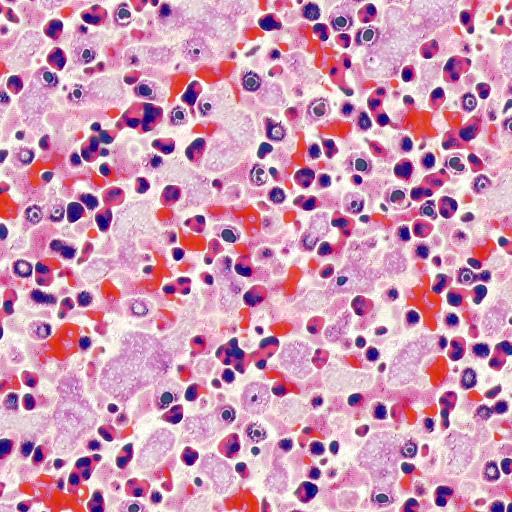}}&
		\raisebox{8mm}{\includegraphics[width=2.2cm]{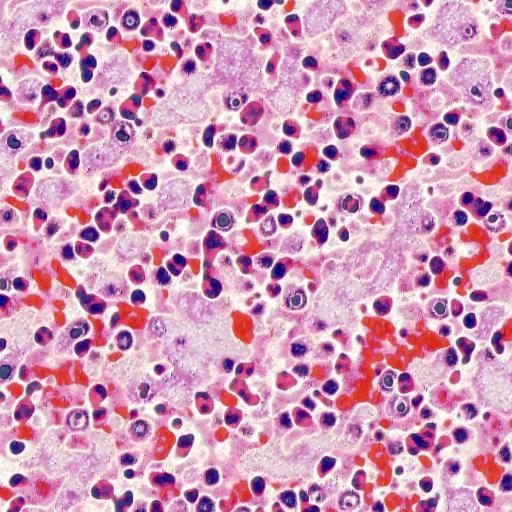}}&
		\raisebox{8mm}{\includegraphics[width=2.2cm]{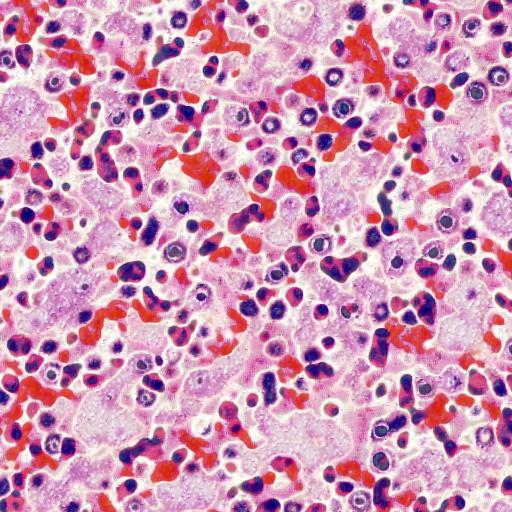}}&
		\raisebox{3mm}{\includegraphics[height=3.111cm, width=2.2cm]{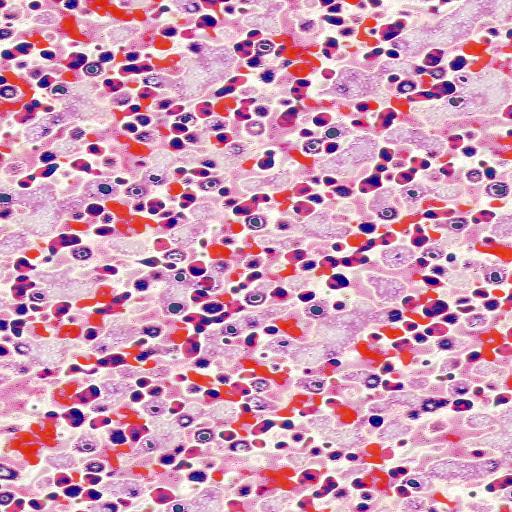}}\\
	\end{tabular}
	\caption{Synthesis of isotropic textures. We train the generator network using the example in the second column of size $512^2$ along $D=3$ directions. The cubes on the first column are generated samples of size $512^3$ built by assembling blocks of $32^3$ voxels generated using on demand evaluation. Subsequent columns show the middle slices of the generated cube across the three considered directions and a slice extracted in an oblique direction with a $45^{\circ}$ angle. The trained models successfully reproduce the visual features of the example in 3D.}
	\label{Fig:Results_isotropic1}
\end{figure*}

\begin{figure*}[!htb]
	\centering
	\begin{tabular}{c p{3.6cm}@{\hspace{2pt}}|@{\hspace{2pt}}p{2.2cm}@{\hspace{2pt}}|@{\hspace{2pt}}p{2.2cm}@{\hspace{4pt}}p{2.2cm}@{\hspace{4pt}}p{2.2cm}@{\hspace{4pt}}p{2.2cm}}
		& \centering {Generated} volume
		& \centering {Examples}
		& \multicolumn{4}{c}{Generated slices}
		\\
		& \centering $v$
		& \centering $u_1=u_2=u_3$ 
		& \centering $v_{1,\frac{N_1}{2}}$ 
		& \centering $v_{2,\frac{N_2}{2}}$ 
		& \centering $v_{3,\frac{N_3}{2}}$ 
		& {\centering oblique ($45^{\circ}$) }
		\\
		\rotatebox{90}{\hspace{1.2cm} pebble}&
		\includegraphics[width=3.6cm]{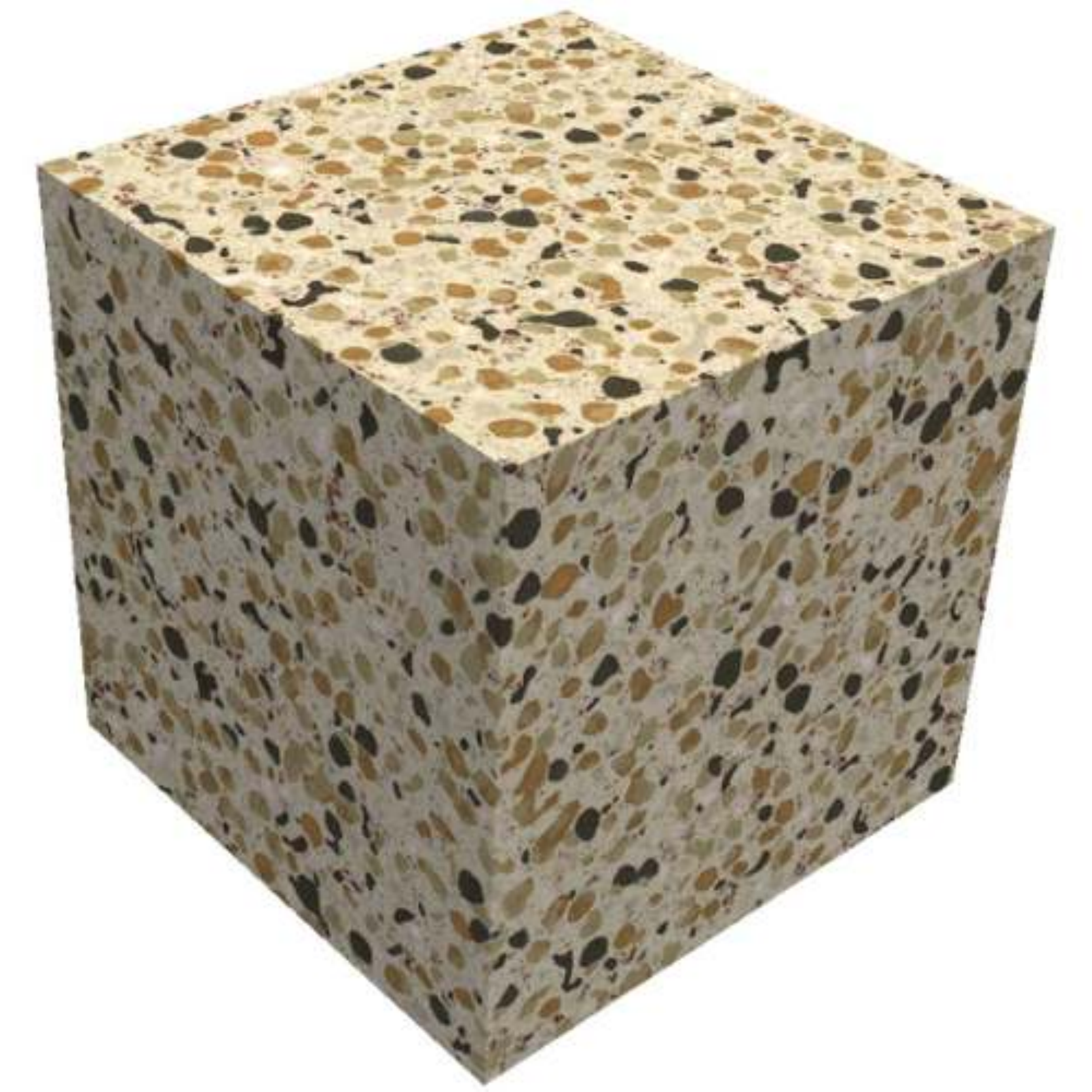}&
		\raisebox{8mm}{\includegraphics[width=2.2cm]{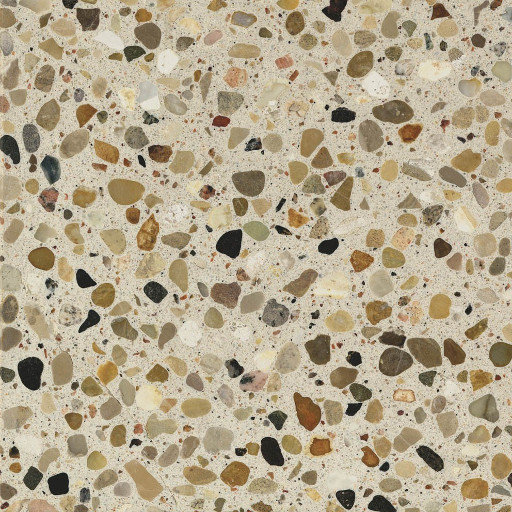}}&
		\raisebox{8mm}{\includegraphics[width=2.2cm]{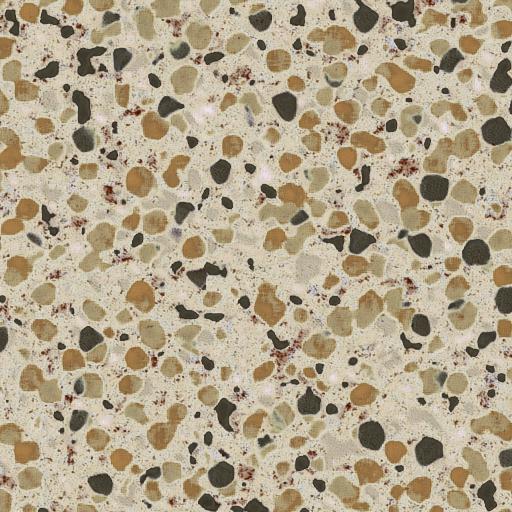}}&
		\raisebox{8mm}{\includegraphics[width=2.2cm]{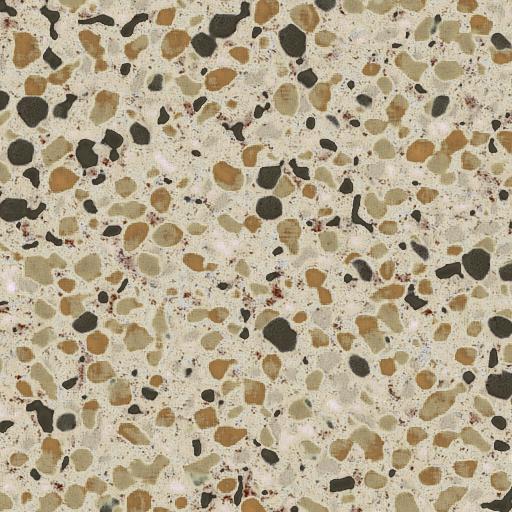}}&
		\raisebox{8mm}{\includegraphics[width=2.2cm]{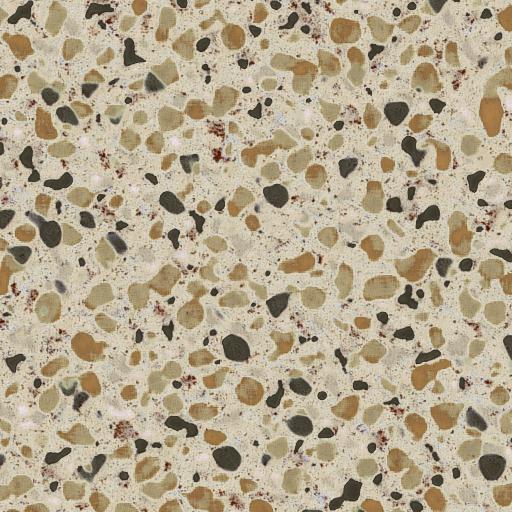}}&
		\raisebox{3mm}{\includegraphics[height=3.111cm, width=2.2cm]{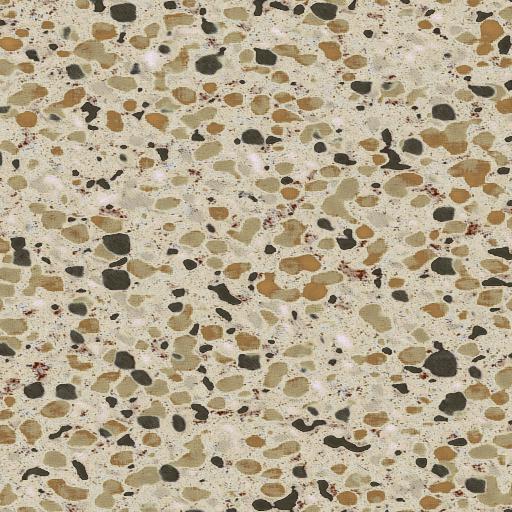}}\\
		\rotatebox{90}{\hspace{1.2cm} sand}&
		\includegraphics[width=3.6cm]{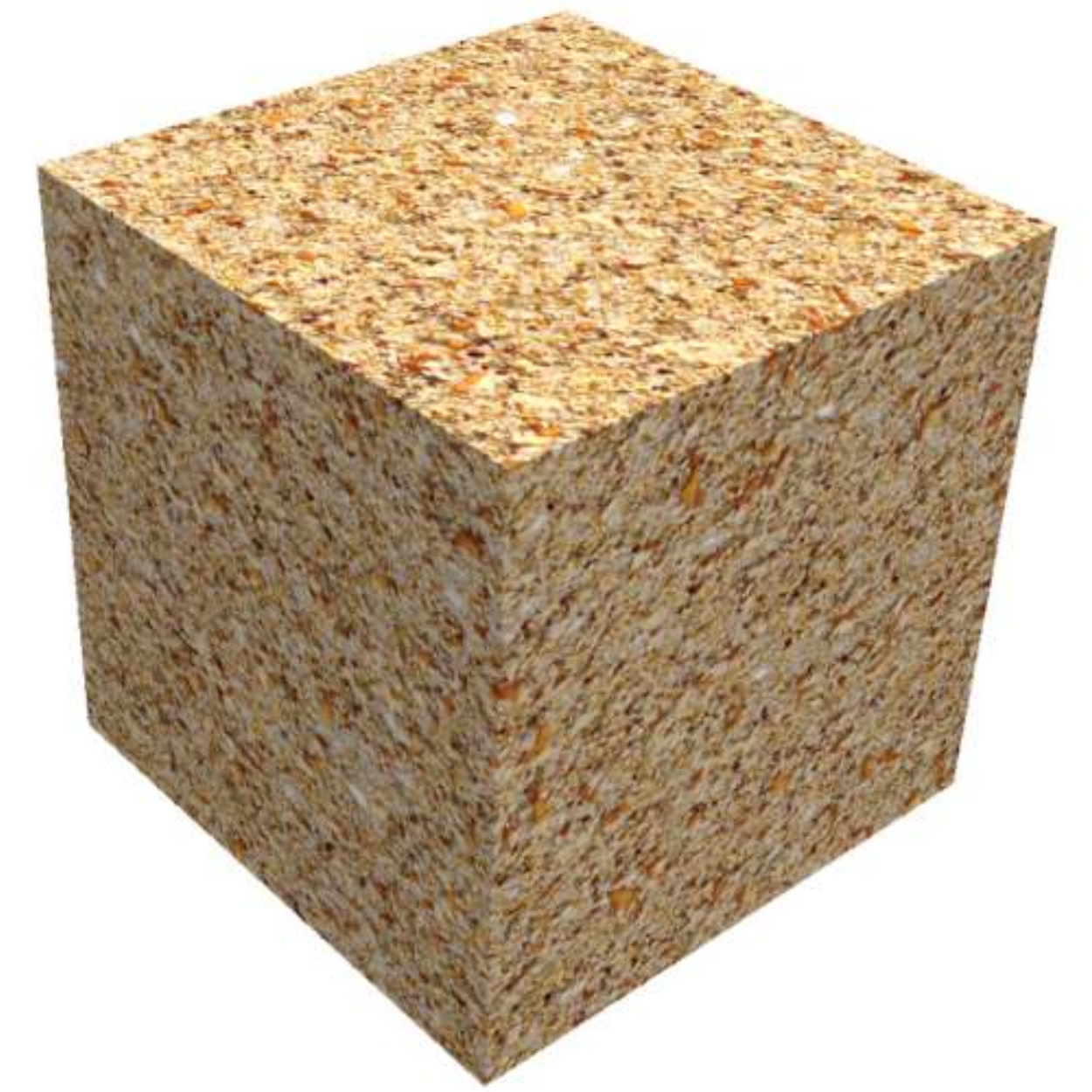}&
		\raisebox{8mm}{\includegraphics[width=2.2cm]{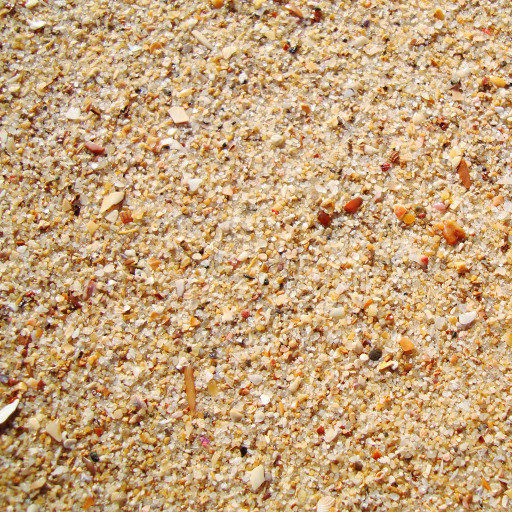}}&
		\raisebox{8mm}{\includegraphics[width=2.2cm]{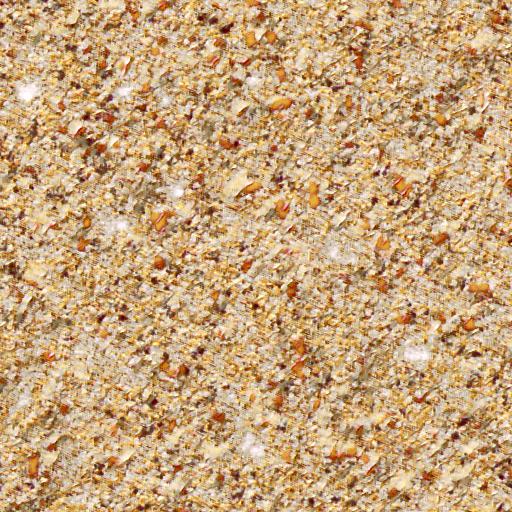}}&
		\raisebox{8mm}{\includegraphics[width=2.2cm]{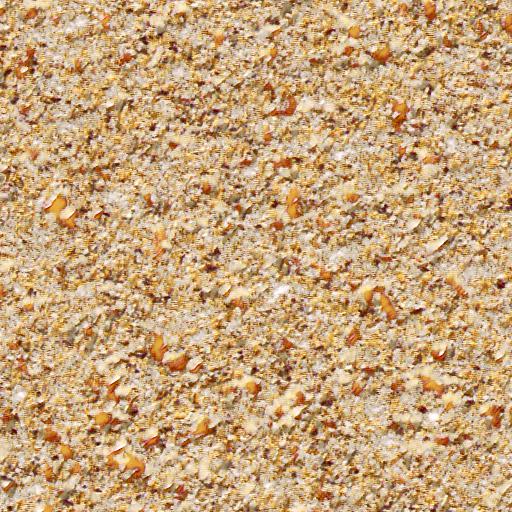}}&
		\raisebox{8mm}{\includegraphics[width=2.2cm]{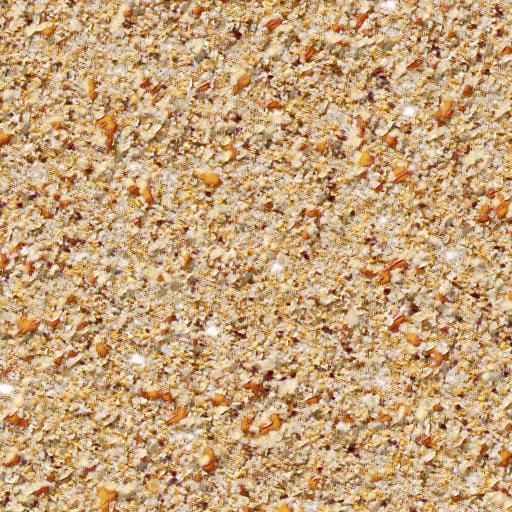}}&
		\raisebox{3mm}{\includegraphics[height=3.111cm, width=2.2cm]{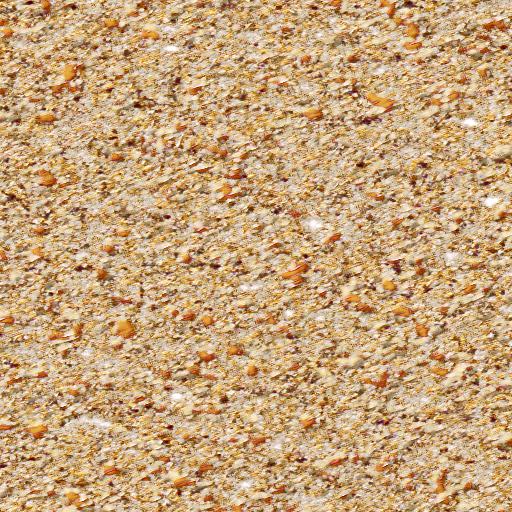}}\\
		\rotatebox{90}{\hspace{1.2cm} grass}&
		\includegraphics[width=3.6cm]{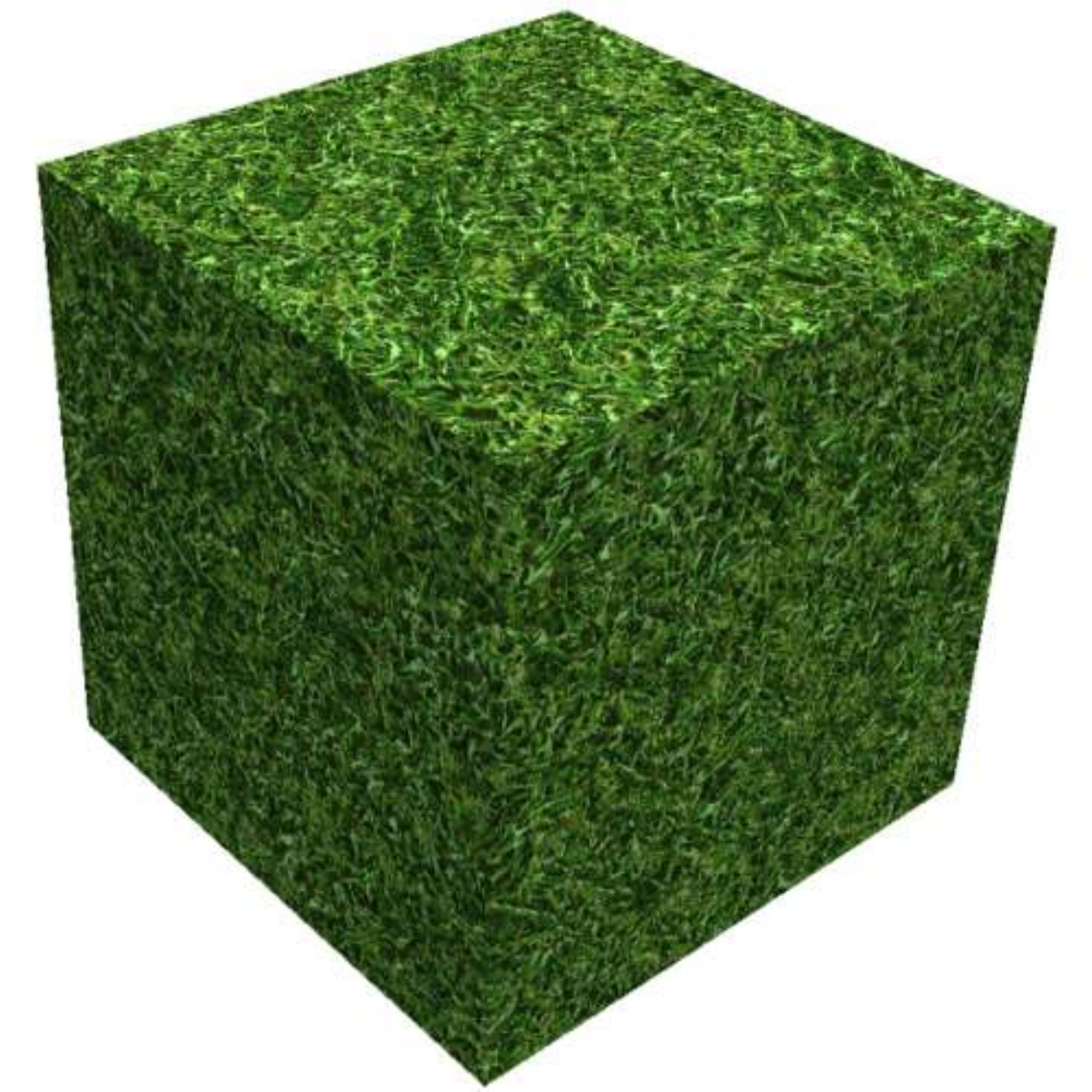}&
		\raisebox{8mm}{\includegraphics[width=2.2cm]{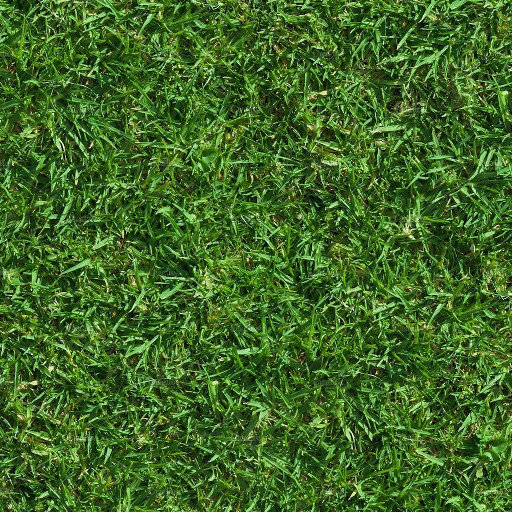}}&
		\raisebox{8mm}{\includegraphics[width=2.2cm]{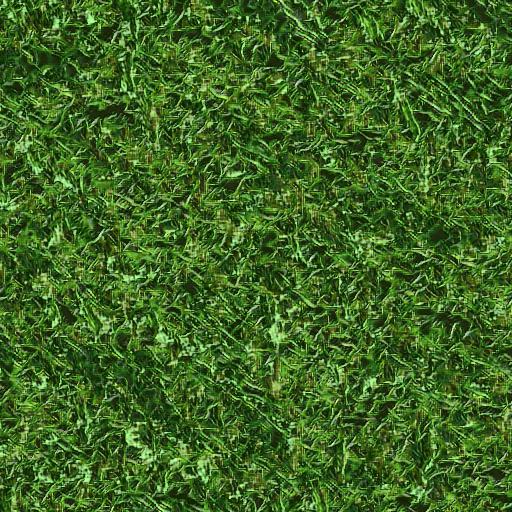}}&
		\raisebox{8mm}{\includegraphics[width=2.2cm]{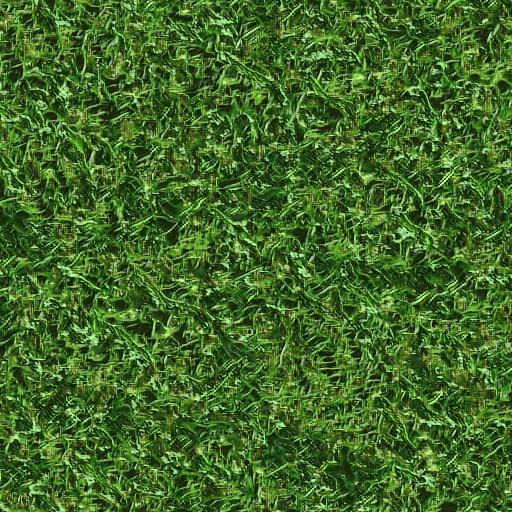}}&
		\raisebox{8mm}{\includegraphics[width=2.2cm]{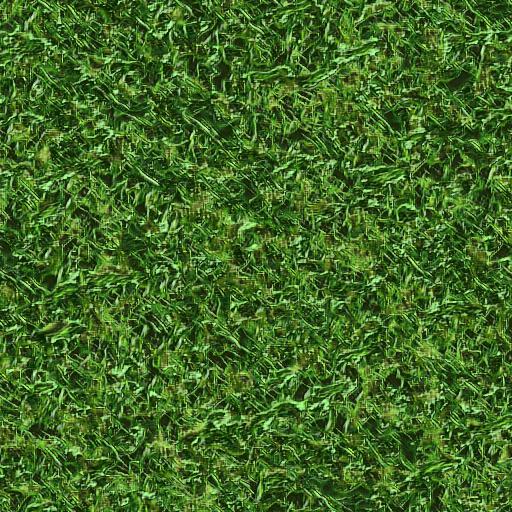}}&
		\raisebox{3mm}{\includegraphics[height=3.111cm, width=2.2cm]{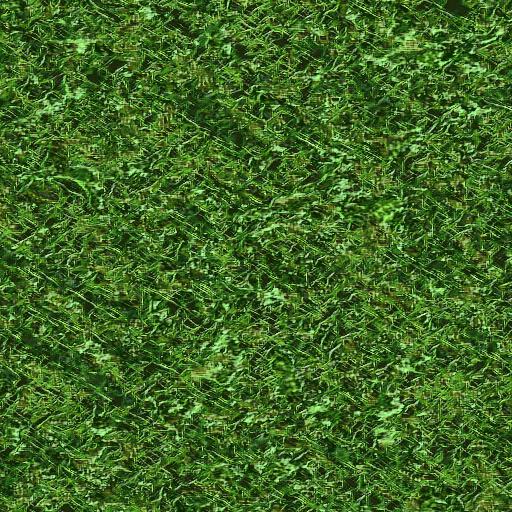}}\\
		\rotatebox{90}{\hspace{1.2cm} lava}&
		\includegraphics[width=3.6cm]{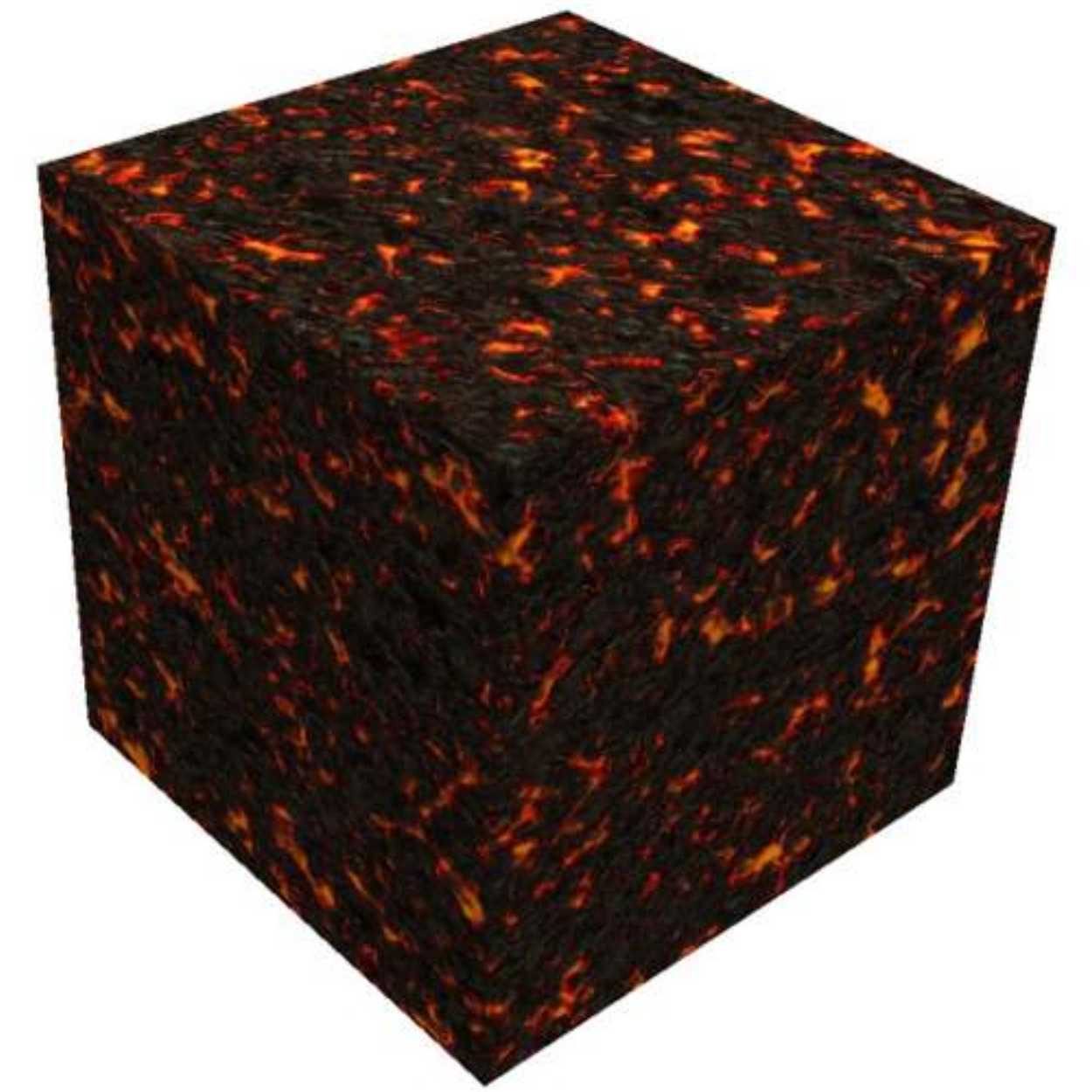}&
		\raisebox{8mm}{\includegraphics[width=2.2cm]{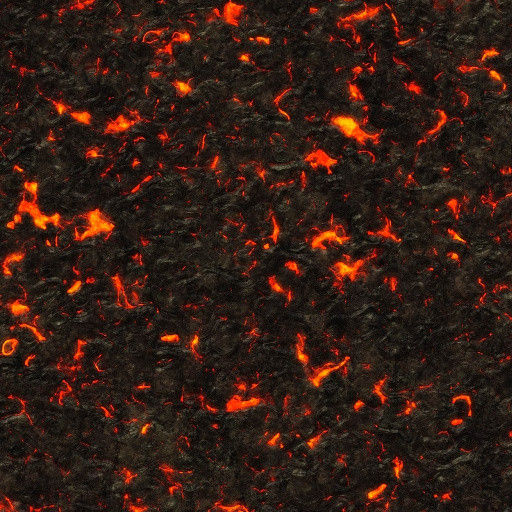}}&
		\raisebox{8mm}{\includegraphics[width=2.2cm]{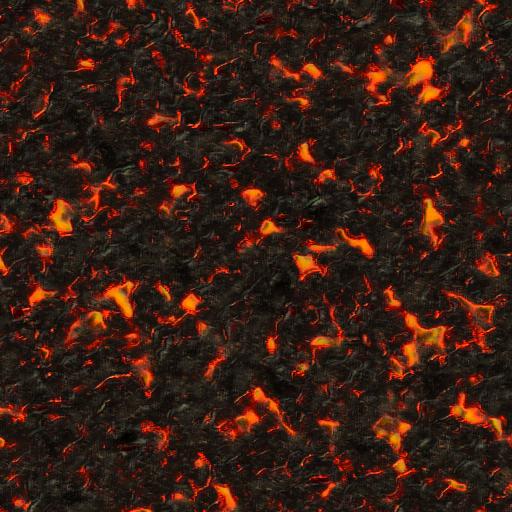}}&
		\raisebox{8mm}{\includegraphics[width=2.2cm]{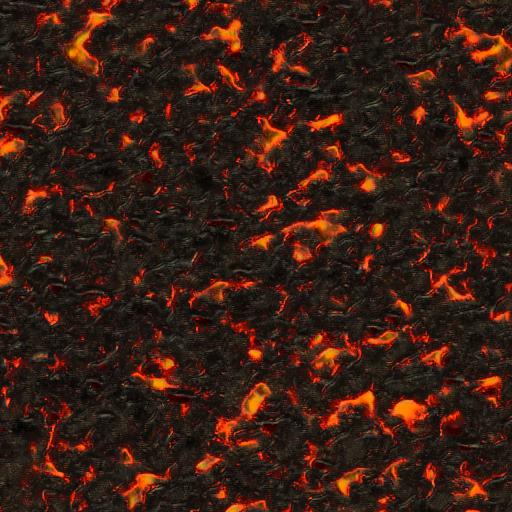}}&
		\raisebox{8mm}{\includegraphics[width=2.2cm]{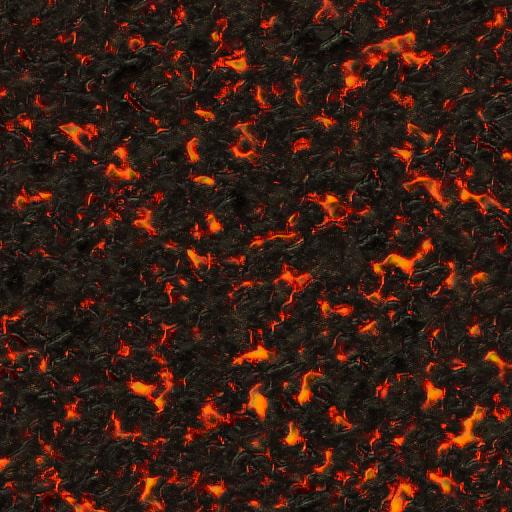}}&
		\raisebox{3mm}{\includegraphics[height=3.111cm, width=2.2cm]{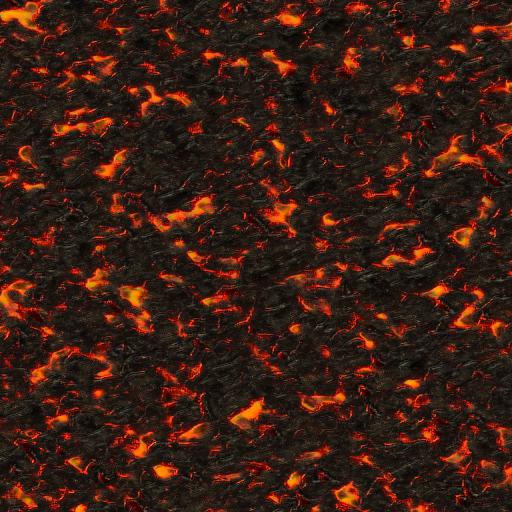}}
	\end{tabular}
	\caption{
	Synthesis of isotropic textures (same presentation as in Figure~\ref{Fig:Results_isotropic1}).
	While generally satisfactory, the examples in the first and third rows have a slightly degraded quality. In the first row the features have more rounded shapes than in the example and in the third row we observe high frequency artifacts.}
	\label{Fig:Results_isotropic2}
\end{figure*}

\begin{figure}[!htb]
	\centering
	\includegraphics[height=5cm]{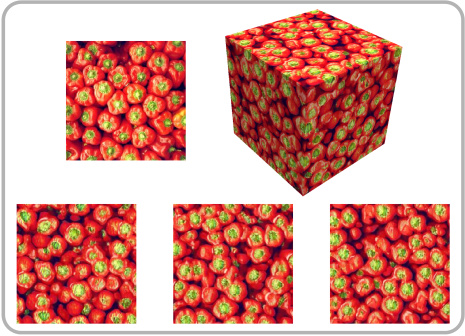}
	\caption{ 
	2D to 3D isotropic patterns. 
	The example texture (top-left) depicts a pattern that is approximately isotropic in 2D, but the material it depicts is not. 
	Training the generator using the example along three orthogonal directions results in solid textures that are isotropic in the three dimensions (top-right). 
	Here, the red and green patterns vary isotropically in the third dimension, this creates a bigger variation on their size and makes some slices contain red patterns that lack the green spot.
	This is a case where the slices of the solid texture (bottom) cannot match exactly the patterns of the 2D example, thus, not complying with the example in the way a 2D algorithm would.
	}
	\label{Fig:Example_peppers}
\end{figure}

\paragraph*{Single example setting}
Figures~\ref{Fig:Results_isotropic1}~and~\ref{Fig:Results_isotropic2} show synthesized samples using our method on a set of examples depicting some physical material. 
Considering their isotropic structure, we train the generator network using a single example to designate the appearance along the three orthogonal directions, \ie $u_1=u_2=u_3$.  
On the first column we show a generated sample of size $512^3$ voxels built by assembling blocks of $32^3$ voxels generated using on demand evaluation. 
The second column is the example image of size $512^2$ pixels, columns 3-5 show the middle slices of the generated cube across the three considered directions and the last column shows a slice extracted in an oblique direction with a $45^{\circ}$ angle.
These examples illustrate the capacity of the model to infer a plausible 3D structure from the 2D features present in isotropic example images. 
Observe that a slice across an  \emph{oblique} direction still displays a conceivable structure given the examples.
They also demonstrate the spatial consistency while using on demand evaluation.  
Regarding the visual quality, notice that the model successfully reproduces the patterns' structure while also capturing the richness of colors and variations.  
The quality of the slices is comparable to that of the state-of-the-art 2D methods \cite{ulyanov2016texturenets,Ulyanov17improved,li2017diversified,gatys2015cnn}, which is striking as solid texture synthesis is a more constrained problem.
However, such successful extrapolation to 3D might not be possible for all textures. Figure~\ref{Fig:Example_peppers} shows a case where the slices of the synthesized solid contain patterns not present in the example texture. This is related to the question of the existence of a solution discussed in the next paragraph.

\paragraph*{Existence of a solution}
The example texture used in Figure~\ref{Fig:Example_peppers} is isotropic (arrangement of red shapes having green spot inside), but the volumetric material it depicts is not (the green stem being outside the pepper). Training the generator using three orthogonal directions assumes 3D isotropy, and thus, the outcome is a solid texture where the patterns are isotropic through the whole volume. This creates some new patterns in the slices, which makes them somewhat different from the example (red shapes without a green spot inside). Actually, the generated volume just does not makes sense physically, as one always obtains full peppers after slicing them.
This example shows that for a given texture example $u$ and a number of directions $D>1$ it is possible that a corresponding 3D texture does not exist, \ie not all the slices in the chosen directions will respect the structure defined by the 2D example. 

\begin{figure}[!h]
	\centering
	\includegraphics[height=5cm]{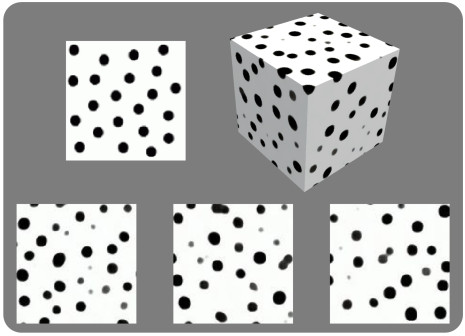}
	\caption{Illustration of a solid texture whose cross sections cannot comply with the example along three directions. Given a 2D example formed by discs of a fixed diameter (upper left) a direct isotropic 3D extrapolation would be a cube formed by spheres of the same diameter. Slicing that cube would result in images with discs of different diameters.
	The cube in the upper right is generated after training our network with the 2D example along the three orthogonal axes. The bottom row shows cross sections along the axes, all of them present discs of varying diameters thus failing to look like the example. }
	\label{Fig:Example_spheres}
\end{figure}

This existence issue is vital for example-based solid texture synthesis as it delineates the limits of this slice-based formulation used by many methods in the literature. It is briefly mentioned in \cite{kopf2007solid,dong2008lazy}  and here we aim to extend the discussion. 
Let us  consider for instance the isotropic example shown in Figure~\ref{Fig:Example_spheres}, where the input image contains only discs at a given scale (\eg a few pixels diameter). It follows that when slicing a volume containing spheres with the same diameter $\delta$, the obtained image will necessarily contain objects with various diameters, ranging from $0$ to $\delta$. This seems to be a paradigm natural to the 2D-3D extrapolation. 
It demonstrates that for some 2D textures an isotropic solid version might be impossible and conversely, that the example texture and the imposed directions must be chosen carefully given the goal 3D structure. 

\begin{figure*}[!htb]
	\centering
	\begin{tabular}{c@{\hspace{4pt}}|@{\hspace{4pt}}c@{\hspace{4pt}}c@{\hspace{4pt}}|@{\hspace{4pt}}c@{\hspace{4pt}}c@{\hspace{4pt}}c|c}
		 Generated volume & Training  & Examples &  \multicolumn{3}{c}{Generated views} & Training\\ 
		 $v$ & configuration & $u_1 = u_2 = u_3$  &  $v_{1,\frac{N_1}{2}}$ & $v_{2,\frac{N_2}{2}}$ & $v_{3,\frac{N_3}{2}}$ & empirical loss\\
		\includegraphics[width=3.2cm]{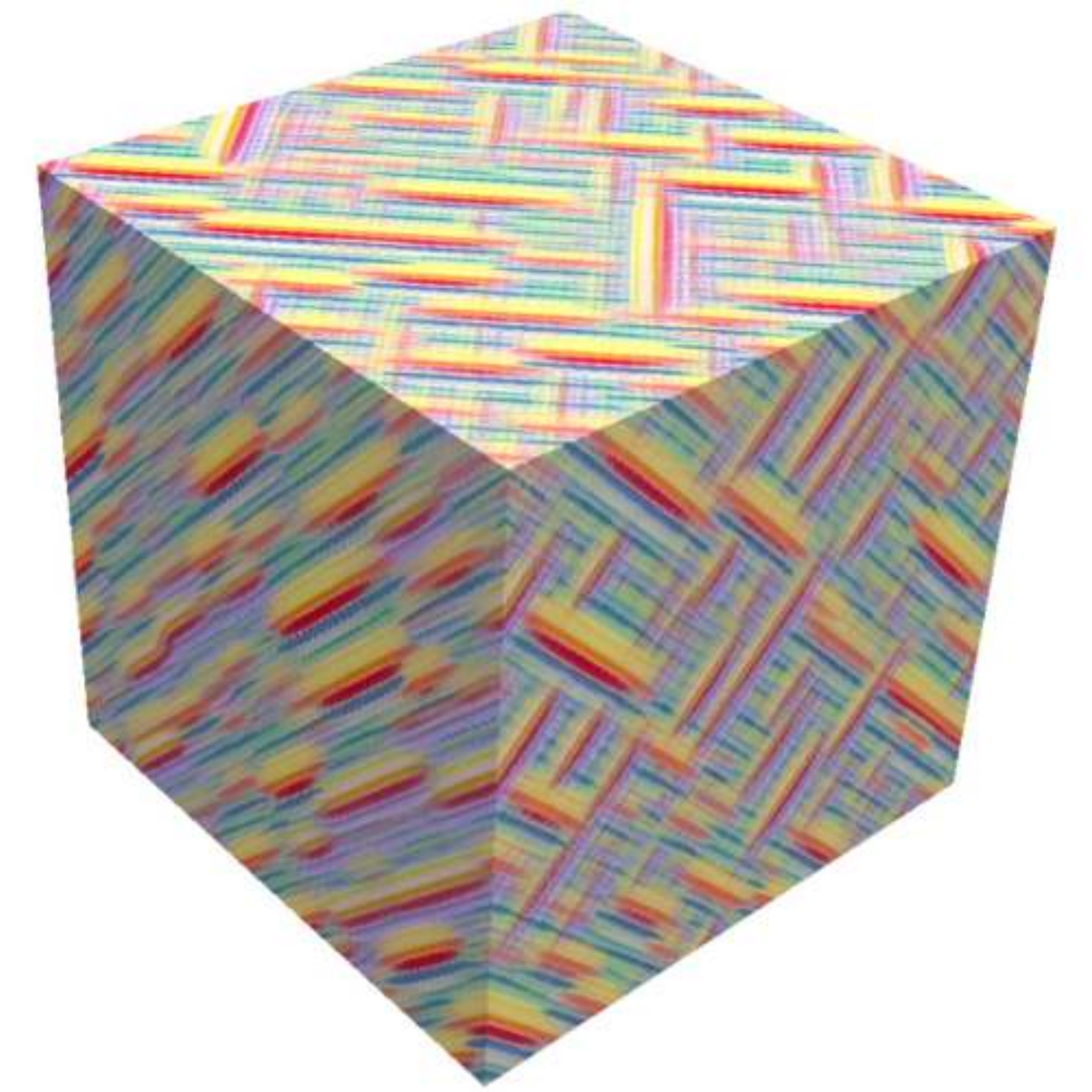}&
		\raisebox{6mm}{\includegraphics[width=2.2cm]{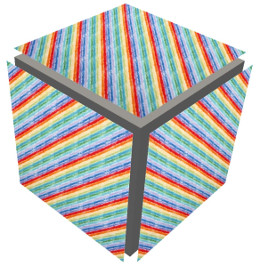}}&
		\raisebox{6mm}{\includegraphics[angle=90,width=2.2cm]{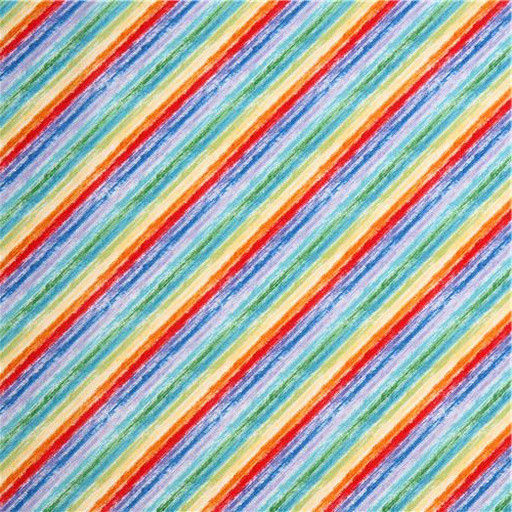}}&
		\raisebox{6mm}{\includegraphics[width=2.2cm]{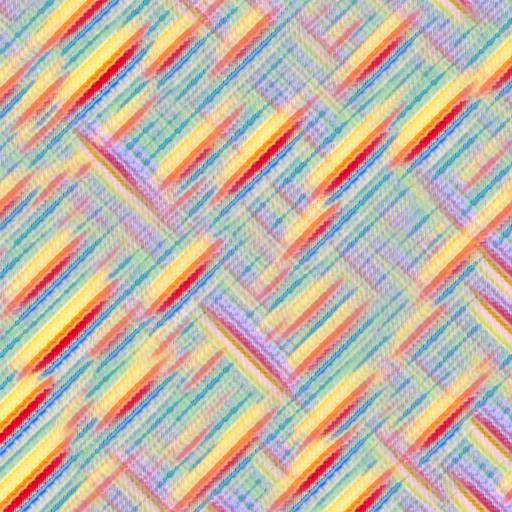}}&
		\raisebox{6mm}{\includegraphics[width=2.2cm]{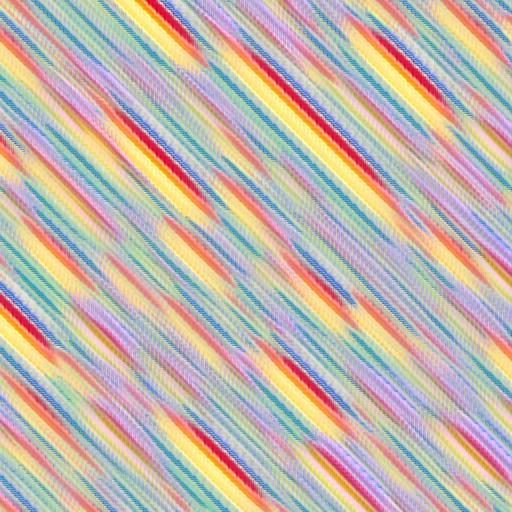}}&
		\raisebox{6mm}{\includegraphics[width=2.2cm]{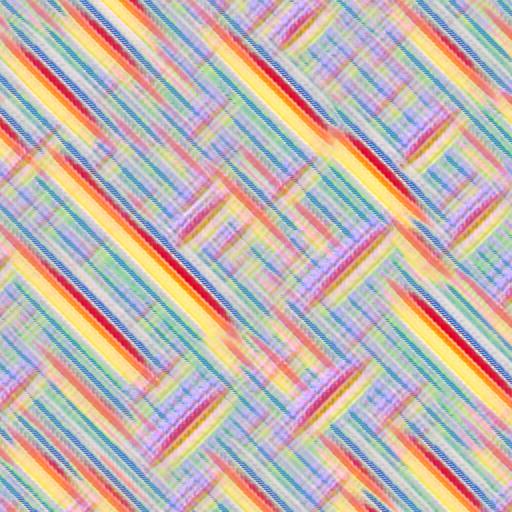}}&
		\raisebox{15mm}{$1183.6$}
		\\

		\includegraphics[width=3.2cm]{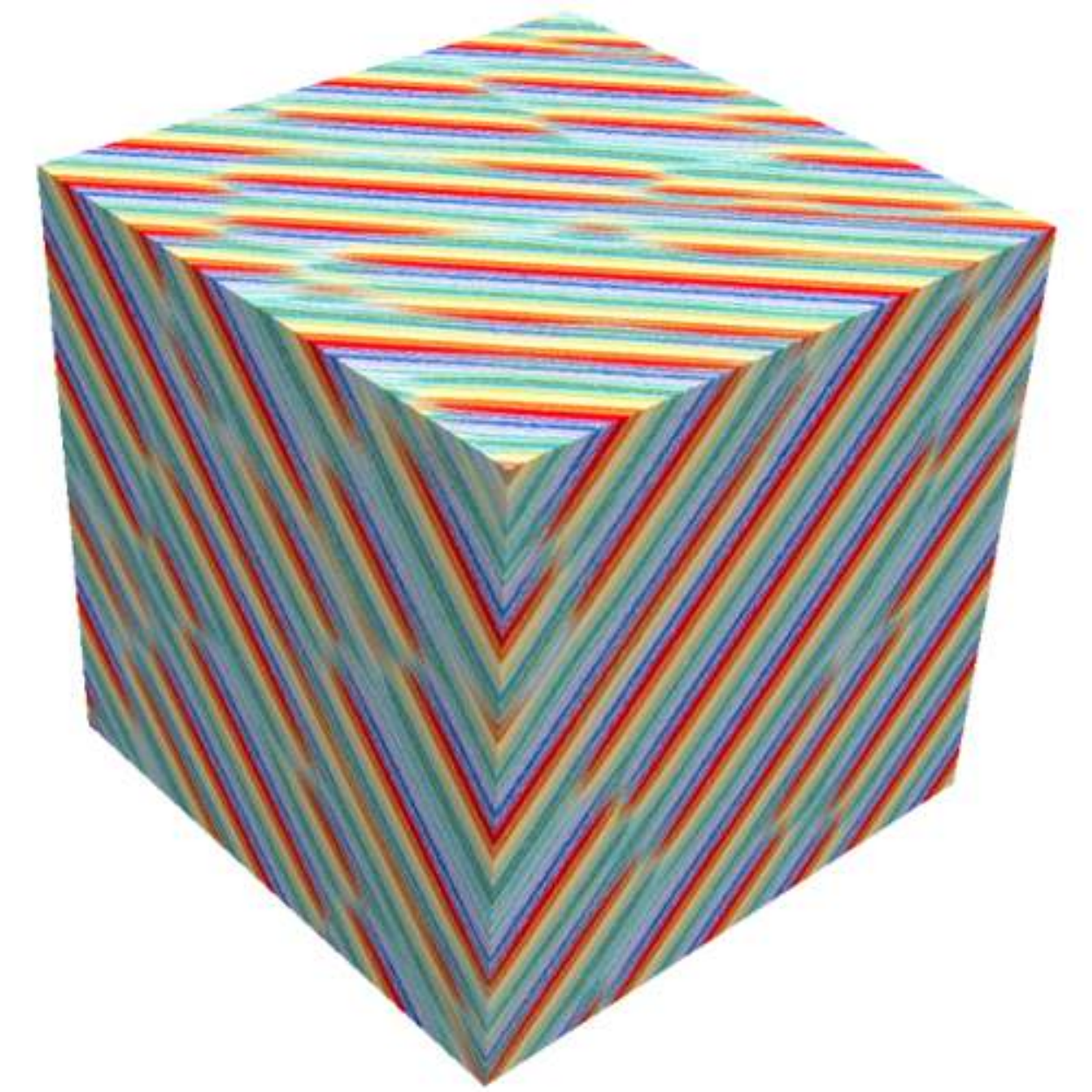}&
		\raisebox{6mm}{\includegraphics[width=2.2cm]{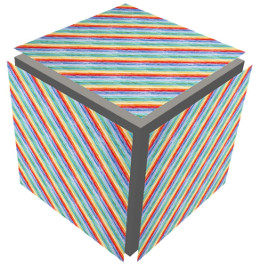}}&
		\raisebox{6mm}{\includegraphics[width=2.2cm]{diagonal-crayons2_512}}&
		\raisebox{6mm}{\includegraphics[width=2.2cm]{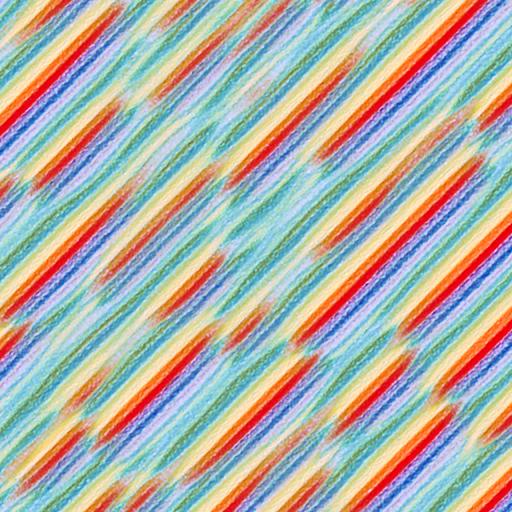}}&
		\raisebox{6mm}{\includegraphics[width=2.2cm]{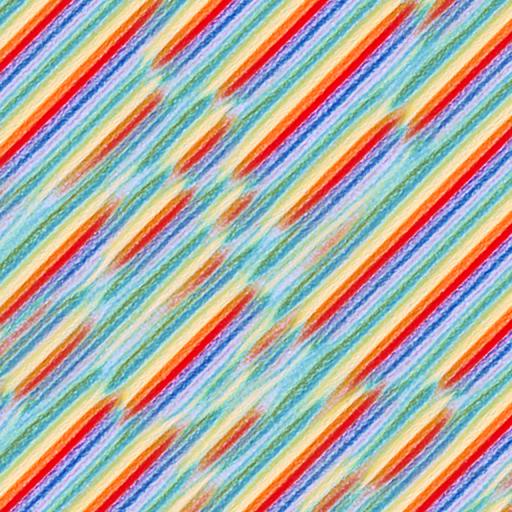}}&
		\raisebox{6mm}{\includegraphics[width=2.2cm]{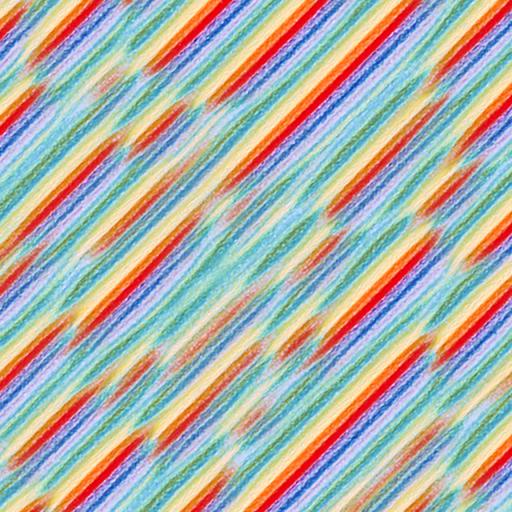}}&
		\raisebox{15mm}{$363.1$}
		\\
	\end{tabular}
	\caption{Importance of the compatibility of examples.
	In this experiment, two generators are trained with the same image along three directions, but for two different orientations.
	The first column shows generated samples of size $512^3$ built by assembling blocks of $32^3$ voxels generated using on demand evaluation.   The second column illustrates the training configuration, \ie for each direction the orientation of the example shown in the third column. Subsequent columns show the middle slices of the generated cube across the three constrained directions. Finally, the rightmost column gives the empirical loss value at the last iteration.
	In the first row, no 3D arrangement of the patterns can comply with the orientations of the three examples. 
	Conversely the configuration on the second row can be reproduced in 3D, thus generating more meaningful results. 
	The lower value of the training loss for this configuration reflects the better reproduction of the patterns in the example. 
	}
	\label{Fig:Results_diagonal_crayons}
\end{figure*}

This also might have dramatic consequences regarding convergence issues during optimization. In global optimization methods \cite{kopf2007solid,chen2010highquality}, where patches from the example are sequentially copied view per view, progressing the synthesis in one direction can create new features in the other directions thus potentially preventing convergence.
In contrast, in our method, we seek for a solid texture whose statistics are as close as possible from the examples without requiring a perfect match. 
This always ensured convergence during training in all of our experiments. 
An example of this is illustrated in the first row of Figure~\ref{Fig:Results_diagonal_crayons},  where the example views are incompatible given the proposed 3D configuration: the optimization procedure converges during training and the trained generator is able to synthesize a solid texture, however, the result is clearly a tradeoff which mixes the different contradictory orientations.
This also contrasts with common statistical texture synthesis approaches that are rather based on constrained optimization to guarantee statistical matching, for instance by projecting patches~\cite{Gutierrez2017}, histogram matching~\cite{rabin2011wasserstein}, or moment matching~\cite{portilla2000parametric}.

\begin{figure*}[!ht]
	\centering
	\begin{tabular}{cc@{\hspace{2pt}}|@{\hspace{2pt}}c@{\hspace{4pt}}c@{\hspace{4pt}}|@{\hspace{4pt}}c@{\hspace{4pt}}c@{\hspace{4pt}}c}
	
		 &Generated volume & Training  & Examples &  \multicolumn{3}{c}{Generated views}\\
		& $v$ & configuration &   &  $v_{1,\frac{N_1}{2}}$ & $v_{2,\frac{N_2}{2}}$ & $v_{3,\frac{N_3}{2}}$\\
		\rotatebox{90}{\hspace{1.0cm} soil ($D=3$)}&
		\includegraphics[width=3.6cm]{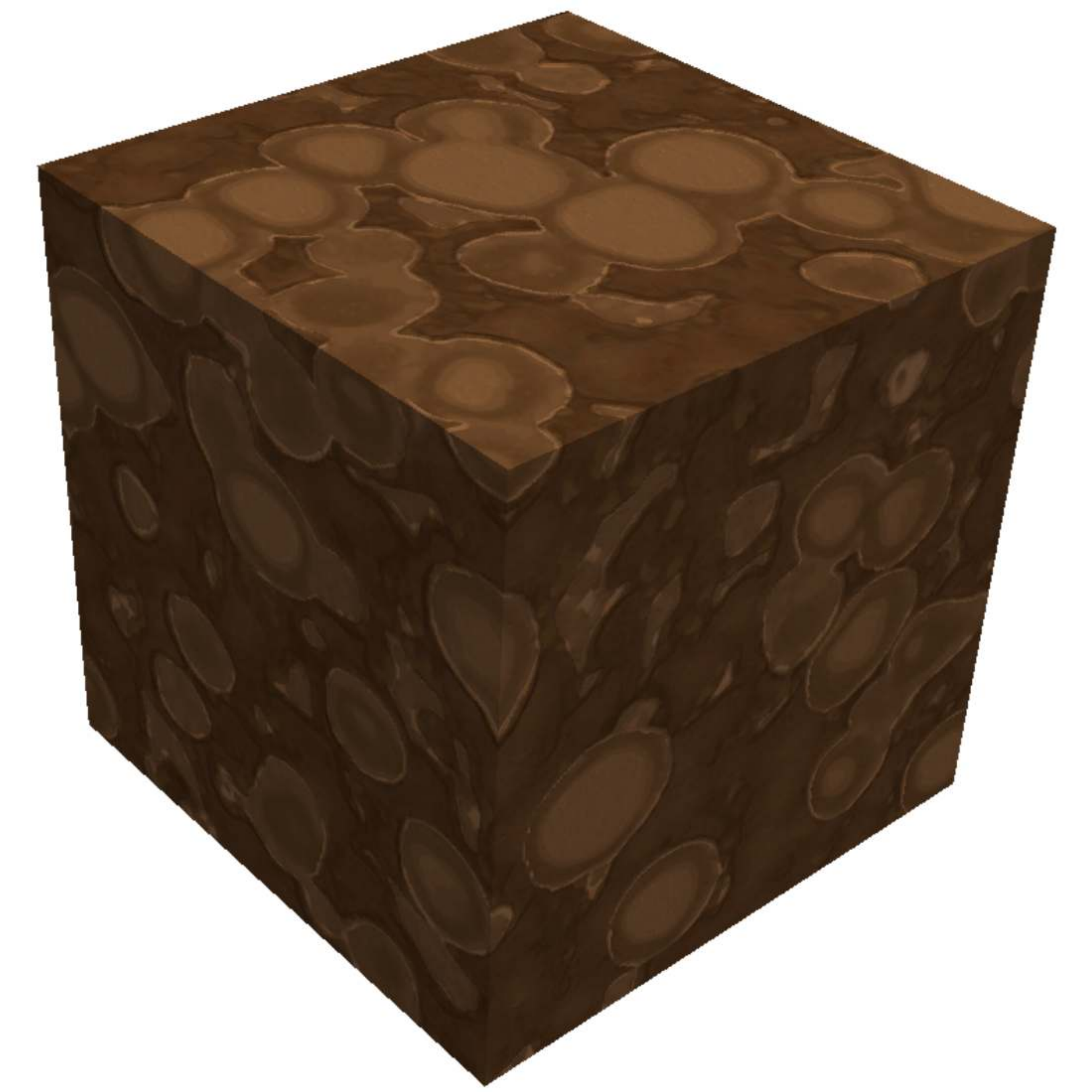}&
		\raisebox{6mm}{\includegraphics[width=2.2cm]{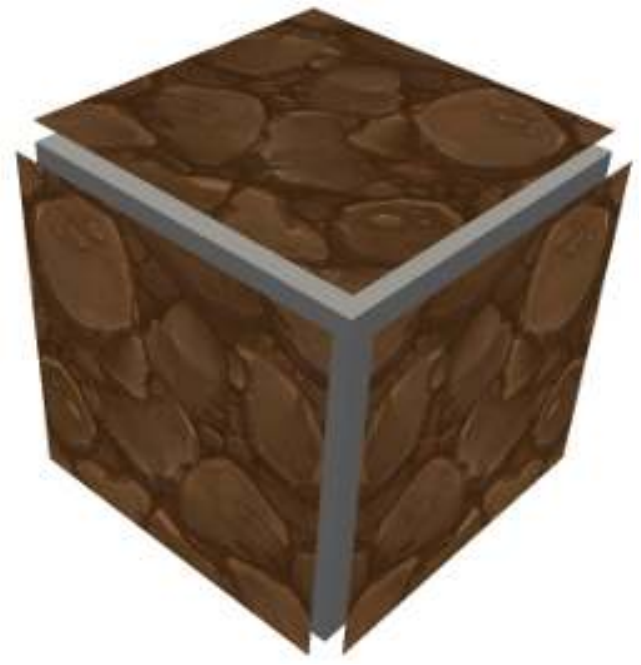}}&
		\raisebox{6mm}{\includegraphics[width=2.2cm]{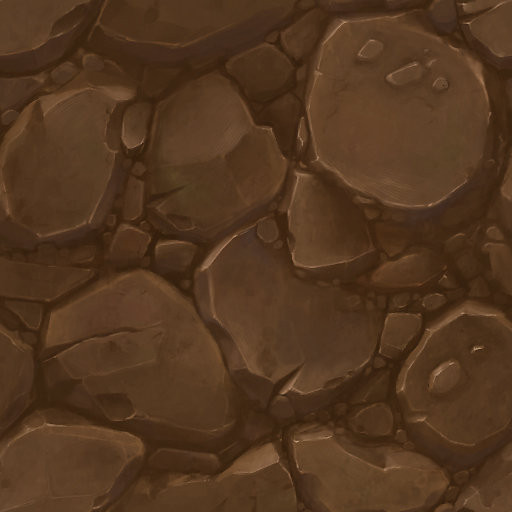}}&
		\raisebox{6mm}{\includegraphics[width=2.2cm]{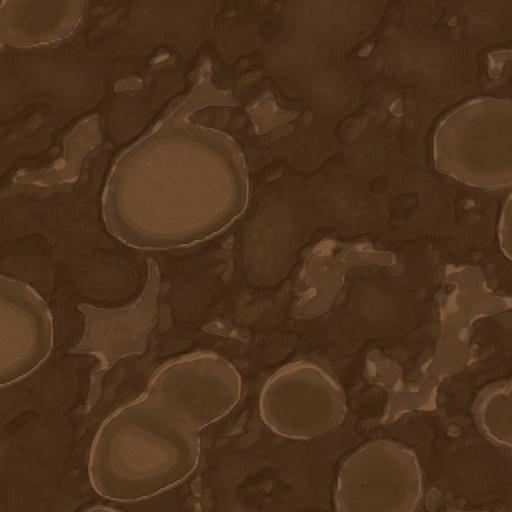}}&
		\raisebox{6mm}{\includegraphics[width=2.2cm]{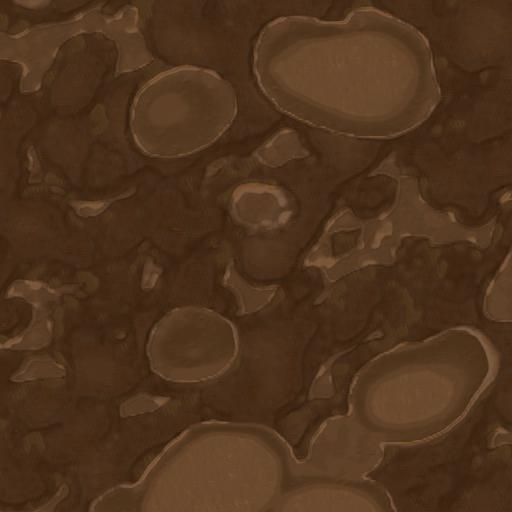}}&
		\raisebox{6mm}{\includegraphics[width=2.2cm]{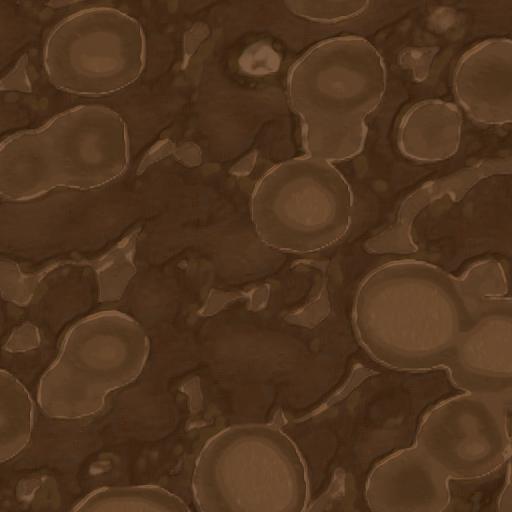}}\\
		\rotatebox{90}{\hspace{1.0cm} soil ($D=2$)}&
		\includegraphics[width=3.6cm]{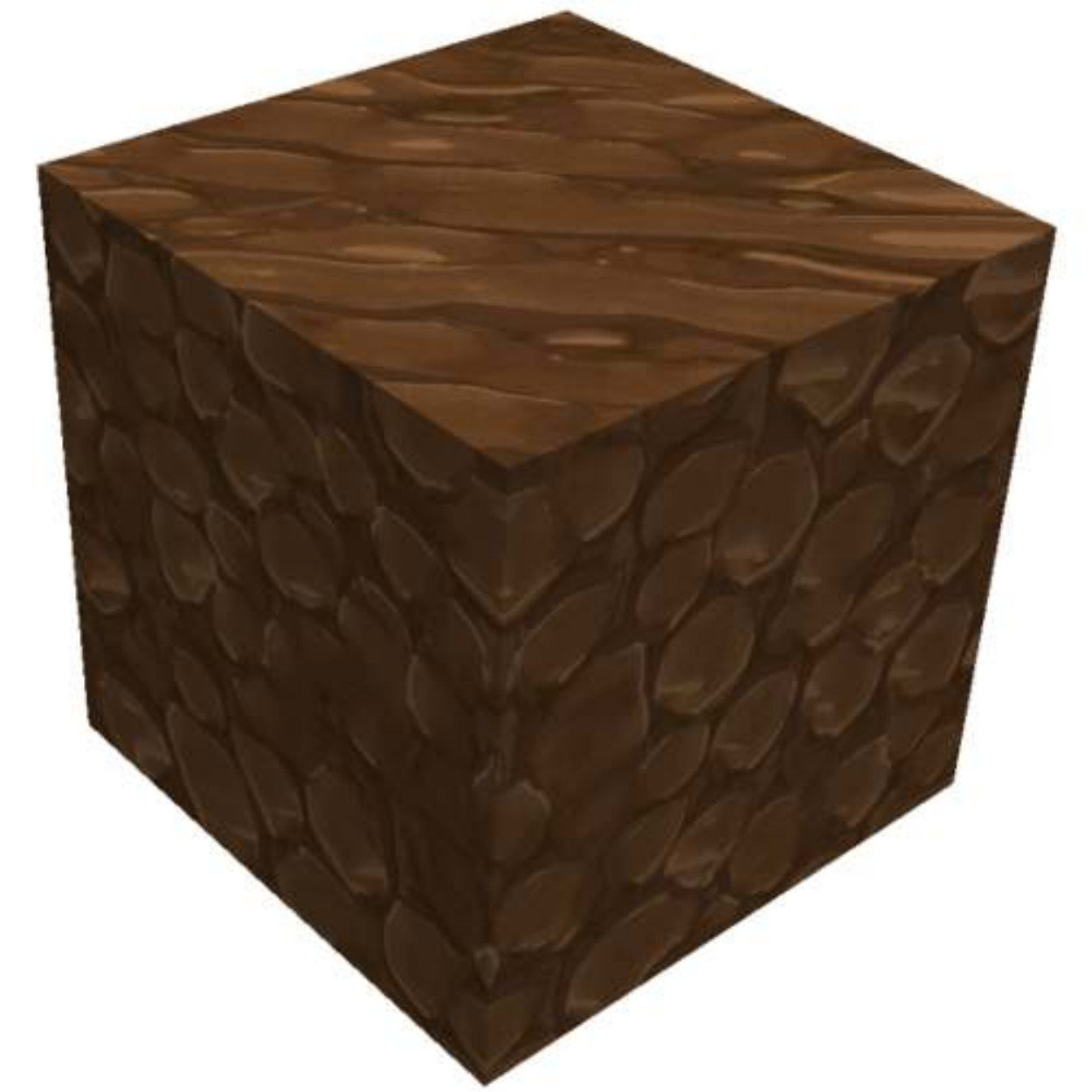}&
		\raisebox{6mm}{\includegraphics[width=2.2cm]{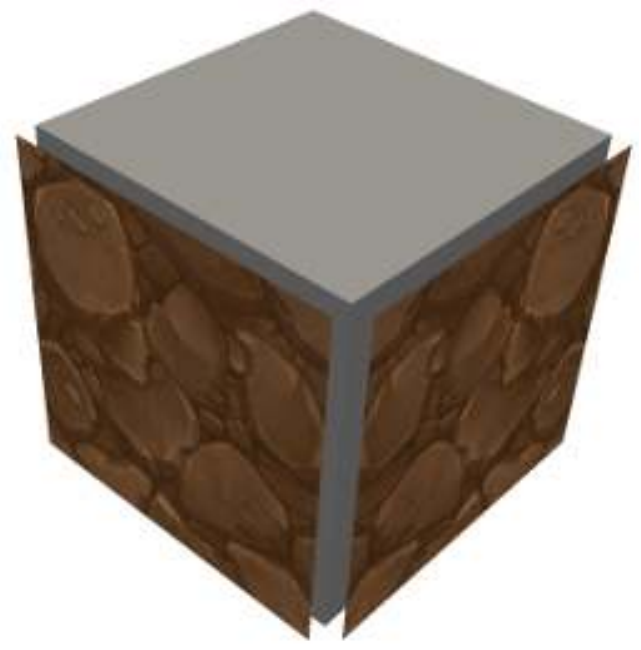}}&
		\raisebox{6mm}{\includegraphics[width=2.2cm]{ulrick-wery-tileableset2-soil}}&
		\raisebox{6mm}{\includegraphics[width=2.2cm]{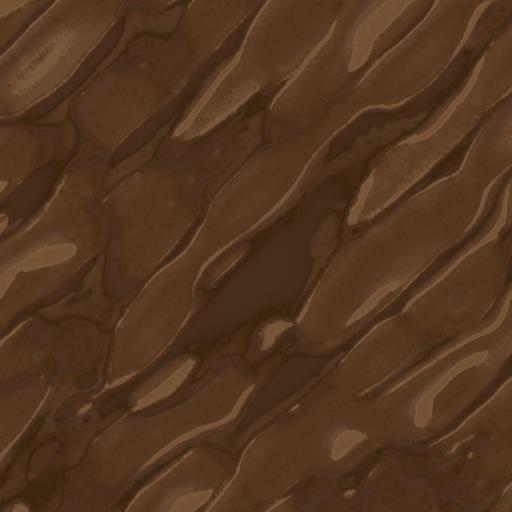}}&
		\raisebox{6mm}{\includegraphics[width=2.2cm]{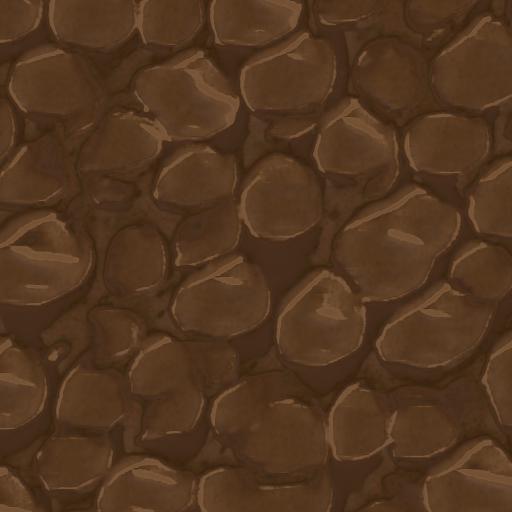}}&
		\raisebox{6mm}{\includegraphics[width=2.2cm]{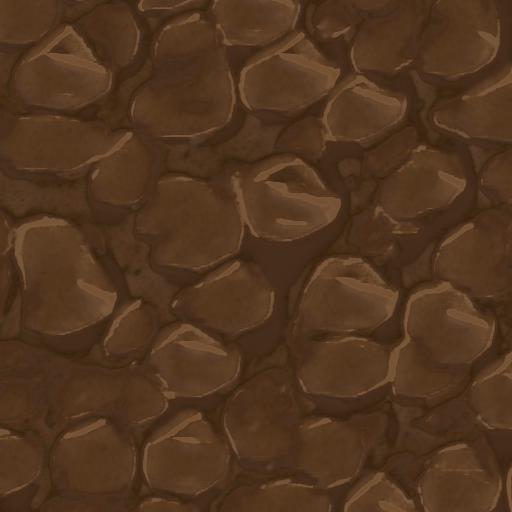}}\\
		\rotatebox{90}{\hspace{1.0cm} brick wall ($D=2$)}&
		\includegraphics[width=3.6cm]{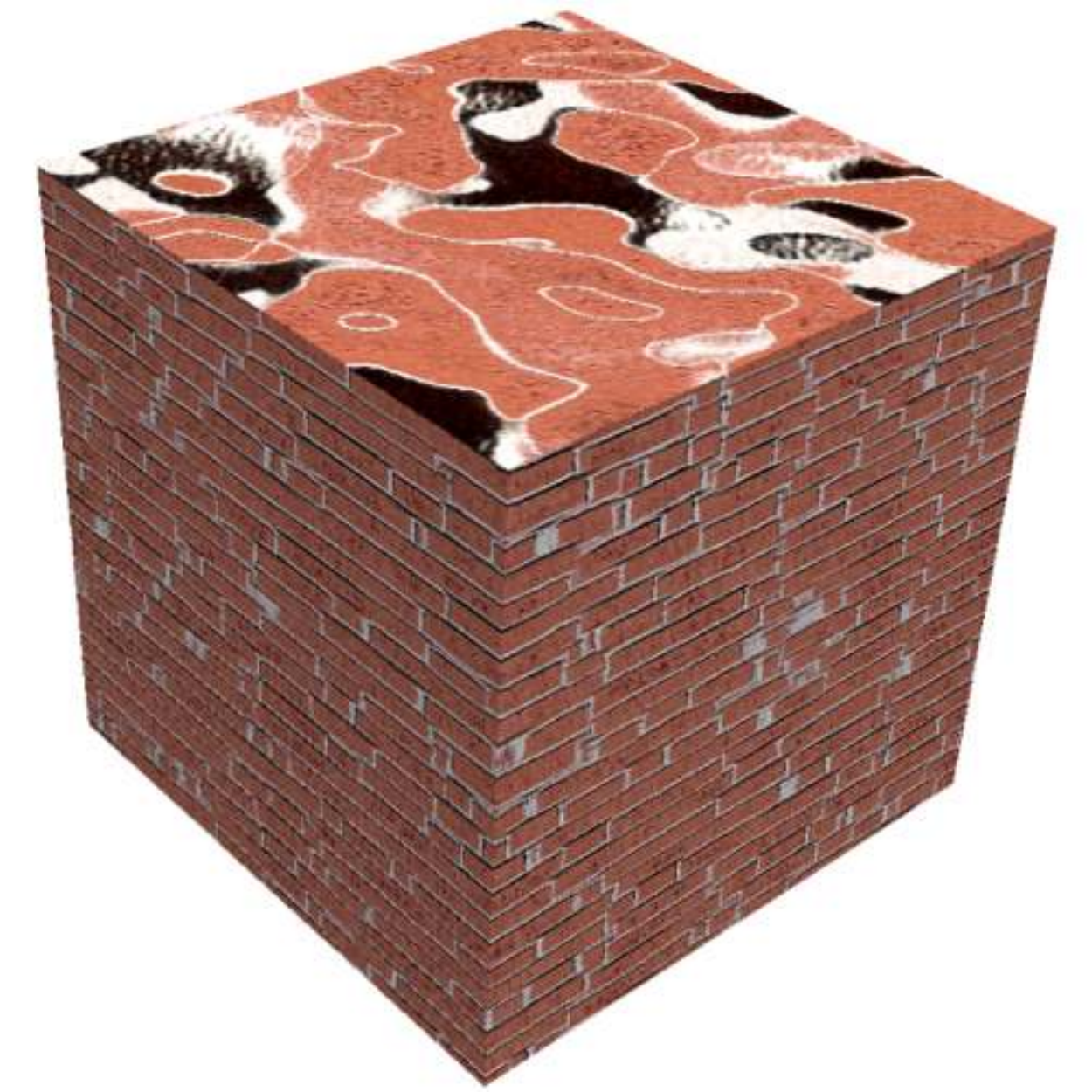}&
		\raisebox{6mm}{\includegraphics[width=2.2cm]{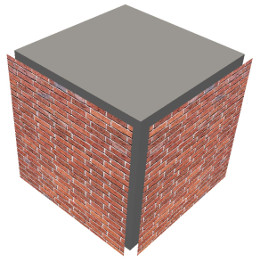}}&
		\raisebox{6mm}{\includegraphics[width=2.2cm]{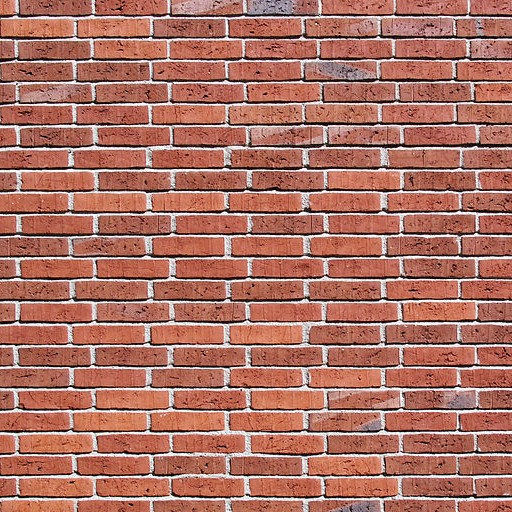}}&
		\raisebox{6mm}{\includegraphics[width=2.2cm]{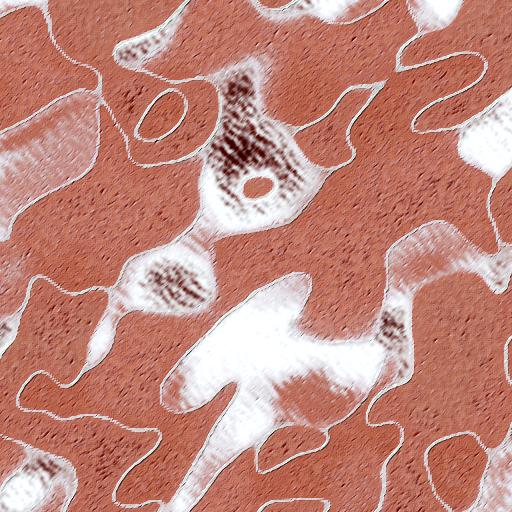}}&
		\raisebox{6mm}{\includegraphics[width=2.2cm]{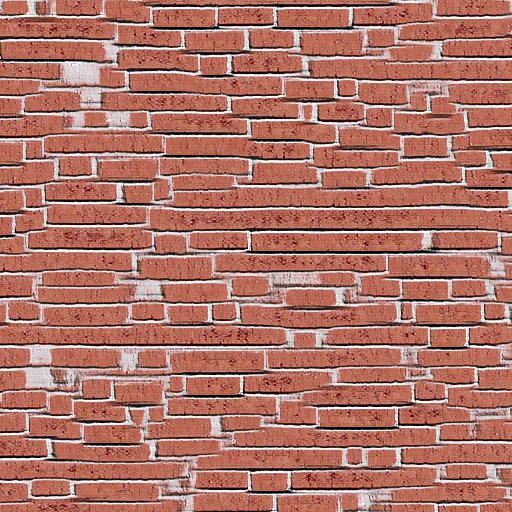}}&
		\raisebox{6mm}{\includegraphics[width=2.2cm]{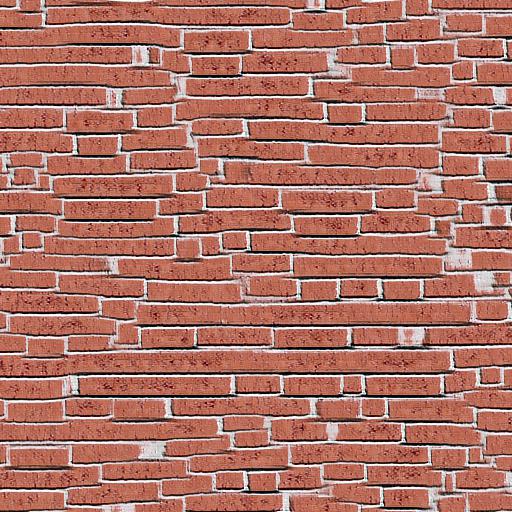}}\\
		\rotatebox{90}{\hspace{1.0cm} cobble wall ($D=2$)}&
		\includegraphics[width=3.6cm]{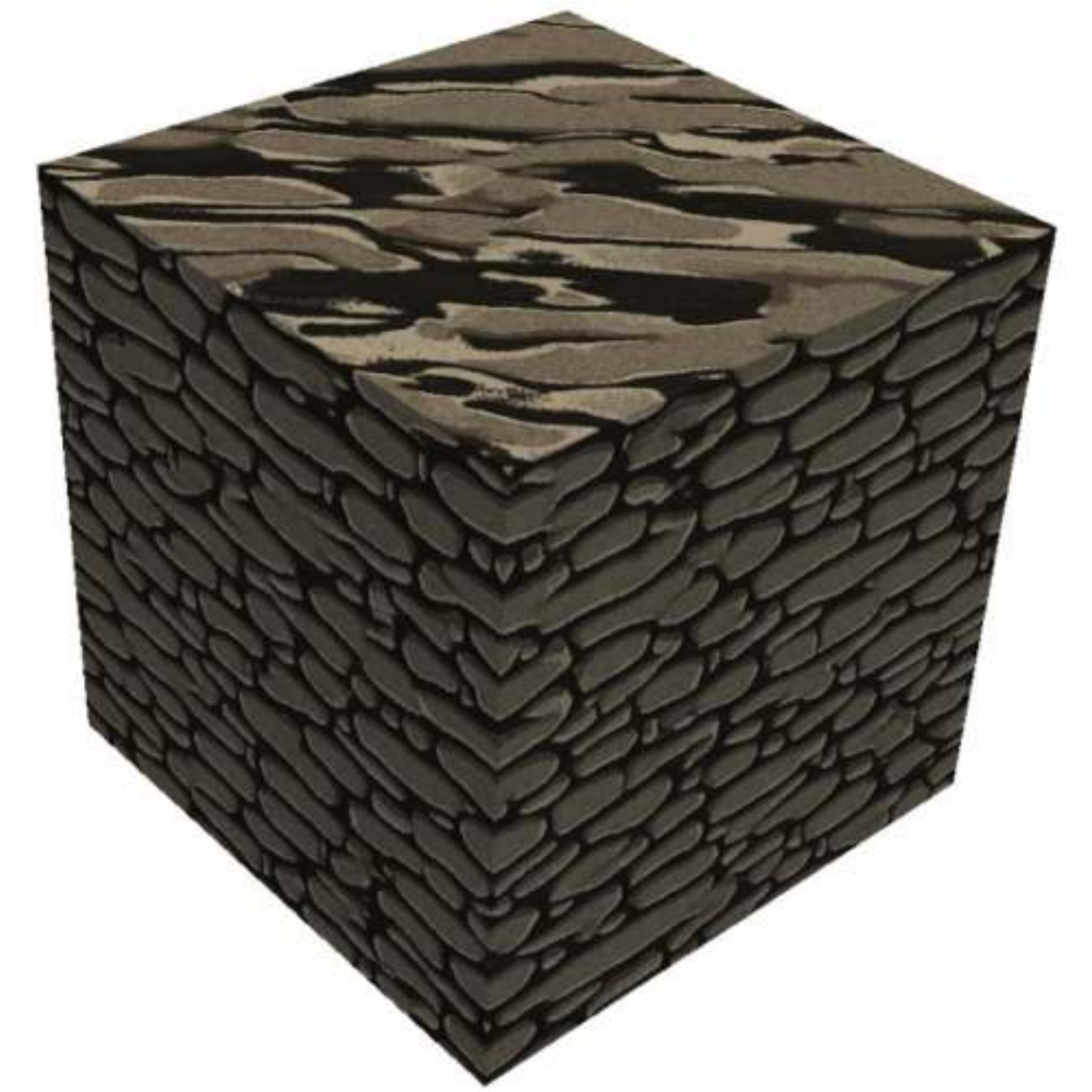}&
		\raisebox{6mm}{\includegraphics[width=2.2cm]{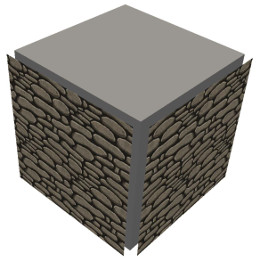}}&
		\raisebox{6mm}{\includegraphics[width=2.2cm]{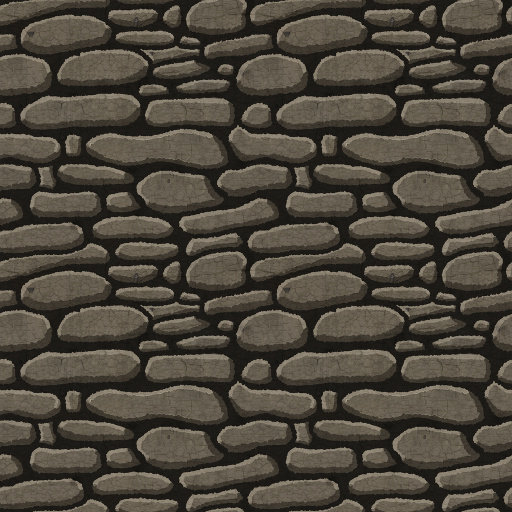}}&
		\raisebox{6mm}{\includegraphics[width=2.2cm]{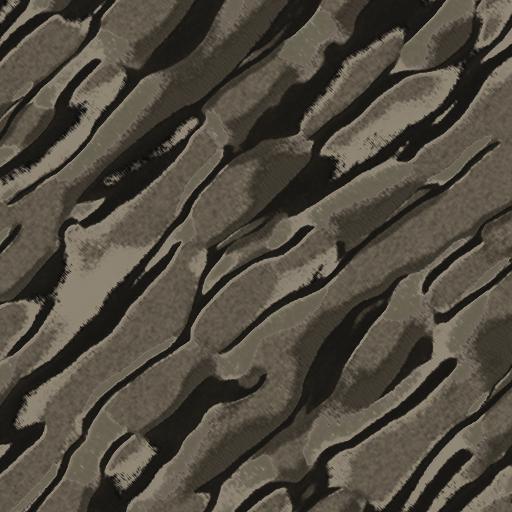}}&
		\raisebox{6mm}{\includegraphics[width=2.2cm]{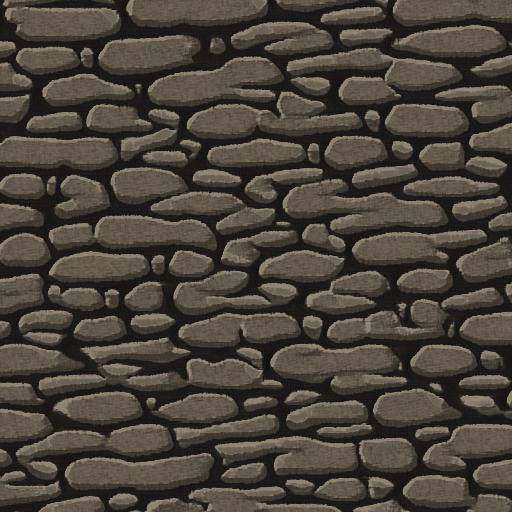}}&
		\raisebox{6mm}{\includegraphics[width=2.2cm]{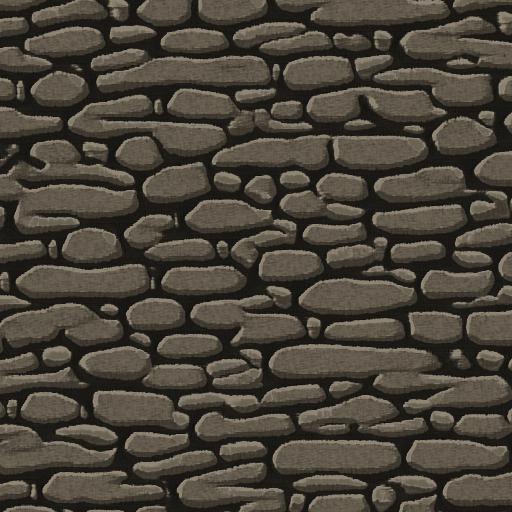}}\\
	\end{tabular}
	\caption{Training the generator using two or three directions for anisotropic textures. The first column shows generated samples of size $512^3$ built by assembling blocks of $32^3$ voxels generated using on demand evaluation. The second column illustrates the training configuration, \ie which axes are considered and the orientation used. Subsequent columns show the middle slices of the generated cube across the three considered directions. The top two rows show that for some examples considering only two directions allow the model to better match the features along the directions considered. The bottom rows show examples where the appearance along one direction might not be important.}
	\label{Fig:Results_views_2_directions}
\end{figure*}

\paragraph*{Constraining directions} While for isotropic textures, constraining $D=3$ directions gives good visual results, for some textures it may be more interesting to consider only two views.
Indeed using $D<3$ might be essential to obtain acceptable results, at least along the considered directions. This acts in accordance to the question of existence of a solution discussed in the previous paragraph. 
We exemplify this in the top two rows of Figure~\ref{Fig:Results_views_2_directions}, where considering only two training directions (second row) results in a solid texture that more closely resembles the example along those directions, compared to using three training directions (first row) which generates a more consistent volume using isotropic shapes that are not present in the example texture.
The brick and cobblestone textures in Figure~\ref{Fig:Results_views_2_directions} highlight the fact that, when depicting an object where the top view is not crucial to the desired appearance, such as a wall, it can be left to the algorithm to infer a coherent structure. 
Of course, not considering a direction during training will generate cross-sections that do not necessarily contain the same patterns as the example texture (see column $v_{1, \tfrac{N_1}{2}}$), but rather color structures that fulfill the visual requirements for the other considered directions.

Additionally, pattern compatibility across different directions is essential to obtain coherent results. In the examples of Figure~\ref{Fig:Results_diagonal_crayons} the generator was trained with the same image but in a different orientation configuration. In the top row example no 3D arrangement of the patterns can comply with the orientations of the three examples. 
Conversely the configuration on the bottom row can be reproduced in 3D thus generating more meaningful results.
All this has to be taken into account when choosing the set of training directions given the expected 3D texture structure.
Observe that the value of the loss at the end of the training gives a hint on which configuration \emph{works} better. This can be exploited for automatically finding the best training configuration for a given set of examples.

These results bring some light to the scope of the slice-based formulation for solid texture synthesis using 2D examples. This formulation is best suited for textures depicting 3D isotropic materials,
for which we obtain an agreement with the example's patterns comparable with 2D state-of-the-art methods. For most anisotropic textures we can usually obtain high quality results by considering only two directions. Finally, for textures with patterns that are only isotropic in 2D, using more than one direction inevitably creates new patterns.

\paragraph*{Diversity}
A required feature of texture synthesis models is that they are able to generate diverse samples. Ideally the generated samples are different from each other and from the example itself while still sharing its visual characteristics. Additionally, depending on the texture it might be desired that the patterns vary spatially inside each single sample. 
Yet, observe that many methods in the literature for 2D texture synthesis generate images that are local copies of the input image \cite{wei_levoy_2000,kwatra2005texture}, which strongly limits the diversity. 
As reported in Gutierrez~\etal~\cite{Gutierrez2017}, the (unwanted) optimal solution of methods based on patch optimization 
is the input image itself. For most of these methods though, the local copies are sufficiently randomized to deliver enough diversity.  
Variability issues have also been reported in the literature for texture generation based on CNN, and~\cite{Ulyanov17improved,li2017diversified} have proposed to use a diversity term in the training loss to fix it. 
Without such diversity term to promote variability, the generated samples are nearly identical from each other, although sufficiently different from the example. 
In these cases it seems that the generative networks \emph{learn} to synthesize a single sample that induces a low value of the perceptual loss while disregarding the random inputs.

\begin{figure*}[!ht]
	\centering
	\includegraphics[width = 16cm]{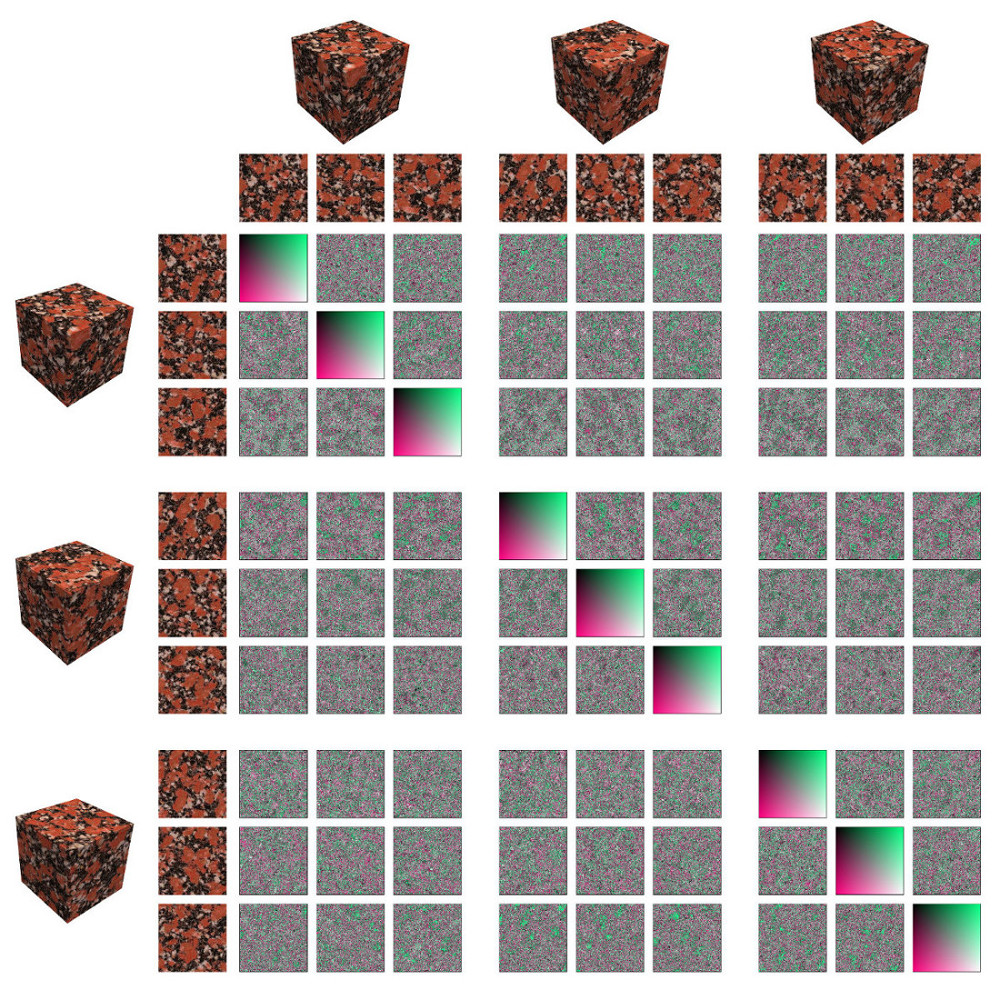}
	\caption{Diversity among generated samples. We compare the middle slices along three axis of three generated textures of $256^3$ voxels. The comparison consists in finding the pixel with the most similar neighborhood (size $4^2$) in the other image and constructing a correspondence map given its coordinates. The smooth result in the diagonal occurs when comparing a slice to itself. The stochasticity in the rest of results means that the compared slices do not share a similar arrangement of patterns and colors.}
	\label{Fig:diversity_corr_maps}
\end{figure*}

When dealing with solid texture synthesis from 2D examples, such a trivial optimal solution only arises when considering one direction, 
where the example itself is copied along such direction.
Yet, there is no theoretical guarantee that prevents the generator network from copying large pieces of the example as it has been shown that deep generative networks can memorize an image in~\cite{ulyanov18deep} .
However, the compactness or the architecture and the stochastic nature of the proposed model make it very unlikely. 
In practice, we do not observe repetition among the samples generated with our trained models, even when pursuing the optimization long after the visual convergence (which generally occurs after $1000$ iterations, see Figure~\ref{Fig:Training_curves}). 
This is consistent with the results of Ulyanov~\etal~\cite{ulyanov2016texturenets}, the 2D architecture that inspired ours, where diversity is not an issue.
One explanation for this difference with other methods may be that the architectures that exhibit a loss of diversity process an input noise that is small compared to the generated output ($0.0025\%$ in \cite{li2017diversified} and $0.13\%$ in \cite{Ulyanov17improved}) and which is easier to ignore.  
On the contrary, our generative network receives an input that accounts for roughly 1.14 times the size of the output. 
Figure~\ref{Fig:diversity_corr_maps} demonstrates the capacity of our model to generate diverse samples from a single trained generator. It shows three generated solid textures along with their middle slices. To facilitate the comparison, it includes an \emph{a posteriori} correspondence map which highlights spatial similarity by forming smooth regions. In all cases we obtain noisy maps which means that the slices do not repeat arrangements of patterns or colors.

\begin{figure}
	\centering
	\newcommand{\cmark}{\text{\ding{51}}} 
	\newcommand{\xmark}{\text{\ding{55}}}
	\begin{tabular}{c@{\hspace{4pt}}c@{\hspace{4pt}}c@{\hspace{4pt}}c@{\hspace{2pt}}|@{\hspace{2pt}}c}
		HM & \multicolumn{2}{c}{examples} &  configuration &  $v$ \\
		\raisebox{8mm}{\xmark}&
		\raisebox{3mm}{\includegraphics[width=1.6cm]{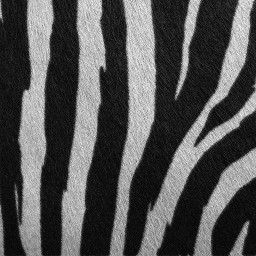}}&
		\raisebox{3mm}{\includegraphics[width=1.6cm]{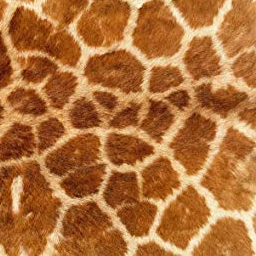}}&
		\raisebox{2mm}{\includegraphics[width=1.8cm]{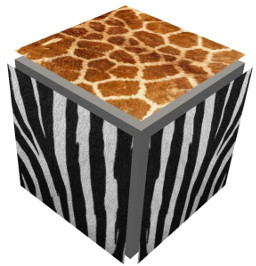}}&
		\includegraphics[width=2.4cm]{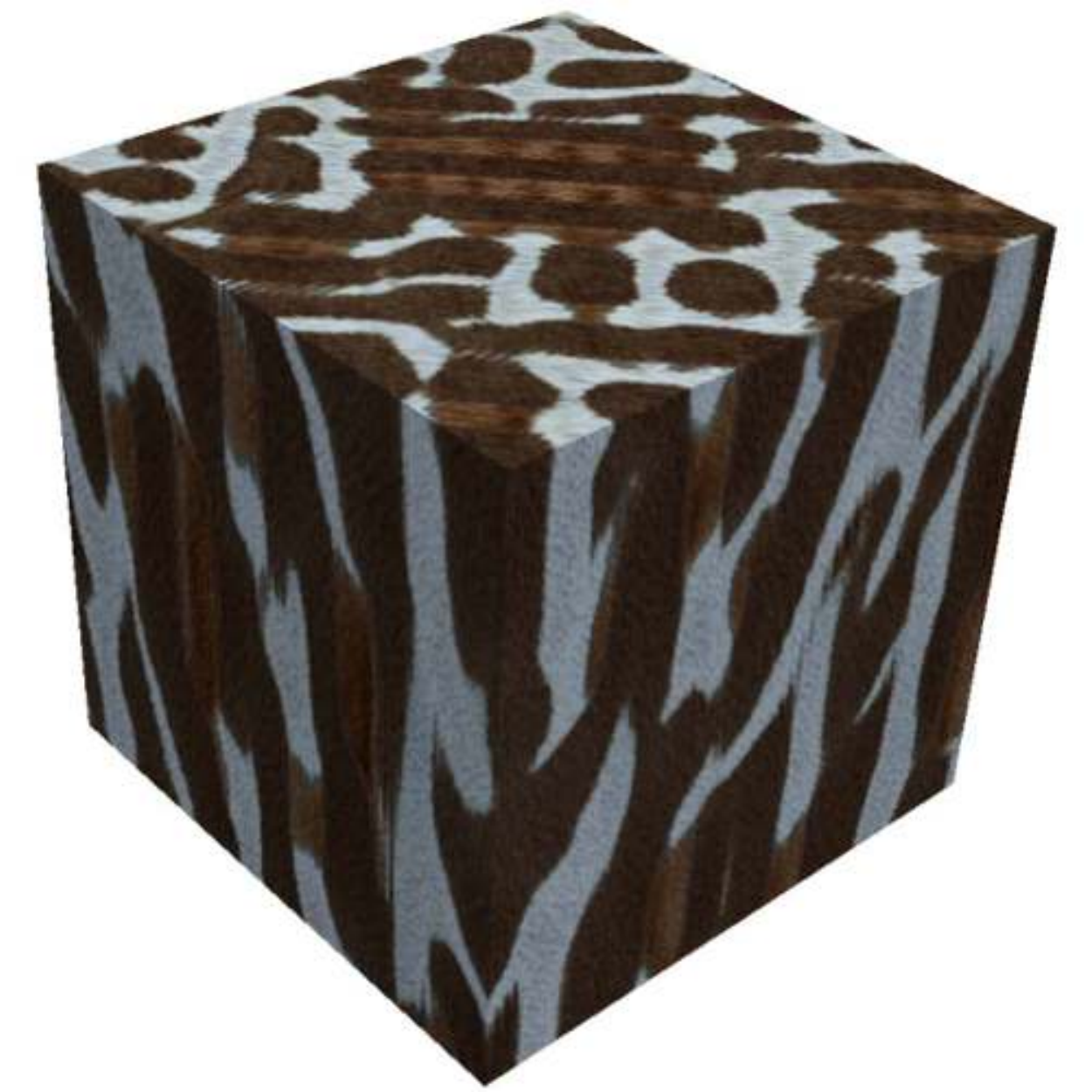}\\
		
		\raisebox{8mm}{\cmark}&
		\raisebox{3mm}{\includegraphics[width=1.6cm]{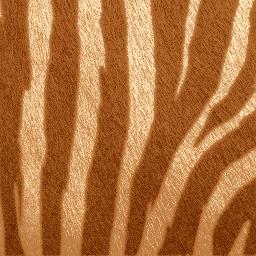}}&
		\raisebox{3mm}{\includegraphics[width=1.6cm]{giraffe}}&
		\raisebox{2mm}{\includegraphics[width=1.8cm]{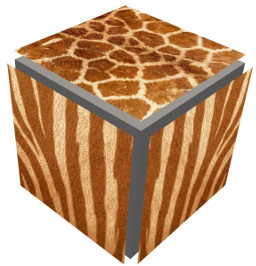}}&
		\includegraphics[width=2.4cm]{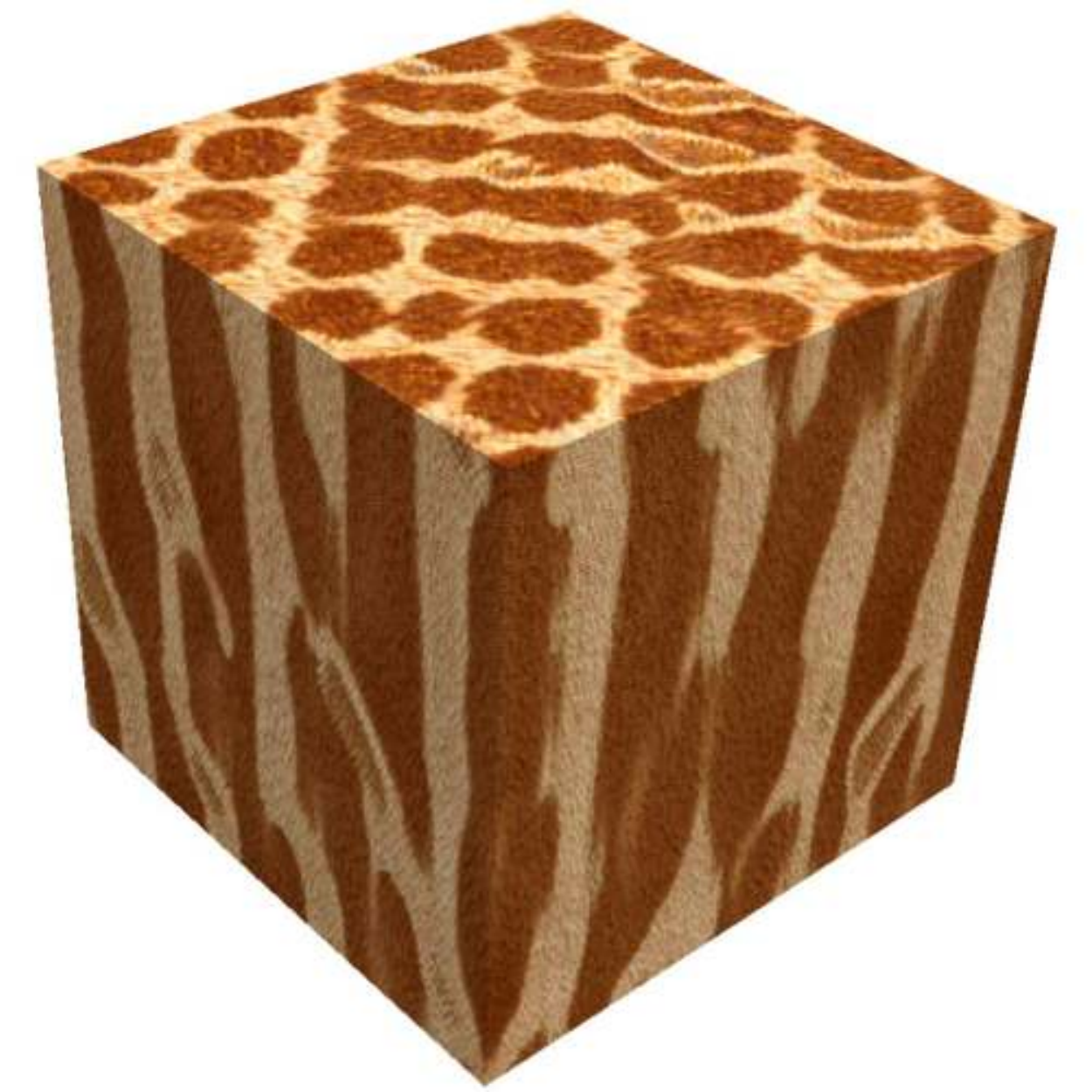}\\
		\hline 
		
		\raisebox{8mm}{\xmark}&
		\raisebox{3mm}{\includegraphics[width=1.6cm]{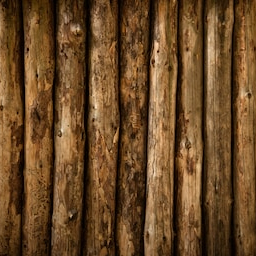}}&
		\raisebox{3mm}{\includegraphics[width=1.6cm]{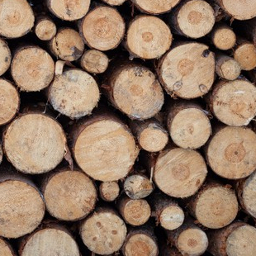}}&
		\raisebox{2mm}{\includegraphics[width=1.8cm]{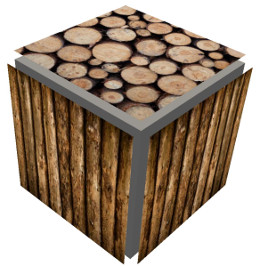}}&
		\includegraphics[width=2.4cm]{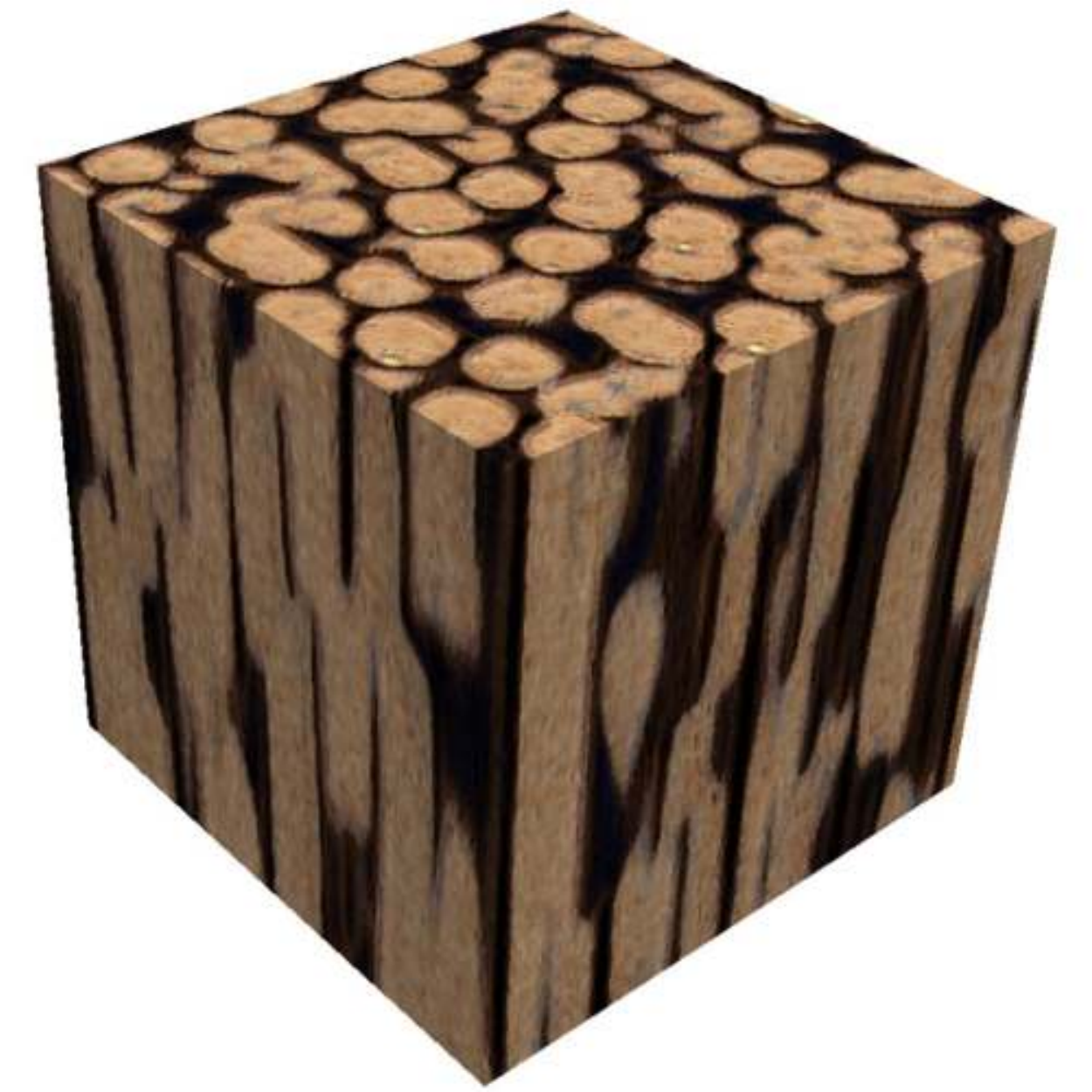}\\
		
		\raisebox{8mm}{\cmark}&
		\raisebox{3mm}{\includegraphics[width=1.6cm]{log-wall}}&
		\raisebox{3mm}{\includegraphics[width=1.6cm]{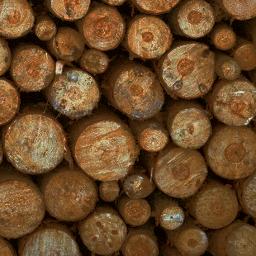}}&
		\raisebox{2mm}{\includegraphics[width=1.8cm]{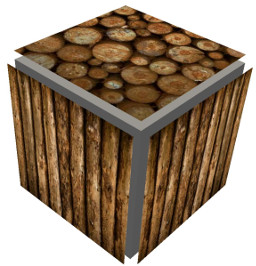}}&
		\includegraphics[width=2.4cm]{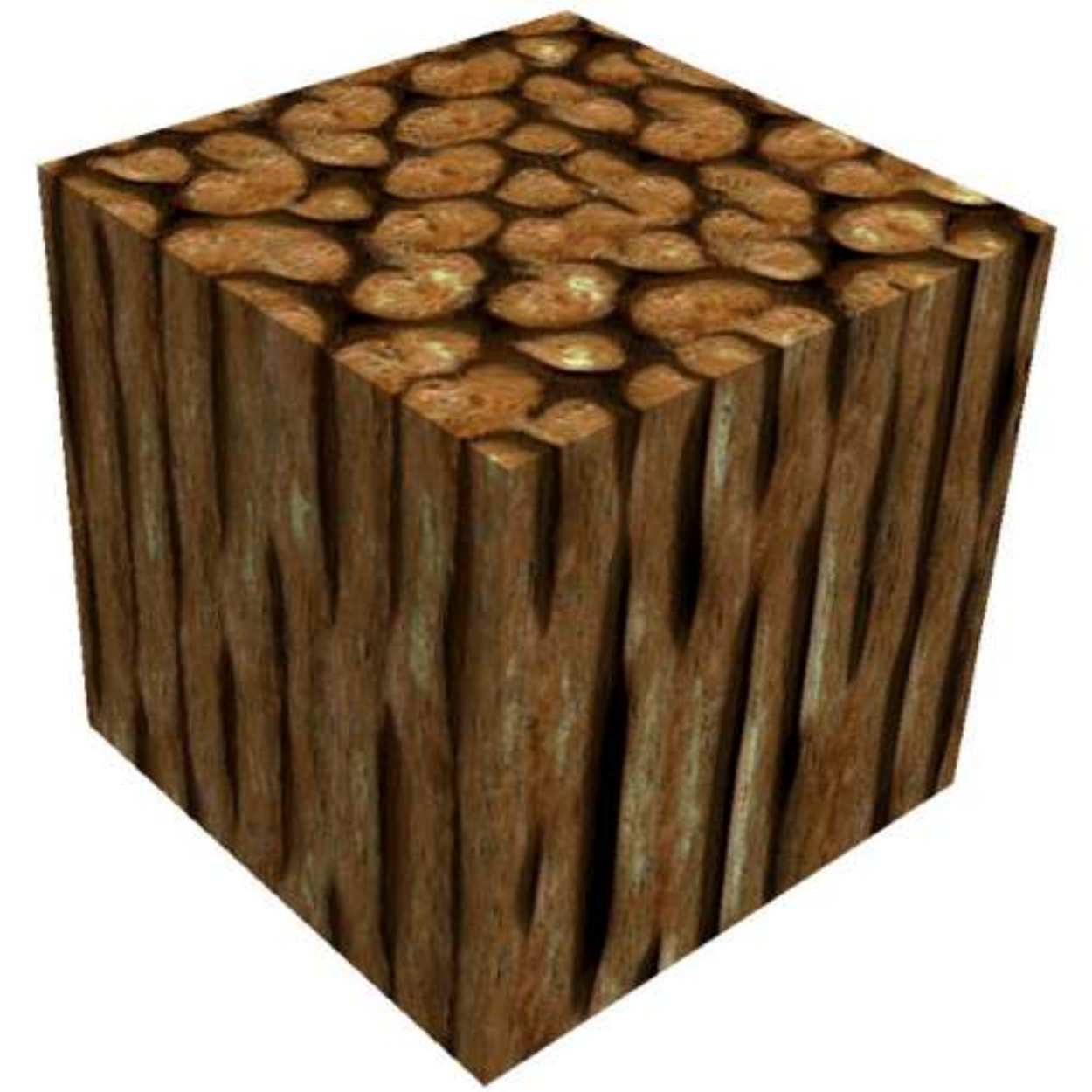}\\
		\hline 
		
		\raisebox{8mm}{\xmark}&
		\raisebox{3mm}{\includegraphics[width=1.6cm]{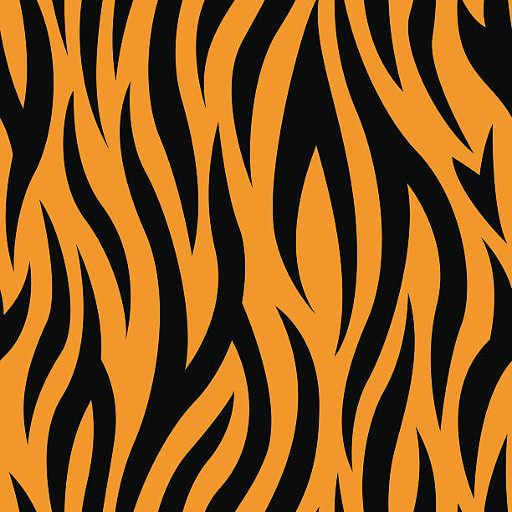}}&
		\raisebox{3mm}{\includegraphics[width=1.6cm]{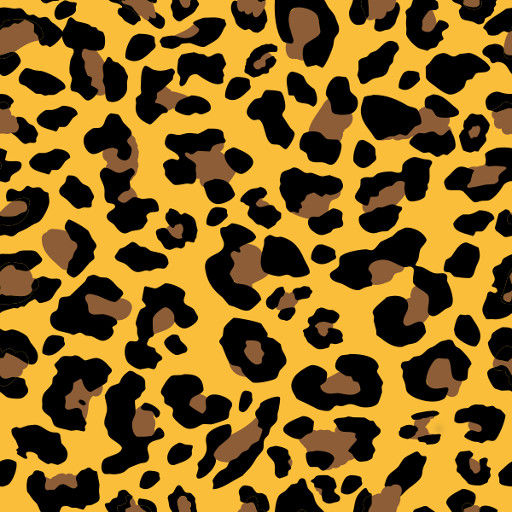}}&
		\raisebox{2mm}{\includegraphics[width=1.8cm]{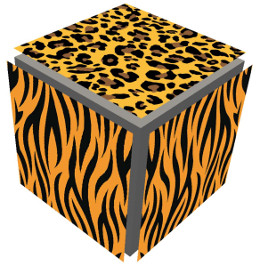}}&
		\includegraphics[width=2.4cm]{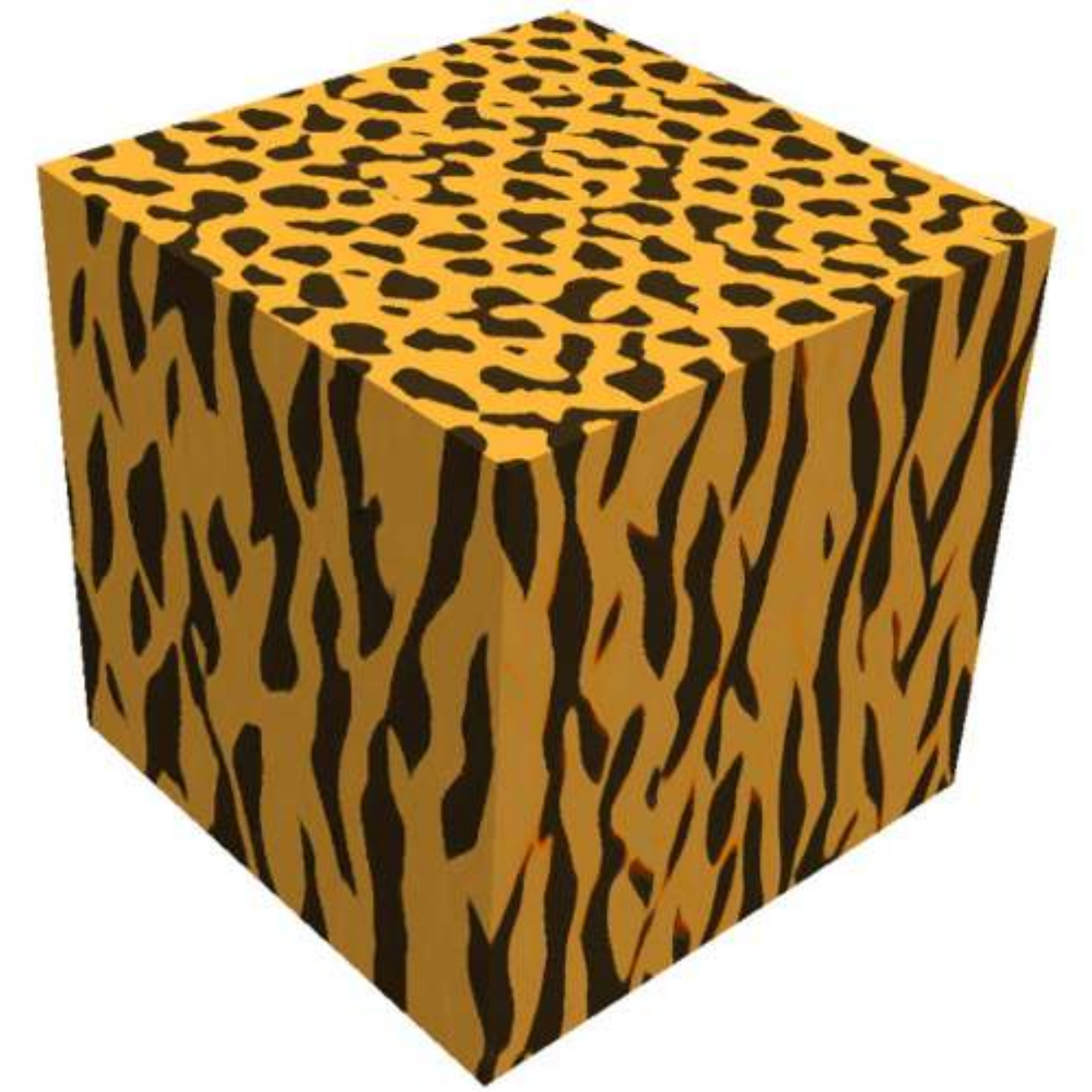}\\
		
		\raisebox{8mm}{\cmark}&
		\raisebox{3mm}{\includegraphics[width=1.6cm]{tiger_cartoon}}&
		\raisebox{3mm}{\includegraphics[width=1.6cm]{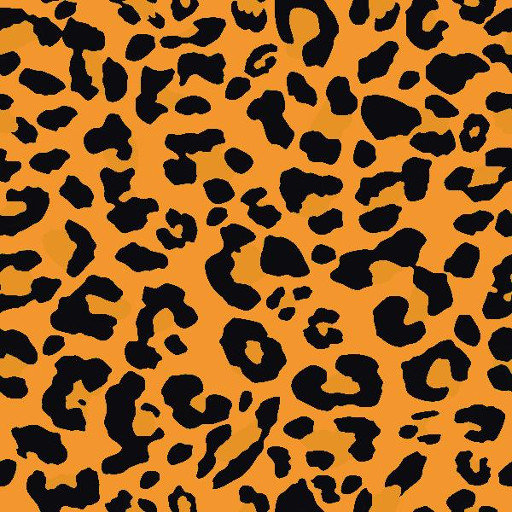}}&
		\raisebox{2mm}{\includegraphics[width=1.8cm]{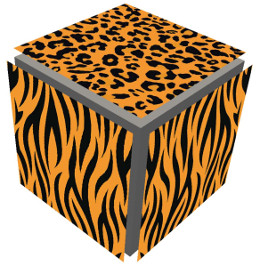}}&
		\includegraphics[width=2.4cm]{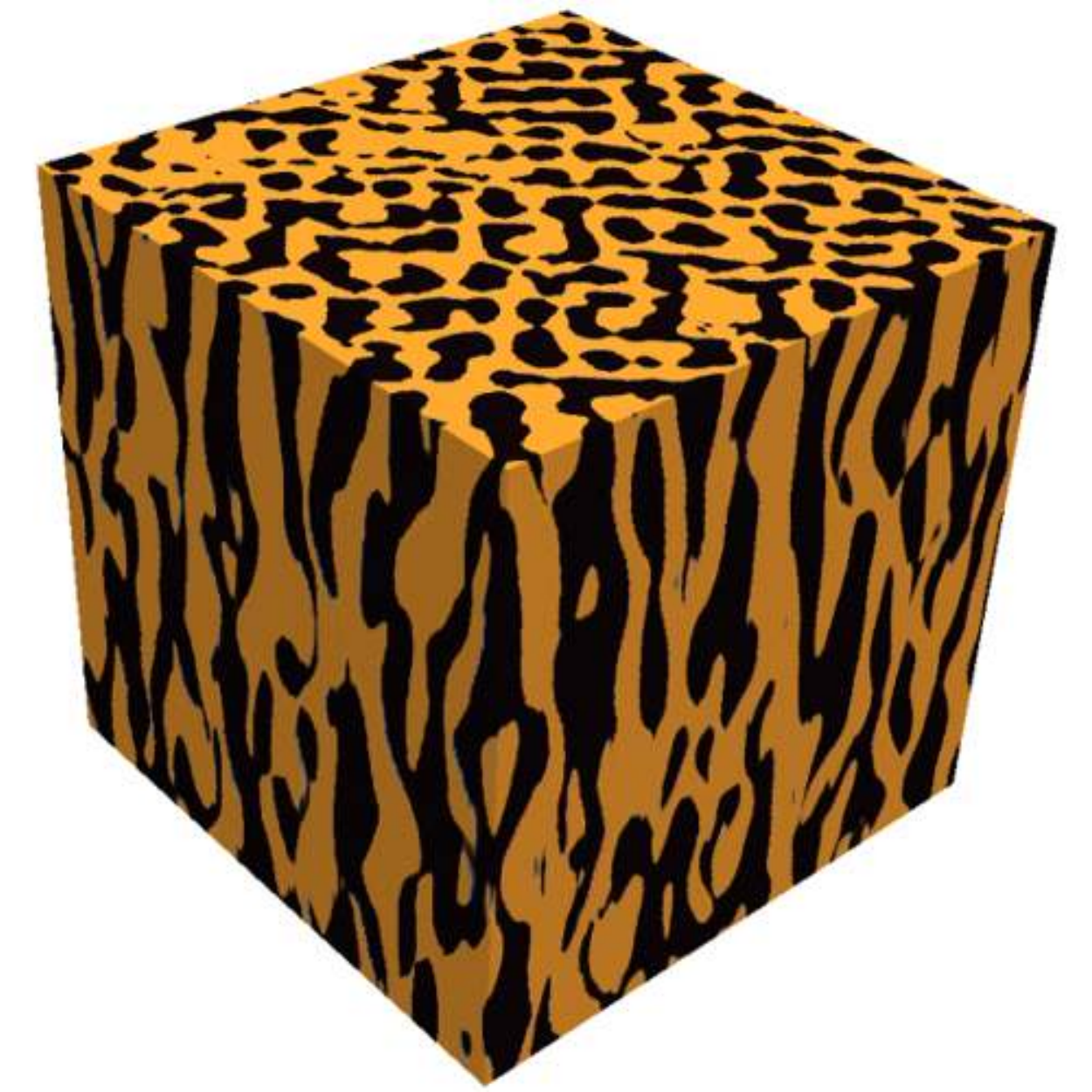}\\
	\end{tabular}
	\caption{
	{Anisotropic} texture synthesis using two examples. 
	The first columns show the two examples used and the training configuration, \ie how images are oriented for each view. 
	Last column shows a sample synthesized using the trained generator.
	For each example, we experiment with (\cmark) and without (\xmark) preprocessing the example images to match color statistics, by performing a histogram matching (HM) on each color channel independently.
	We observe favorable results particularly when the colors of both examples are close.}
	\label{Fig:Results_2differentinputs}
\end{figure}

\paragraph*{Multiple examples setting}
As already discussed in the work of Kopf~\etal~\cite{kopf2007solid} and earlier, it appears that most solid textures can only be modeled from a single example along different directions.
In the literature, to the best of our knowledge, only one success case of a 3D texture using two different 
examples has been proposed~\cite{kopf2007solid,dong2008lazy}.
This is due to the fact that the two examples have to share similar features such as color distribution and compatible geometry as already shown in Figure~\ref{Fig:Results_diagonal_crayons}.
Figure~\ref{Fig:Results_2differentinputs} illustrates this phenomenon, for each example we experiment with and without performing a histogram matching (independently for each color channel) to the input examples. We observe favorable results particularly when the colors of both examples are close. Although the patterns are not perfectly reproduced, the 3D structure is coherent and close to the examples. 

\begin{figure*}
	\centering
	\begin{tabular}{ccccc}
	\raisebox{10mm}{Example $u$} & 
	\raisebox{5mm}{\includegraphics[width=1.4cm]{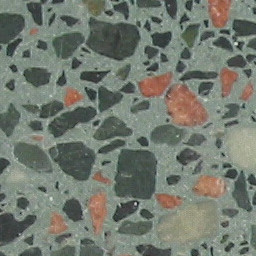}} &
	\raisebox{5mm}{\includegraphics[width=1.4cm]{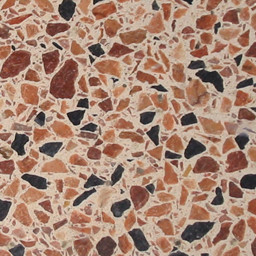}} &
	\raisebox{5mm}{\includegraphics[width=1.4cm]{brown016_exemplar}} &
	\raisebox{5mm}{\includegraphics[width=1.4cm]{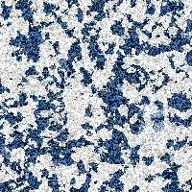}}\\
	\raisebox{15mm}{\cite{kopf2007solid}} &
	\includegraphics[width=2.8cm]{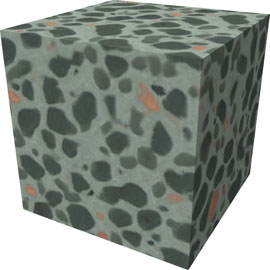} &
	\includegraphics[width=2.8cm]{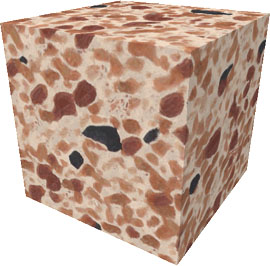} &
	&
	\\
	\raisebox{15mm}{\cite{chen2010highquality}} &
	\includegraphics[width=2.8cm]{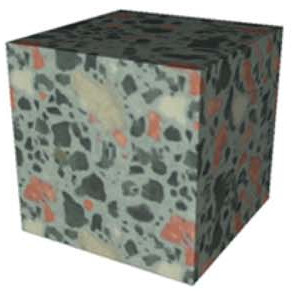} & 
	\includegraphics[width=2.8cm]{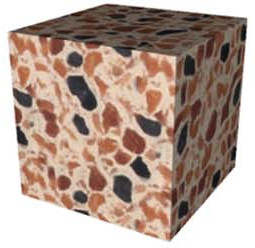} &
	\includegraphics[width=2.8cm]{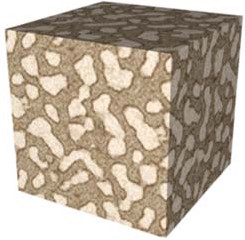} &
	\includegraphics[width=2.8cm]{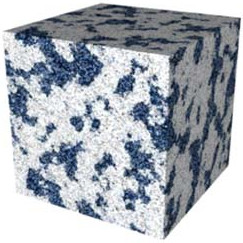}
	\\
	\raisebox{15mm}{Ours} &
	\includegraphics[width=3.0cm]{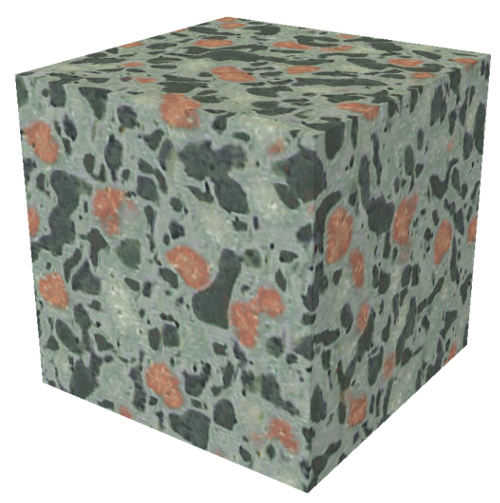}& 
	\includegraphics[width=3.0cm]{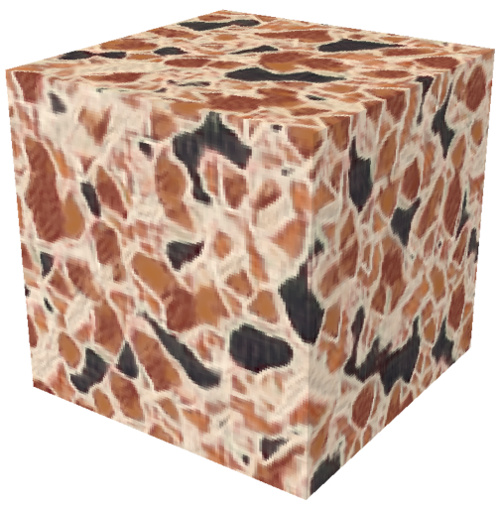}&
	\includegraphics[width=3.0cm]{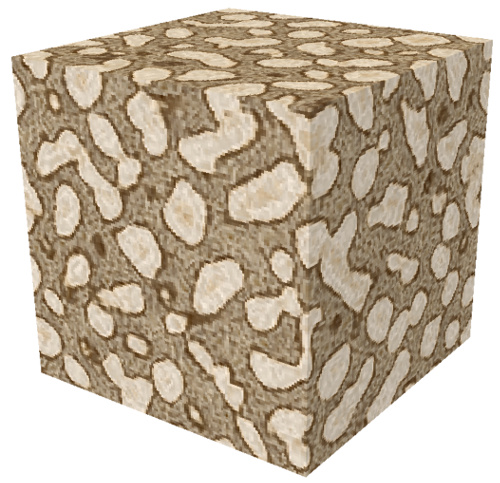}&
	\includegraphics[width=3.0cm]{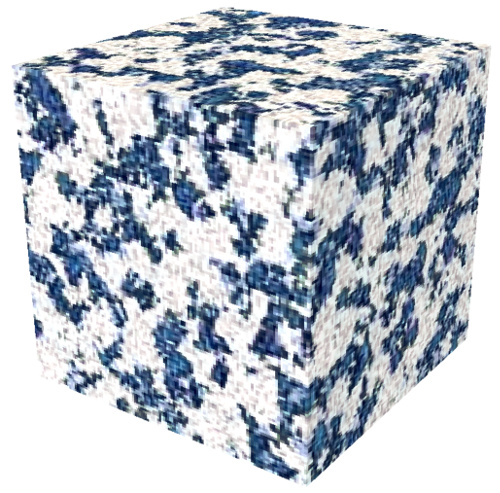} 
	\end{tabular}
	\caption{Comparison with the existing methods that produce the best visual quality.
	The last row show the results using the proposed method. Our method is better at reproducing the statistics of the example compared to \cite{kopf2007solid} and is better at capturing high frequencies compared to both methods.
	}	
	\label{Fig:Comparison}
\end{figure*}

\subsection{Comparison with state-of-the-art}
The results in Figures \ref{Fig:Results_isotropic1} and \ref{Fig:Results_isotropic2} prove the capability of the proposed model to synthesize photo-realistic textures. This is an important improvement with respect to classical on demand methods based on Gabor/LRP-noise.
High resolution examples are important in order to obtain more detailed textures. In previous high quality methods \cite{kopf2007solid,chen2010highquality} the computation times increase substantially with the size of the example, so examples were limited to $256^2$ pixels. Furthermore, the empirical histogram matching steps create a bottleneck for parallel computation.
Our method takes a step forward by allowing higher resolution example textures, depending only on the memory of the GPU used. 

We compare the visual quality of our results with the two existing methods that seem to produce the best results: Kopf~\etal~\cite{kopf2007solid} and  Chen~\etal~\cite{chen2010highquality}. 
Figure~\ref{Fig:Comparison} shows some samples obtained from the respective articles or websites side by side with results using our method. 
The most salient advantage of our method is the ability to better capture high frequency information, making the structures in the samples sharper and more photo-realistic. Considering voxels' statistics, \ie capturing the richness of the example, both our method and that of Chen~\etal~\cite{chen2010highquality} seem to obtain a better result than the method of Kopf~\etal~\cite{kopf2007solid}. 
The examples used in Figure~\ref{Fig:Comparison} have resolutions of either $128^2$ or $256^2$ pixels.
We observe that the visual quality of the textures generated with our method deteriorate when using small examples. This can be due to the descriptor network  which is pre-trained using bigger images.

We do not consider the method of Dong~\etal~\cite{dong2008lazy} for a visual quality comparison as their pre-computation of candidates limits the richness of information, which yields lower quality results. However, thanks to the on-demand evaluation, this model greatly surpasses the computation speeds of the other methods. 
Yet, as detailed before, our method is faster during synthesis while achieving better visual quality. Besides, our computation time does not depend on the resolution of the examples.

\begin{figure}
	\centering
	\begin{tabular}{ccc}
	\raisebox{10mm}{\includegraphics[width=0.85cm]{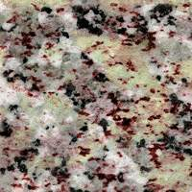}} & 
	\includegraphics[width=2.4cm]{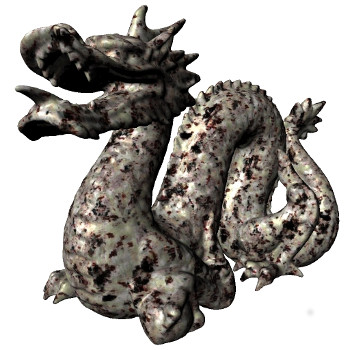} & 
	\includegraphics[width=2.4cm]{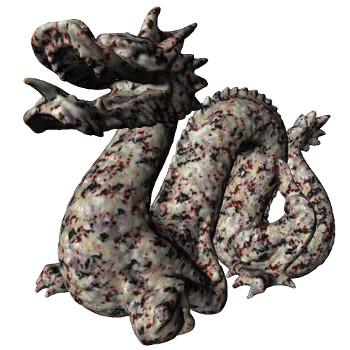}\\
	
	\raisebox{10mm}{\includegraphics[width=0.85cm]{jagnow_ex}}  & 
	\raisebox{10mm}{\includegraphics[width=0.85cm]{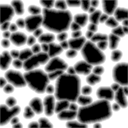}}
	\includegraphics[width=2.4cm]{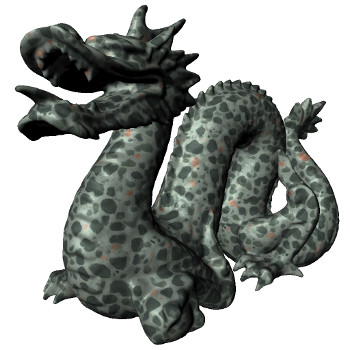} &
	\includegraphics[width=2.4cm]{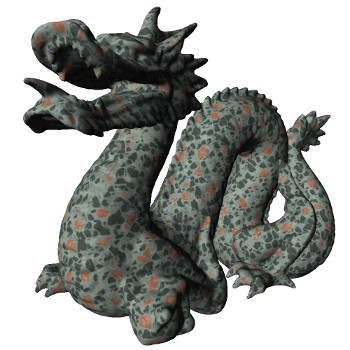}  \\
	
	\raisebox{10mm}{\includegraphics[width=0.85cm]{brown016_exemplar}} &  
	\raisebox{10mm}{\includegraphics[width=0.85cm]{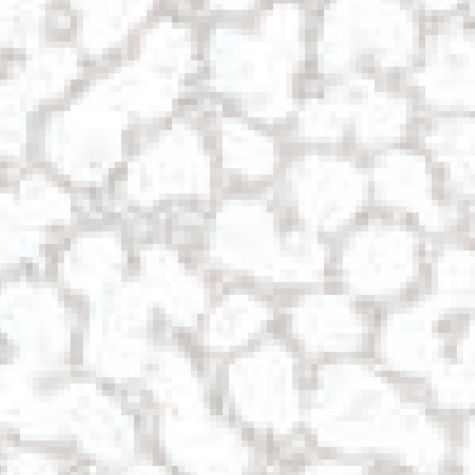}} 
	\includegraphics[width=2.4cm]{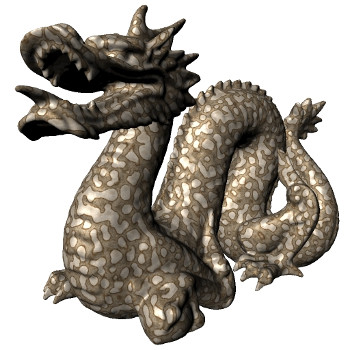} & 
	\includegraphics[width=2.4cm]{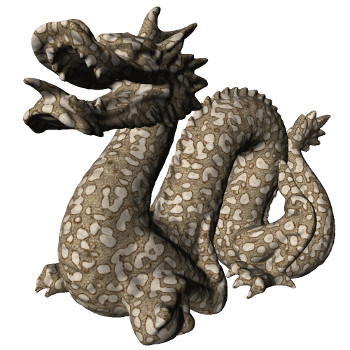}\\
	
	\raisebox{10mm}{\includegraphics[width=0.85cm]{jagnow2_exemplar}} & 
	\raisebox{10mm}{\includegraphics[width=0.85cm]{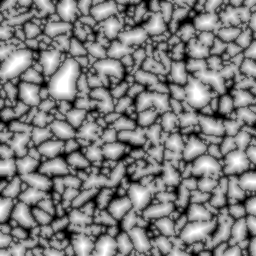}}
	\includegraphics[width=2.4cm]{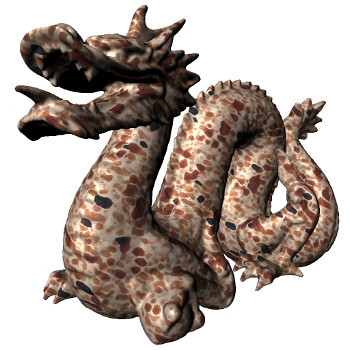} &
	\includegraphics[width=2.4cm]{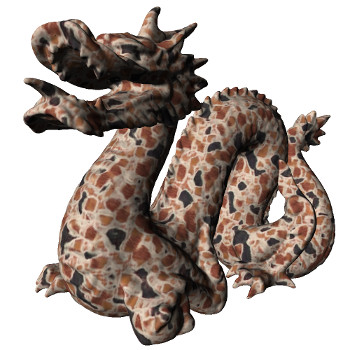}\\
	
	\raisebox{10mm}{\includegraphics[width=0.85cm]{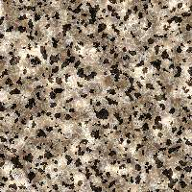}} & 
	\includegraphics[width=2.4cm]{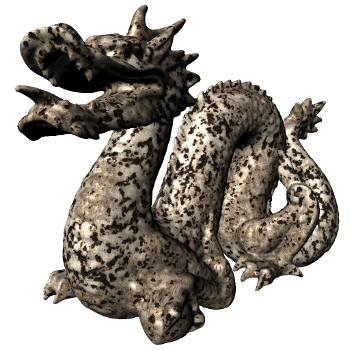} &
	\includegraphics[width=2.4cm]{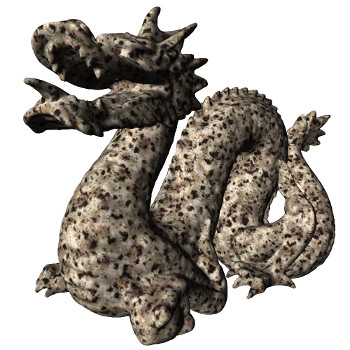}\\

	\raisebox{10mm}{\includegraphics[width=0.85cm]{RGran_Cool_1_192}} &  
	\includegraphics[width=2.4cm]{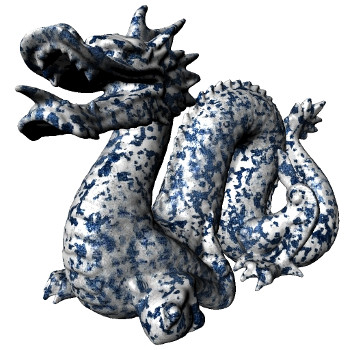} & 
	\includegraphics[width=2.4cm]{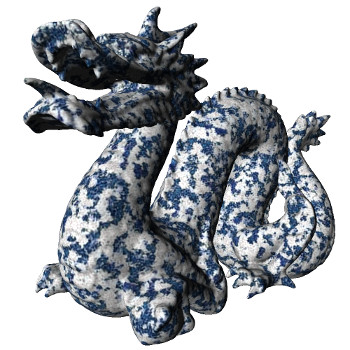}\\
	\raisebox{10mm}{\includegraphics[width=0.85cm]{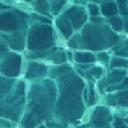}} & 
	\includegraphics[width=2.4cm]{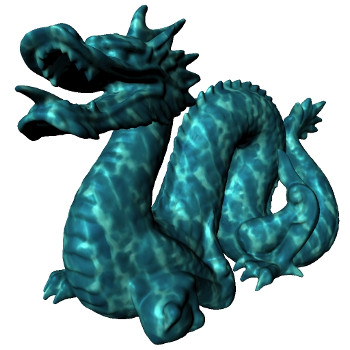} & 
	\includegraphics[width=2.4cm]{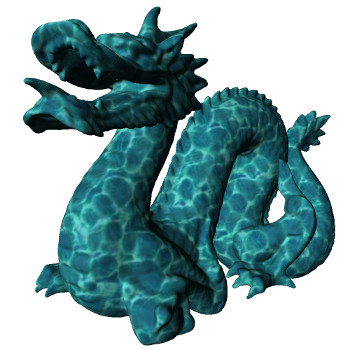}\\
	
	\end{tabular}
	\caption{Comparison of our approach with Kopf~\etal~\cite{kopf2007solid}. Both methods generate a solid texture that is then simply interpolated and intersected with a surface mesh (without parametrization as required for texture mapping). The first column shows the example texture. The second column shows results from~\cite{kopf2007solid}, some of which obtained with additional information (a feature map for the second and fourth rows, a specularity map for the third one). The last column illustrates that our approach, using only the example for training, is able to produce fine scale details. The resolution of the first three rows are $640^3$ and $512^3$ for the rest.
	}
	\label{Fig:Results_dragon}
\end{figure}

On Figure~\ref{Fig:Results_dragon} we show some of our results used for texturing a complex surface and we compare them to the results of Kopf~\etal~\cite{kopf2007solid}. Here the higher frequencies successfully reproduced with our method cause a more realistic impression.   

Finally we would like to point out that although deep learning related models are often thought to produce good results only thanks to a colossal amount of parameters, our method  (with $\sim8.5\times10^4$ parameters to store) stands close to the memory footprint of a patch-based approach working with a $170^2$ color pixels input (\ie $8.67\times10^4$ parameters if all the patches are used).

\section{Limitations and future work}\label{Sec:Limitations}

\paragraph*{Long distance correlation}
As it can be observed in the brick wall texture on Figure~\ref{Fig:Results_views_2_directions} and in the diagonal texture in Figure~\ref{Fig:Results_diagonal_crayons} our model is less successful at preserving the alignment of long patterns in the texture. 
This limitation is also observed in the second row of Figure~\ref{Fig:Results_views_2_directions} where the objects size in the synthesized samples do not match the one in the example, again due to the overlooked long distance correlation. 
One possible explanation comes from the fixed receptive field of VGG as descriptor network. It is likely that it only sees some local patterns which results in breaking long patterns into pieces.
A possible solution could be to use more scales in the generator network, similarly to use larger patches in patch based methods.  
Another possible improvement to explore is to explicitly construct our 2D loss $\mathcal{L}_2$ incorporating those long distance correlations as in \cite{liu2016texture,sendik2017deep}. 

\paragraph*{Constrained directions}
We observed that training the generator with two instead of three constrained directions results in unsatisfying texture along the unconsidered direction, while improving visual quality along the two constrained directions for anisotropic textures (see Figure~\ref{Fig:Results_views_2_directions}).
It would be interesting to explore a middle point between letting the algorithm infer the structure along one direction and constraining it.

\paragraph*{Visual quality}
Although our method delivers high quality results for a varied set of textures, it still presents some visual flaws that we think are independent of the \emph{existence issue}. In textures like the pebble and grass of Figure~\ref{Fig:Results_isotropic2} the synthesized sample presents oversimplified versions of the example's features. 
Although not detailed in the articles, the available codes for~\cite{ulyanov2016texturenets,Ulyanov17improved,li2017diversified} make use of an empirical normalization of the gradients during training. This technique normalizes the gradient of the generator network with respect to the loss at each layer before continuing the back propagation to the generator network's parameters. 
In practice it sometimes leads to a slightly closer reproduction of the patterns' structure of the example. It is however difficult to anticipate which textures can benefit from this technique.

Additionally, our results present some visual artifacts that are typical to generative methods based on deep networks. The most salient are the high frequency checker-board effects, see for instance \cite{johnson2016Perceptual} where a total variation term is used to try to mitigate the artifact.

\paragraph*{Non-stationary textures}
The perceptual loss in Equation~\eqref{eq:slice_loss} stands out for traditional texture synthesis, where the examples are stationary.
An interesting problem is to consider non-stationary textures, such as the recent method of Zhou~\etal~\cite{zhou2018nonstationary}, which uses an auto-encoder to extend the example texture twice. This problem is specifically challenging in our setting in absence of any 3D examples of such a texture.

\paragraph*{Real time rendering}
The trained generator can be integrated in a fragment shader to generate the visible values of a 3D model thanks to its on demand capability. Note however that on-the-fly filtering of the generated solid texture is a challenging problem that is not addressed in this work.

\section{Conclusion}\label{Sec:Conclusion}

The main goal of this paper was to address the problem of example based solid texture synthesis by the means of a convolutional neural network.
First, we presented a simple and compact generative network capable of synthesizing portions of infinitely extendable solid texture.   
The parameters of this 3D generator are stochastically optimized using a pre-trained 2D descriptor network and a slice-based 3D objective function.
The complete framework is efficient both during training and at evaluation time. 
The training can be performed at high resolution, and textures of arbitrary size can be synthesized on demand.  
This method is capable of achieving high quality results on a wide set of textures. We showed the outcome on textures with varying levels of structure and on isotropic and anisotropic arrangements. 
We demonstrate that, although solid texture synthesis from a single example image is an intricate problem, our method delivers compelling results given the desired look imposed via the 3D loss function.

The second aim of this study was to achieve on demand synthesis for which, to the best of our knowledge, no other method based on neural networks is capable of. 
The on demand evaluation capability of the generator allows for it to be integrated with a 3D graphics renderer to replace the use of 2D textures on surfaces and thus eliminating the possible accompanying artifacts.
The proposed techniques during training and evaluation can be extended to any fully convolutional generative network. 
We observed some limitations of our method mainly in the lack of control over the directions not considered in the training. Using multiple examples could complement the training by giving information of the desired aspect along different directions.
We aim to further study the limits of solid texture synthesis from multiple sources with the goal of obtaining an upgraded framework better capable of simulating real life objects.   


\section*{Aknowledgments} 
We acknowledge the support of the Natural Sciences and Engineering Research Council of Canada (NSERC), RGPIN-2015-06025.
This project has been carried out with support from the French
State, managed by the French National Research Agency (ANR-16-CE33-0010-01).
This study has been also carried out with financial support from 
the CNRS for supporting grants PEPS 2018 I3A "3DTextureNets". 
In like manner, we acknowledge the support of CONACYT.
Bruno Galerne acknowledges the support of NVIDIA Corporation with the donation of a Titan Xp GPU used for this research.
The authors would like to thank Loïc Simon for fruitful discussions on 3D rendering, and Guillaume-Alexandre Bilodeau for giving us access to one of his GPUs.

\bibliographystyle{eg-alpha-doi}

\bibliography{cgf2018.bib}

\newcommand{\etalchar}[1]{$^{#1}$}
\begin{thebibliography}{\uppercase{GPAM{\etalchar{*}}14}}

\bibitem[BJV17]{bergmann17apsGAN}
\textsc{Bergmann U., Jetchev N., Vollgraf R.}:
\newblock Learning texture manifolds with the periodic spatial {GAN}.
\newblock In \emph{ICML} (2017), pp.~469--477.

\bibitem[BK88]{bourlard1988auto}
\textsc{Bourlard H., Kamp Y.}:
\newblock Auto-association by multilayer perceptrons and singular value
  decomposition.
\newblock \emph{Biological cybernetics}, 4 (1988), 291--294.
\newblock \href {http://dx.doi.org/10.1007/BF00332918}
  {\path{doi:10.1007/BF00332918}}.

\bibitem[BM17]{berger2017incorporating}
\textsc{Berger G., Memisevic R.}:
\newblock Incorporating long-range consistency in cnn-based texture generation.
\newblock \emph{ICLR} (2017).

\bibitem[CW10]{chen2010highquality}
\textsc{Chen J., Wang B.}:
\newblock High quality solid texture synthesis using position and index
  histogram matching.
\newblock \emph{Vis. Comput. 26}, 4 (2010), 253--262.
\newblock \href {http://dx.doi.org/10.1007/s00371-009-0408-3}
  {\path{doi:10.1007/s00371-009-0408-3}}.

\bibitem[{De }97]{debonet1997multiresolution}
\textsc{{De Bonet} J.~S.}:
\newblock Multiresolution sampling procedure for analysis and synthesis of
  texture images.
\newblock In \emph{SIGGRAPH} (1997), ACM, pp.~361--368.
\newblock \href {http://dx.doi.org/10.1145/258734.258882}
  {\path{doi:10.1145/258734.258882}}.

\bibitem[DGF98]{dischler1998anisotropic}
\textsc{Dischler J.~M., Ghazanfarpour D., Freydier R.}:
\newblock Anisotropic solid texture synthesis using orthogonal 2d views.
\newblock \emph{Computer Graphics Forum 17}, 3 (1998), 87--95.
\newblock \href {http://dx.doi.org/10.1111/1467-8659.00256}
  {\path{doi:10.1111/1467-8659.00256}}.

\bibitem[DLTD08]{dong2008lazy}
\textsc{Dong Y., Lefebvre S., Tong X., Drettakis G.}:
\newblock Lazy solid texture synthesis.
\newblock In \emph{EGSR} (2008), pp.~1165--1174.
\newblock \href {http://dx.doi.org/10.1111/j.1467-8659.2008.01254.x}
  {\path{doi:10.1111/j.1467-8659.2008.01254.x}}.

\bibitem[GCC{\etalchar{*}}17]{gwak2017weakly2}
\textsc{Gwak J., Choy C.~B., Chandraker M., Garg A., Savarese S.}:
\newblock Weakly supervised 3d reconstruction with adversarial constraint.
\newblock In \emph{International Conference on 3D Vision} (2017), pp.~263--272.
\newblock \href {http://dx.doi.org/10.1109/3DV.2017.00038}
  {\path{doi:10.1109/3DV.2017.00038}}.

\bibitem[GD95]{ghazanfarpour1995Spectral}
\textsc{Ghazanfarpour D., Dischler J.}:
\newblock Spectral analysis for automatic 3-d texture generation.
\newblock \emph{Computers \& Graphics 19}, 3 (1995), 413--422.
\newblock \href {http://dx.doi.org/10.1016/0097-8493(95)00011-Z}
  {\path{doi:10.1016/0097-8493(95)00011-Z}}.

\bibitem[GEB15]{gatys2015cnn}
\textsc{Gatys L., Ecker A.~S., Bethge M.}:
\newblock Texture synthesis using convolutional neural networks.
\newblock In \emph{NIPS} (2015), pp.~262 -- 270.

\bibitem[GEB16]{Gatys_2016_CVPR}
\textsc{Gatys L.~A., Ecker A.~S., Bethge M.}:
\newblock Image style transfer using convolutional neural networks.
\newblock In \emph{CVPR} (2016), pp.~2414--2423.

\bibitem[GLLD12]{GLLD_Gabor_noise_by_example_2012}
\textsc{Galerne B., Lagae A., Lefebvre S., Drettakis G.}:
\newblock {G}abor noise by example.
\newblock \emph{ACM Trans. Graph. 31}, 4 (2012), 73:1--73:9.
\newblock \href {http://dx.doi.org/10.1145/2185520.2185569}
  {\path{doi:10.1145/2185520.2185569}}.

\bibitem[GLM17]{galerne2017texton}
\textsc{Galerne B., Leclaire A., Moisan L.}:
\newblock Texton noise.
\newblock \emph{Computer Graphics Forum 36}, 8 (2017), 205--218.
\newblock \href {http://dx.doi.org/10.1111/cgf.13073}
  {\path{doi:10.1111/cgf.13073}}.

\bibitem[GLR18]{galerne2018texture}
\textsc{Galerne B., Leclaire A., Rabin J.}:
\newblock A texture synthesis model based on semi-discrete optimal transport in
  patch space.
\newblock \emph{SIAM 11}, 4 (2018), 2456--2493.
\newblock \href {http://dx.doi.org/10.1137/18M1175781}
  {\path{doi:10.1137/18M1175781}}.

\bibitem[GPAM{\etalchar{*}}14]{goodfellow2014generative}
\textsc{Goodfellow I., Pouget-Abadie J., Mirza M., Xu B., Warde-Farley D.,
  Ozair S., Courville A., Bengio Y.}:
\newblock Generative adversarial nets.
\newblock In \emph{NIPS} (2014), pp.~2672--2680.

\bibitem[GRGH17]{Gutierrez2017}
\textsc{Gutierrez J., Rabin J., Galerne B., Hurtut T.}:
\newblock Optimal patch assignment for statistically constrained texture
  synthesis.
\newblock In \emph{SSVM} (2017), pp.~172--183.
\newblock \href {http://dx.doi.org/10.1007/978-3-319-58771-4_14}
  {\path{doi:10.1007/978-3-319-58771-4_14}}.

\bibitem[GSV{\etalchar{*}}14]{Gilet_et_al_2014_local_random_phase_noise_2014}
\textsc{Gilet G., Sauvage B., Vanhoey K., Dischler J.-M., Ghazanfarpour D.}:
\newblock Local random-phase noise for procedural texturing.
\newblock \emph{ACM Trans. Graph. 33}, 6 (2014), 195:1--195:11.
\newblock \href {http://dx.doi.org/10.1145/2661229.2661249}
  {\path{doi:10.1145/2661229.2661249}}.

\bibitem[HB95]{heeger1995pyramid}
\textsc{Heeger D.~J., Bergen J.~R.}:
\newblock Pyramid-based texture analysis/synthesis.
\newblock In \emph{SIGGRAPH} (1995), ACM, pp.~229--238.
\newblock \href {http://dx.doi.org/10.1145/218380.218446}
  {\path{doi:10.1145/218380.218446}}.

\bibitem[JAFF16]{johnson2016Perceptual}
\textsc{Johnson J., Alahi A., Fei-Fei L.}:
\newblock Perceptual losses for real-time style transfer and super-resolution.
\newblock In \emph{European Conference on Computer Vision} (2016),
  pp.~694--711.
\newblock \href {http://dx.doi.org/10.1007/978-3-319-46475-6_43}
  {\path{doi:10.1007/978-3-319-46475-6_43}}.

\bibitem[JREM{\etalchar{*}}16]{jimenez2016unsupervised}
\textsc{Jimenez~Rezende D., Eslami S. M.~A., Mohamed S., Battaglia P.,
  Jaderberg M., Heess N.}:
\newblock Unsupervised learning of 3d structure from images.
\newblock In \emph{NIPS} (2016), pp.~4996--5004.

\bibitem[KB15]{kingma2014adam}
\textsc{Kingma D.~P., Ba J.}:
\newblock Adam: {A} method for stochastic optimization.
\newblock In \emph{ICLR} (2015).

\bibitem[KEBK05]{kwatra2005texture}
\textsc{Kwatra V., Essa I., Bobick A., Kwatra N.}:
\newblock Texture optimization for example-based synthesis.
\newblock In \emph{SIGGRAPH} (2005), ACM, pp.~795--802.
\newblock \href {http://dx.doi.org/10.1145/1186822.1073263}
  {\path{doi:10.1145/1186822.1073263}}.

\bibitem[KFCO{\etalchar{*}}07]{kopf2007solid}
\textsc{Kopf J., Fu C., Cohen-Or D., Deussen O., Lischinski D., Wong T.}:
\newblock Solid texture synthesis from 2d exemplars.
\newblock In \emph{SIGGRAPH} (2007), ACM.
\newblock \href {http://dx.doi.org/10.1145/1275808.1276380}
  {\path{doi:10.1145/1275808.1276380}}.

\bibitem[LeC87]{yann1987modeles}
\textsc{LeCun Y.}:
\newblock \emph{Modeles connexionnistes de lapprentissage}.
\newblock PhD thesis, Universite Paris 6, 1987.

\bibitem[LFY{\etalchar{*}}17a]{li2017diversified}
\textsc{Li Y., Fang C., Yang J., Wang Z., Lu X., Yang M.}:
\newblock Diversified texture synthesis with feed-forward networks.
\newblock In \emph{CVPR} (2017), pp.~266--274.
\newblock \href {http://dx.doi.org/10.1109/CVPR.2017.36}
  {\path{doi:10.1109/CVPR.2017.36}}.

\bibitem[LFY{\etalchar{*}}17b]{li2017universalstyle}
\textsc{Li Y., Fang C., Yang J., Wang Z., Lu X., Yang M.-H.}:
\newblock Universal style transfer via feature transforms.
\newblock In \emph{NIPS} (2017), pp.~386--396.

\bibitem[LGX16]{liu2016texture}
\textsc{Liu G., Gousseau Y., Xia G.}:
\newblock Texture synthesis through convolutional neural networks and spectrum
  constraints.
\newblock In \emph{ICPR} (2016), pp.~3234--3239.
\newblock \href {http://dx.doi.org/10.1109/ICPR.2016.7900133}
  {\path{doi:10.1109/ICPR.2016.7900133}}.

\bibitem[LH05]{lefebvre2005parallel}
\textsc{Lefebvre S., Hoppe H.}:
\newblock Parallel controllable texture synthesis.
\newblock In \emph{SIGGRAPH} (2005), ACM, pp.~777--786.
\newblock \href {http://dx.doi.org/10.1145/1186822.1073261}
  {\path{doi:10.1145/1186822.1073261}}.

\bibitem[LVU18]{ulyanov18deep}
\textsc{Lempitsky V., Vedaldi A., Ulyanov D.}:
\newblock Deep image prior.
\newblock In \emph{CVPR} (2018), pp.~9446--9454.
\newblock \href {http://dx.doi.org/10.1109/CVPR.2018.00984}
  {\path{doi:10.1109/CVPR.2018.00984}}.

\bibitem[Mar03]{marsaglia2003xorshift}
\textsc{Marsaglia G.}:
\newblock Xorshift rngs.
\newblock \emph{Journal of Statistical Software 8}, 14 (2003), 1--6.
\newblock \href {http://dx.doi.org/10.18637/jss.v008.i14}
  {\path{doi:10.18637/jss.v008.i14}}.

\bibitem[Pea85]{peachey1985solid}
\textsc{Peachey D.~R.}:
\newblock Solid texturing of complex surfaces.
\newblock \emph{SIGGRAPH} (1985), 279--286.
\newblock \href {http://dx.doi.org/10.1145/325165.325246}
  {\path{doi:10.1145/325165.325246}}.

\bibitem[Per85]{perlin1985image}
\textsc{Perlin K.}:
\newblock An image synthesizer.
\newblock \emph{SIGGRAPH} (1985), 287--296.
\newblock \href {http://dx.doi.org/10.1145/325165.325247}
  {\path{doi:10.1145/325165.325247}}.

\bibitem[PS00]{portilla2000parametric}
\textsc{Portilla J., Simoncelli E.~P.}:
\newblock A parametric texture model based on joint statistics of complex
  wavelet coefficients.
\newblock \emph{IJCV 40}, 1 (2000), 49--71.
\newblock \href {http://dx.doi.org/10.1023/A:1026553619983}
  {\path{doi:10.1023/A:1026553619983}}.

\bibitem[QhY07]{qin2007aura}
\textsc{Qin X., h.~Yang Y.}:
\newblock Aura 3d textures.
\newblock \emph{Transactions on Visualization and Computer Graphics 13}, 2
  (2007), 379--389.
\newblock \href {http://dx.doi.org/10.1109/TVCG.2007.31}
  {\path{doi:10.1109/TVCG.2007.31}}.

\bibitem[RMC16]{radford2016unsupervised}
\textsc{Radford A., Metz L., Chintala S.}:
\newblock Unsupervised representation learning with deep convolutional
  generative adversarial networks.
\newblock \emph{ICLR} (2016).

\bibitem[RPDB12]{rabin2011wasserstein}
\textsc{Rabin J., Peyr\'{e} G., Delon J., Bernot M.}:
\newblock {W}asserstein barycenter and its application to texture mixing.
\newblock In \emph{SSVM} (2012), pp.~435--446.
\newblock \href {http://dx.doi.org/10.1007/978-3-642-24785-9_37}
  {\path{doi:10.1007/978-3-642-24785-9_37}}.

\bibitem[SCO17]{sendik2017deep}
\textsc{Sendik O., Cohen-Or D.}:
\newblock Deep correlations for texture synthesis.
\newblock \emph{ACM Trans. Graph. 36}, 5 (2017), 161:1--161:15.
\newblock \href {http://dx.doi.org/10.1145/3015461}
  {\path{doi:10.1145/3015461}}.

\bibitem[SZ14]{Simonyan14c}
\textsc{Simonyan K., Zisserman A.}:
\newblock Very deep convolutional networks for large-scale image recognition.
\newblock \emph{CoRR} (2014).
\newblock \href {http://arxiv.org/abs/1409.1556} {\path{arXiv:1409.1556}}.

\bibitem[TBD18]{tesfaldet2018}
\textsc{Tesfaldet M., Brubaker M.~A., Derpanis K.~G.}:
\newblock Two-stream convolutional networks for dynamic texture synthesis.
\newblock In \emph{CVPR} (2018), pp.~6703--6712.
\newblock \href {http://dx.doi.org/10.1109/CVPR.2018.00701}
  {\path{doi:10.1109/CVPR.2018.00701}}.

\bibitem[ULVL16]{ulyanov2016texturenets}
\textsc{Ulyanov D., Lebedev V., Vedaldi A., Lempitsky V.}:
\newblock Texture networks: Feed-forward synthesis of textures and stylized
  images.
\newblock In \emph{ICML} (2016), pp.~1349--1357.

\bibitem[UVL17]{Ulyanov17improved}
\textsc{Ulyanov D., Vedaldi A., Lempitsky V.}:
\newblock Improved texture networks: Maximizing quality and diversity in
  feed-forward stylization and texture synthesis.
\newblock In \emph{CVPR} (2017), pp.~4105--4113.
\newblock \href {http://dx.doi.org/10.1109/CVPR.2017.437}
  {\path{doi:10.1109/CVPR.2017.437}}.

\bibitem[Wei03]{wei2003multiplesources}
\textsc{Wei L.-Y.}:
\newblock Texture synthesis from multiple sources.
\newblock In \emph{SIGGRAPH} (2003), ACM, pp.~1--1.
\newblock URL: \url{http://doi.acm.org/10.1145/965400.965507}, \href
  {http://dx.doi.org/10.1145/965400.965507} {\path{doi:10.1145/965400.965507}}.

\bibitem[WL00]{wei_levoy_2000}
\textsc{Wei L.~Y., Levoy M.}:
\newblock Fast texture synthesis using tree-structured vector quantization.
\newblock In \emph{SIGGRAPH} (2000), ACM, pp.~479--488.
\newblock \href {http://dx.doi.org/10.1145/344779.345009}
  {\path{doi:10.1145/344779.345009}}.

\bibitem[WL03]{wei2003order}
\textsc{Wei L.-Y., Levoy M.}:
\newblock Order-independent texture synthesis.
\newblock \emph{Tech. Rep. TR-2002-01, Computer Science Department, Stanford
  University} (2003).

\bibitem[WRB17]{wilmot2017stable}
\textsc{Wilmot P., Risser E., Barnes C.}:
\newblock Stable and controllable neural texture synthesis and style transfer
  using histogram losses.
\newblock \emph{CoRR} (2017).
\newblock \href {http://arxiv.org/abs/1701.08893} {\path{arXiv:1701.08893}}.

\bibitem[YBS{\etalchar{*}}19]{yu2019texture}
\textsc{Yu N., Barnes C., Shechtman E., Amirghodsi S., Lukac M.}:
\newblock Texture mixer: A network for controllable synthesis and interpolation
  of texture.
\newblock In \emph{CVPR} (June 2019).

\bibitem[YYY{\etalchar{*}}16]{yan2016perspective}
\textsc{Yan X., Yang J., Yumer E., Guo Y., Lee H.}:
\newblock Perspective transformer nets: Learning single-view 3d object
  reconstruction without 3d supervision.
\newblock In \emph{NIPS} (2016), pp.~1696--1704.

\bibitem[ZDL{\etalchar{*}}11]{zhang2011sketch}
\textsc{Zhang G.-X., Du S.-P., Lai Y.-K., Ni T., Hu S.-M.}:
\newblock Sketch guided solid texturing.
\newblock \emph{Graphical Models 73}, 3 (2011), 59--73.
\newblock \href {http://dx.doi.org/https://doi.org/10.1016/j.gmod.2010.10.006}
  {\path{doi:https://doi.org/10.1016/j.gmod.2010.10.006}}.

\bibitem[ZZB{\etalchar{*}}18]{zhou2018nonstationary}
\textsc{Zhou Y., Zhu Z., Bai X., Lischinski D., Cohen-Or D., Huang H.}:
\newblock Non-stationary texture synthesis by adversarial expansion.
\newblock \emph{ACM Trans. Graph. 37}, 4 (2018), 49:1--49:13.
\newblock \href {http://dx.doi.org/10.1145/3197517.3201285}
  {\path{doi:10.1145/3197517.3201285}}.

\end{thebibliography}

\end{document}